\documentclass[aps,preprint,showpacs,superscriptaddress,groupedaddress]{revtex4-1}  

\usepackage{graphicx}  
\usepackage{dcolumn}   
\usepackage{bm}        
\usepackage{amssymb}   
\usepackage{amsmath}   
\usepackage{listings}
\usepackage{array}
\usepackage{graphicx}
\usepackage{lineno}
\usepackage{dcolumn}
\usepackage{color}
\usepackage{overpic}
\usepackage{multirow}
\usepackage{hyperref}

\newcommand{\bo}{\mathcal{O}}
\newcommand{\ttil}{\tilde{t}}
\newcommand{\qtil}{\tilde{q}}
\newcommand{\nbins}{N_{\text{bins}}}

\newcommand{\mL}{\mathcal{L}}

\newcommand{\hatmu}{\hat{\mu}}

\newcommand{\bV}{\bold{V}}
\newcommand{\bH}{\bold{H}}
\newcommand{\bA}{\bold{A}}
\newcommand{\bI}{\mathbf{1}}
\newcommand{\bx}{\bold{x}}

\newcommand{\bB}{\bold{B}}

\newcommand{\mup}{\mu^\prime}
\newcommand{\sigmap}{\sigma^\prime}
\newcommand{\muH}{\mu_{H}}
\newcommand{\nsmall}{n_{\text{small}}}
\newcommand{\ls}{\text{SS}}
\newcommand{\LS}{\text{LS}}
\newcommand{\wald}{\text{Wald}}
\newcommand{\tot}{\text{tot.}}
\newcommand{\classic}{\text{classic}}
\newcommand{\binned}{\text{binned}}
\newcommand{\nsysts}{N_\text{systs}}

\begin{document}

\title{Improved Asymptotic Formulae for Statistical Interpretation Based on Likelihood Ratio Tests}
\author{Li-Gang Xia\footnote{ligang.xia@cern.ch}, Yan Zhang}
\affiliation{School of Physics, Nanjing University \\ No. 22 Hankou Road, Nanjing, China}

\begin{abstract}
    In this work, we attempt to refine the classic asymptotic formulae to describe the probability distribution of likelihood-ratio statistical tests. The idea is to split the probability distribution function into two parts.
    One part is universal and described by the asymptotic formulae.
    The other part is case-dependent and is estimated explicitly using a 6-bin model proposed in this work. The latter is similar to performing toy simulations and can therefore predict the discrete structures in the probability distributions.
    The new asymptotic formulae provide a much better differential description of the test statistics. This improved performance is demonstrated in two toy examples for common likelihood ratio statistics.
\end{abstract}
\maketitle 

\section{Introduction}
Searching for new physics is always the goal for most experimenters in particle physics, especially after the discovery of the Higgs boson~\cite{higgs1,higgs2}. Once a measurement is completed, it is important to report the results in a precise and well-accepted way. If no significant signal is observed, one often reports two things. One is the probability that the observation is due to the fluctuation of known backgrounds. 
This represents the statistical significance of a signal to establish its discovery. The other is the parameter space about the new signal that the measurement can exclude for a given confidence level (C.L.).
We usually build a test statistic based on the likelihood ratio to interpret the results because it is the most powerful discriminant. To determine the statistical significance and the exclusion limits, we need to know the probability distribution of the statistical test
under many hypotheses with different signal strengths or other parameters of interest (POI). We can resort to toy Monte Carlo (MC) simulation. However, this method is usually computationally expensive.

Fortunately, asymptotic formulae have been found in Ref.~\cite{asimov} to describe the distribution of the likelihood ratio tests when the sample size is big enough. Therefore, one can easily obtain the expected statistical significance and exclusion limits for a new signal based on the idea of ``Asimov'' dataset~\cite{asimov}. The validity of these formulae is based on a theorem by Wald~\cite{Wald} and the condition is that the sample size is sufficiently large. 
Recently, one of the authors has finished a study on the feasibility of searching for leptoquarks in Pb-Pb ultra-peripheral collisions~\cite{LQxia} and the background level in this case is very low (the expected number of background events is much less than one). The current work is motivated by the desire to improve the asymptotic formulae via refining the approximation of the test statistics' distributions. 
Ref.~\cite{higher_order} reports another effort to make the asymptotic formulae work better via modifying the test statistics themselves.

In Sec.~\ref{sec:review}, we will have a brief review about the test statistic and the classic asymptotic formulae. In Sec.~\ref{sec:new}, we will elaborate the idea for improvement and present the new formulae. 
The classic and new asymptotic formulae are compared using two examples in Sec.~\ref{sec:example}. 
Sec.~\ref{sec:summary} is a short summary. We note that Appendix~\ref{app:real_phys_app} presents the applications of the new method to two actual physics analyses with some supporting codes in our GitHub repository~\cite{githubxia}. 

\section{Review of the test statistic and the asymptotic formulae}\label{sec:review}
We will review the test statistics and the asymptotic formulae according to Ref.~\cite{asimov}. To test a hypothesis with the signal strength $\mu$, we consider the likelihood ratio
\begin{equation}
    \lambda(\mu) = \frac{\mL(\mu,\hat{\hat{\bm{\theta}}}(\mu))}{\mL(\hatmu,\hat{\bm{\theta}})} \: , 
\end{equation}
where $\bm{\theta}$ denotes a set of nuisance parameters; $\hatmu$ and $\hat{\bm{\theta}}$ are the optimal values to maximize the likelihood function; $\hat{\hat{\bm{\theta}}}(\mu)$ are the optimal values with $\mu$ fixed and can be seen as functions of $\mu$.
Based on this ratio, six test statistics~\cite{asimov}, namely, $t_0, q_0, t_\mu, \ttil_\mu, q_\mu$ and $\qtil_\mu$, are defined for different purposes. They are summarized in Table~\ref{tab:6t}.

\begin{figure}[htbp]
     \centering
     \includegraphics[width=0.45\textwidth]{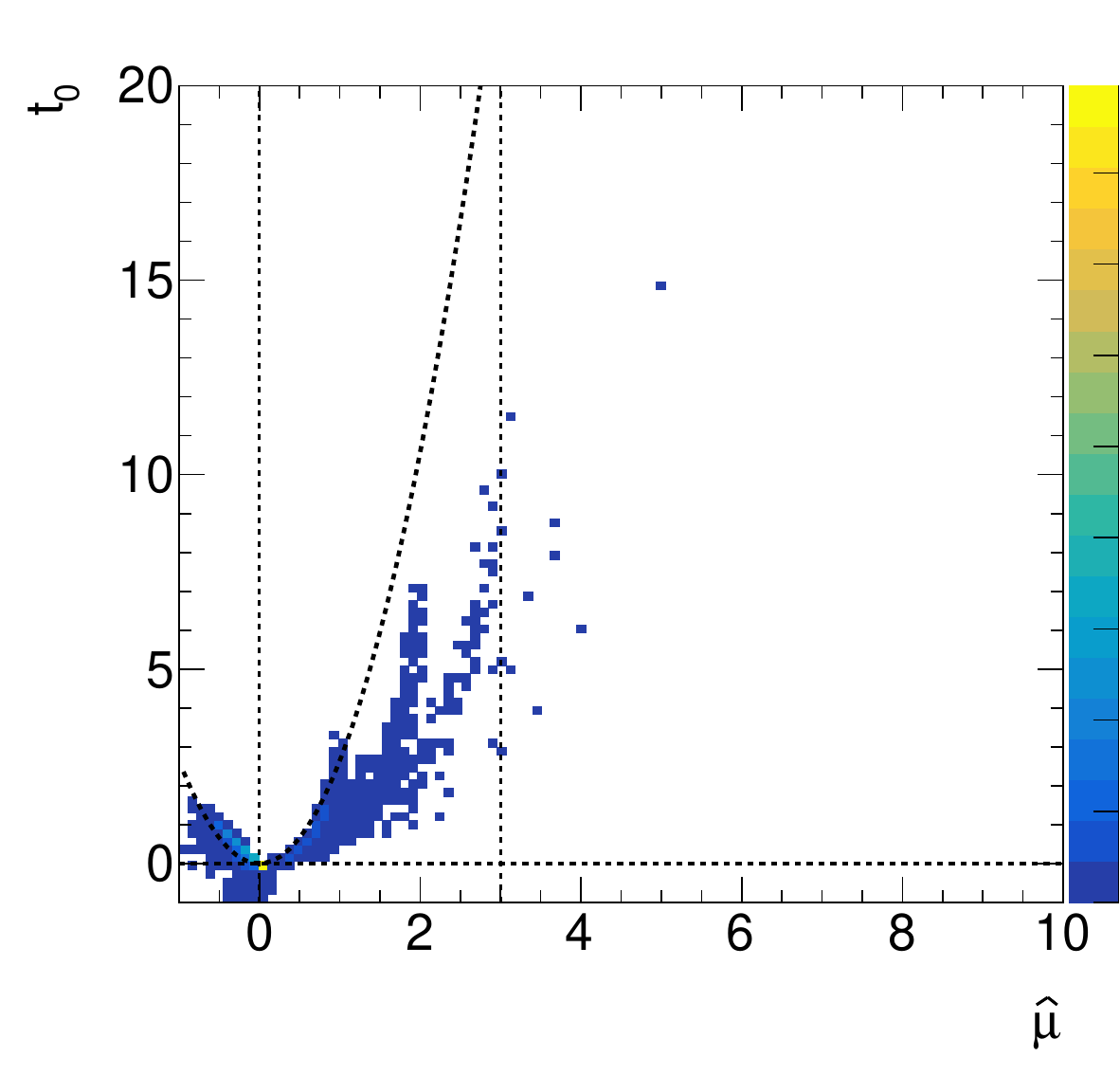}
     \includegraphics[width=0.45\textwidth]{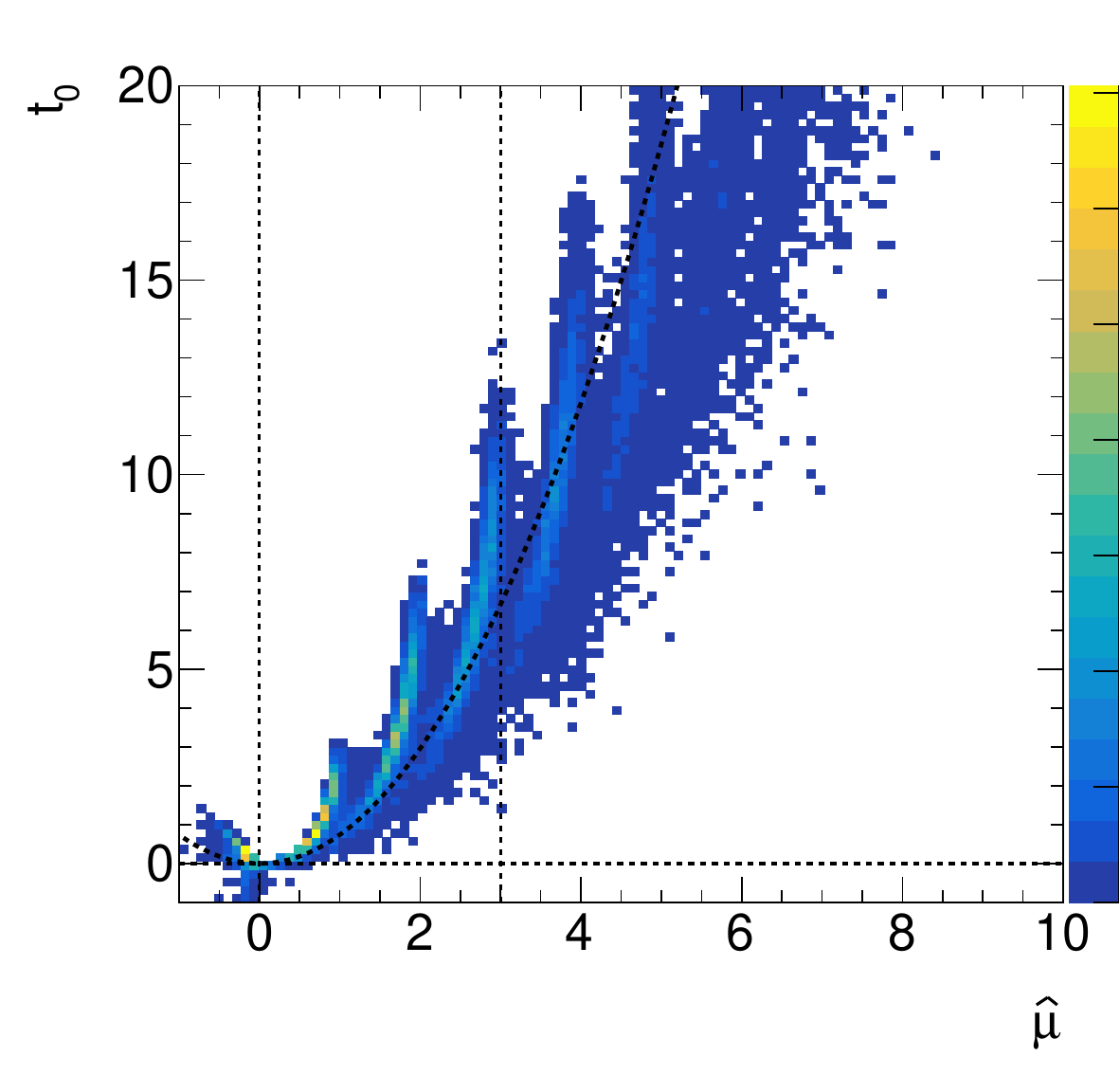}
     \caption{\label{fig:T1}
     The distribution of $t_0:\hatmu$ in Ex.~0 of Sec.~\ref{sec:example} for the hypothesized signal strength $\mu_H=0$ (L) and $\mu_H=3$ (R). The bold dashed curves represent the asymptotic relation from Wald's approximation. 
     }
 \end{figure}

\begin{figure}[htbp]
     \centering
     \includegraphics[width=0.45\textwidth]{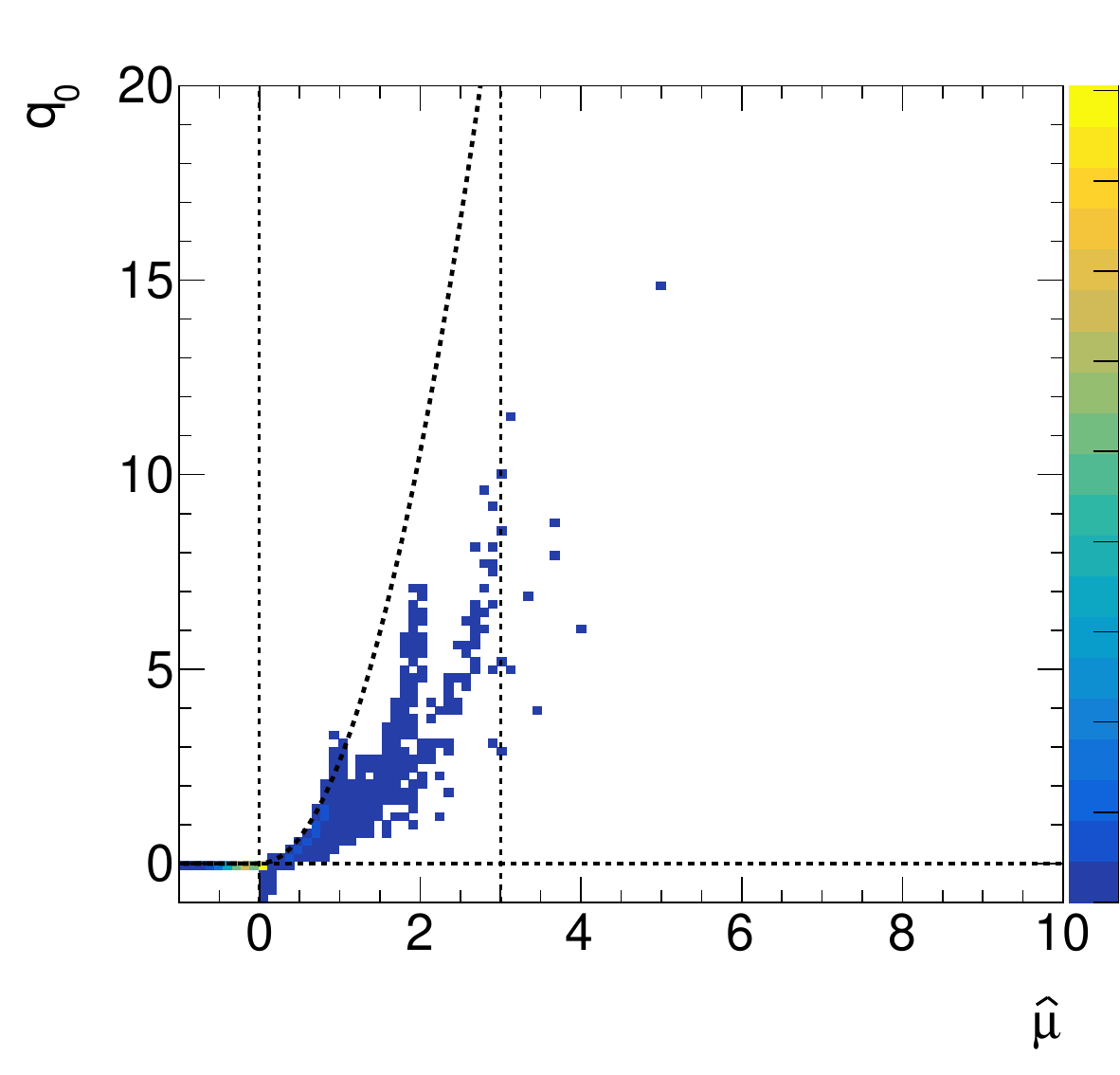}
     \includegraphics[width=0.45\textwidth]{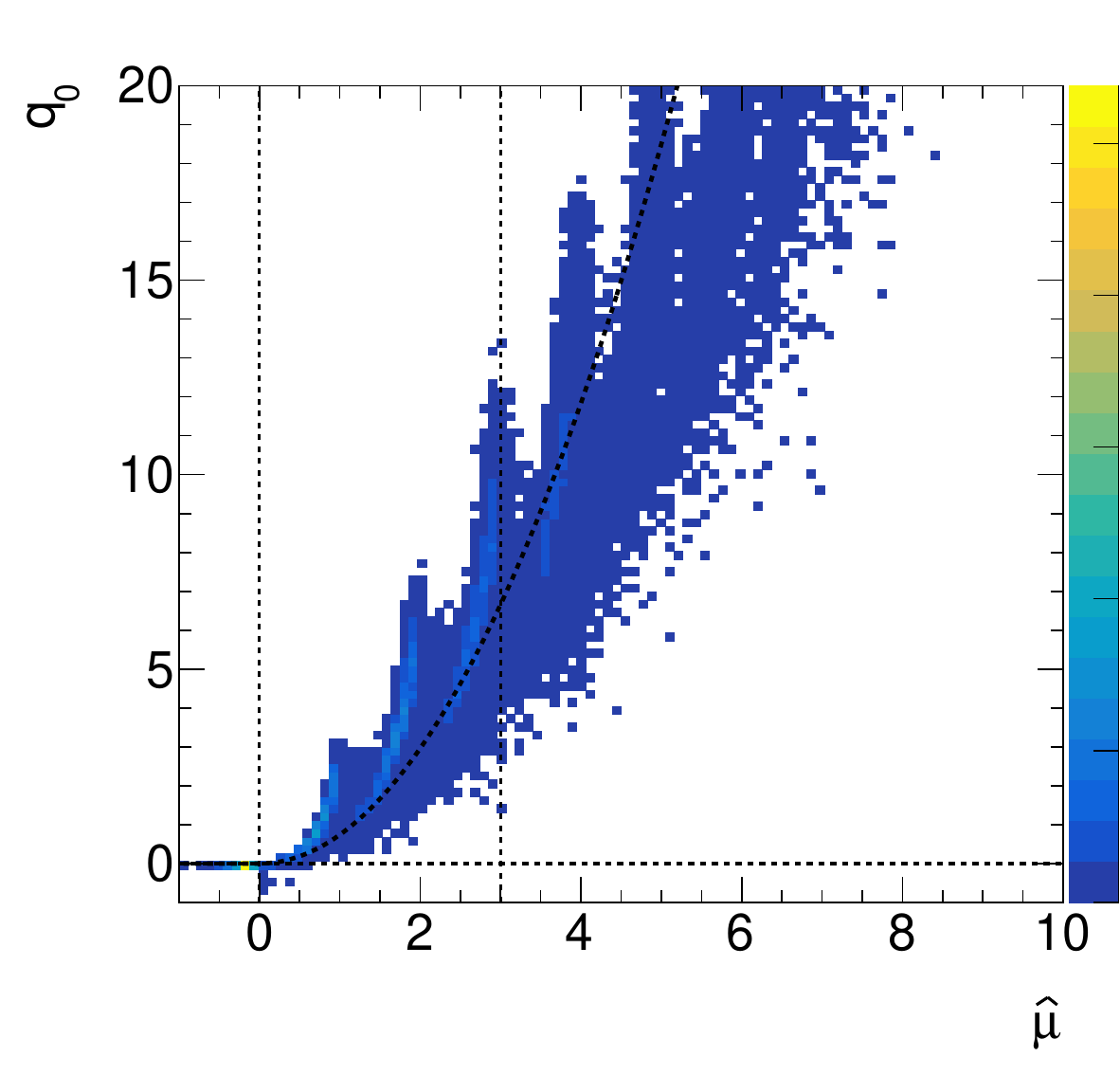}
     \caption{\label{fig:T4}
     The distribution of $q_0:\hatmu$ in Ex.~0 of Sec.~\ref{sec:example} for the hypothesized signal strength $\mu_H=0$ (L) and $\mu_H=3$ (R). The bold dashed curves represent the asymptotic relation from Wald's approximation. 
     }
 \end{figure}

\begin{figure}[htbp]
     \centering
     \includegraphics[width=0.45\textwidth]{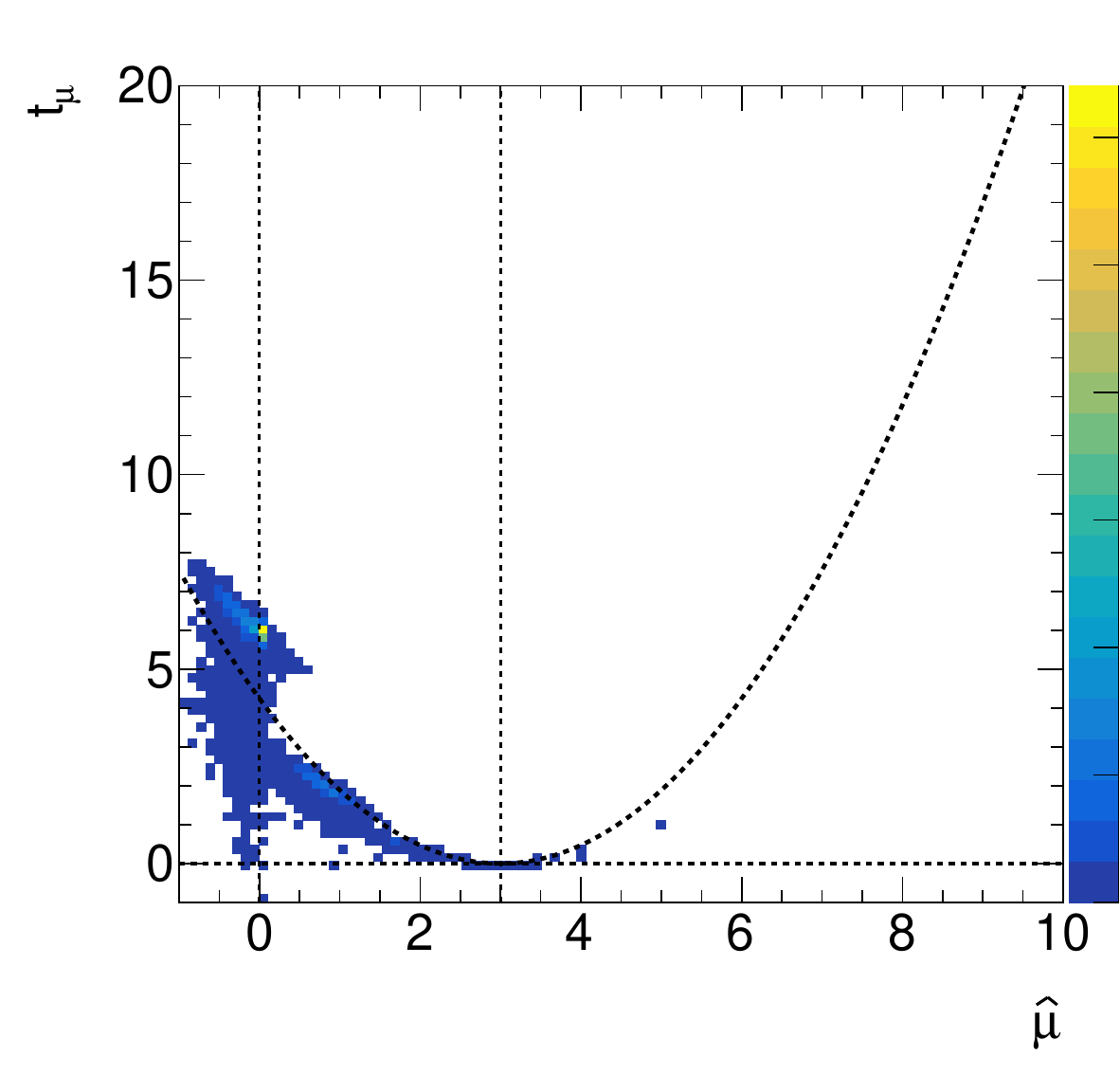}
     \includegraphics[width=0.45\textwidth]{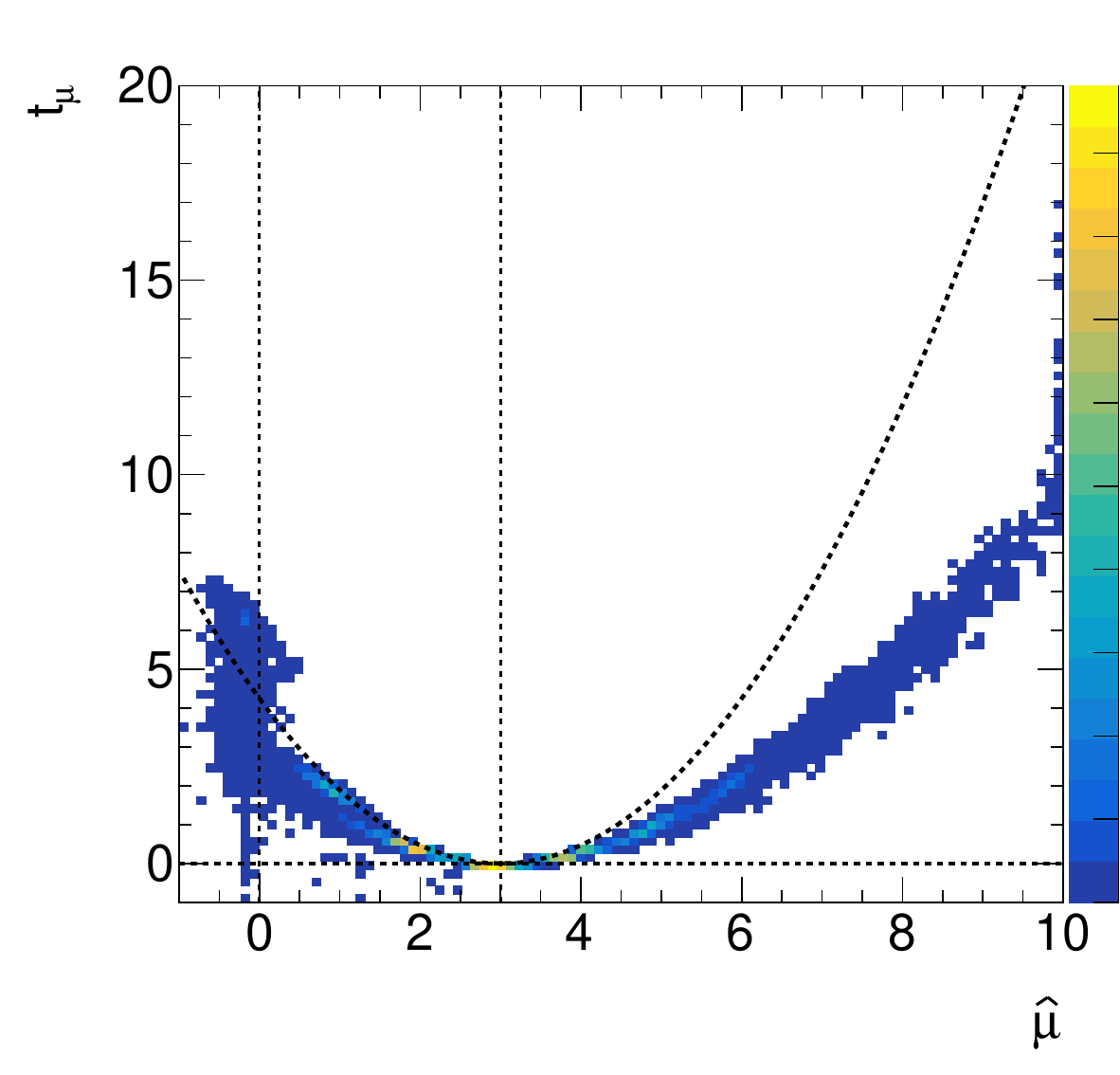}
     \caption{\label{fig:T2}
     The distribution of $t_{\mu}:\hatmu$ in Ex.~0 of Sec.~\ref{sec:example} for the hypothesized signal strength $\mu_H=0$ (L) and $\mu_H=3$ (R). The bold dashed curves represent the asymptotic relation from Wald's approximation. 
     }
 \end{figure}

\begin{figure}[htbp]
     \centering
     \includegraphics[width=0.45\textwidth]{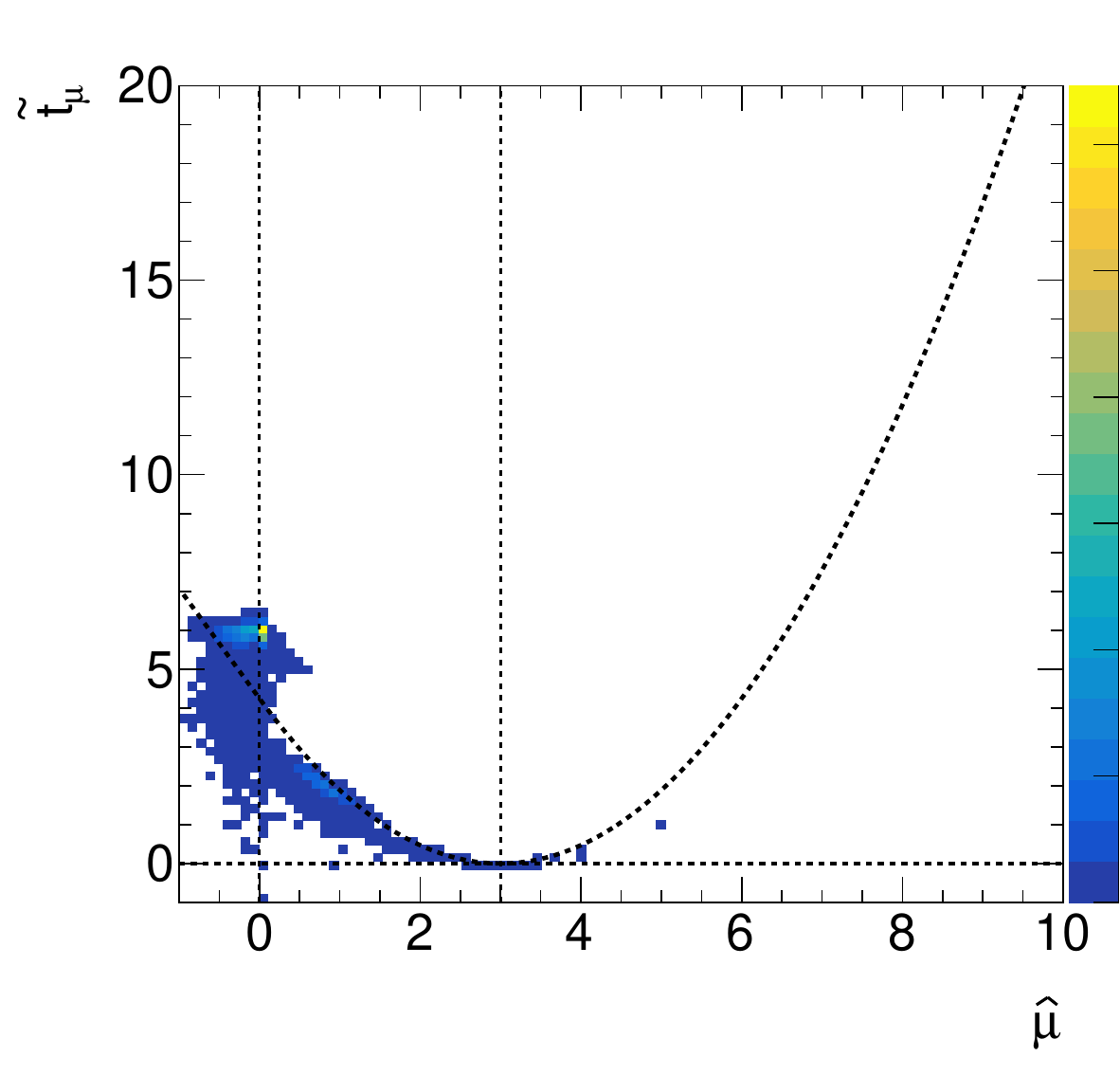}
     \includegraphics[width=0.45\textwidth]{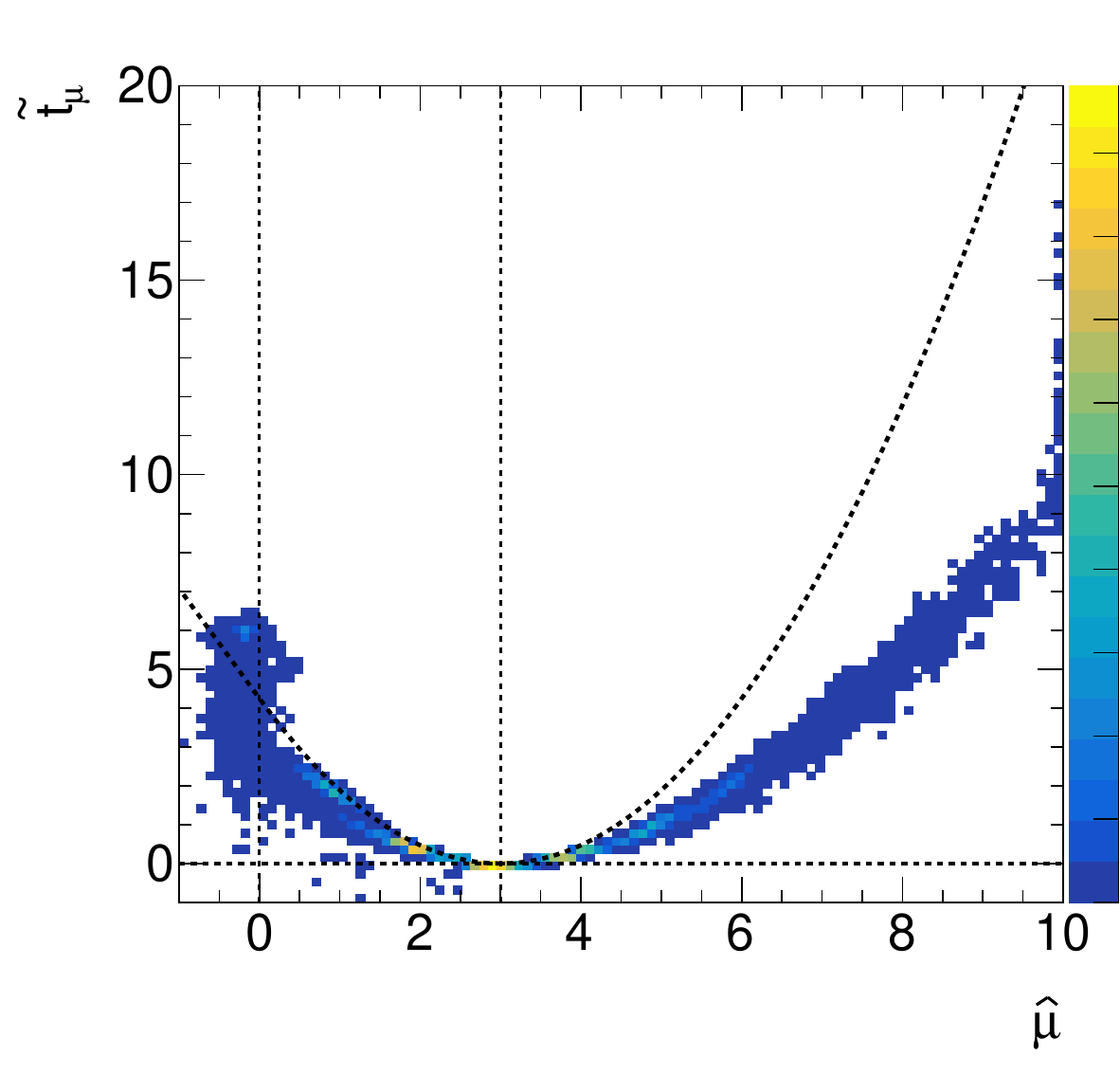}
     \caption{\label{fig:T5}
     The distribution of $\ttil_{\mu}:\hatmu$ in Ex.~0 of Sec.~\ref{sec:example} for the hypothesized signal strength $\mu_H=0$ (L) and $\mu_H=3$ (R). The bold dashed curves represent the asymptotic relation from Wald's approximation. 
     }
 \end{figure}

\begin{figure}[htbp]
     \centering
     \includegraphics[width=0.45\textwidth]{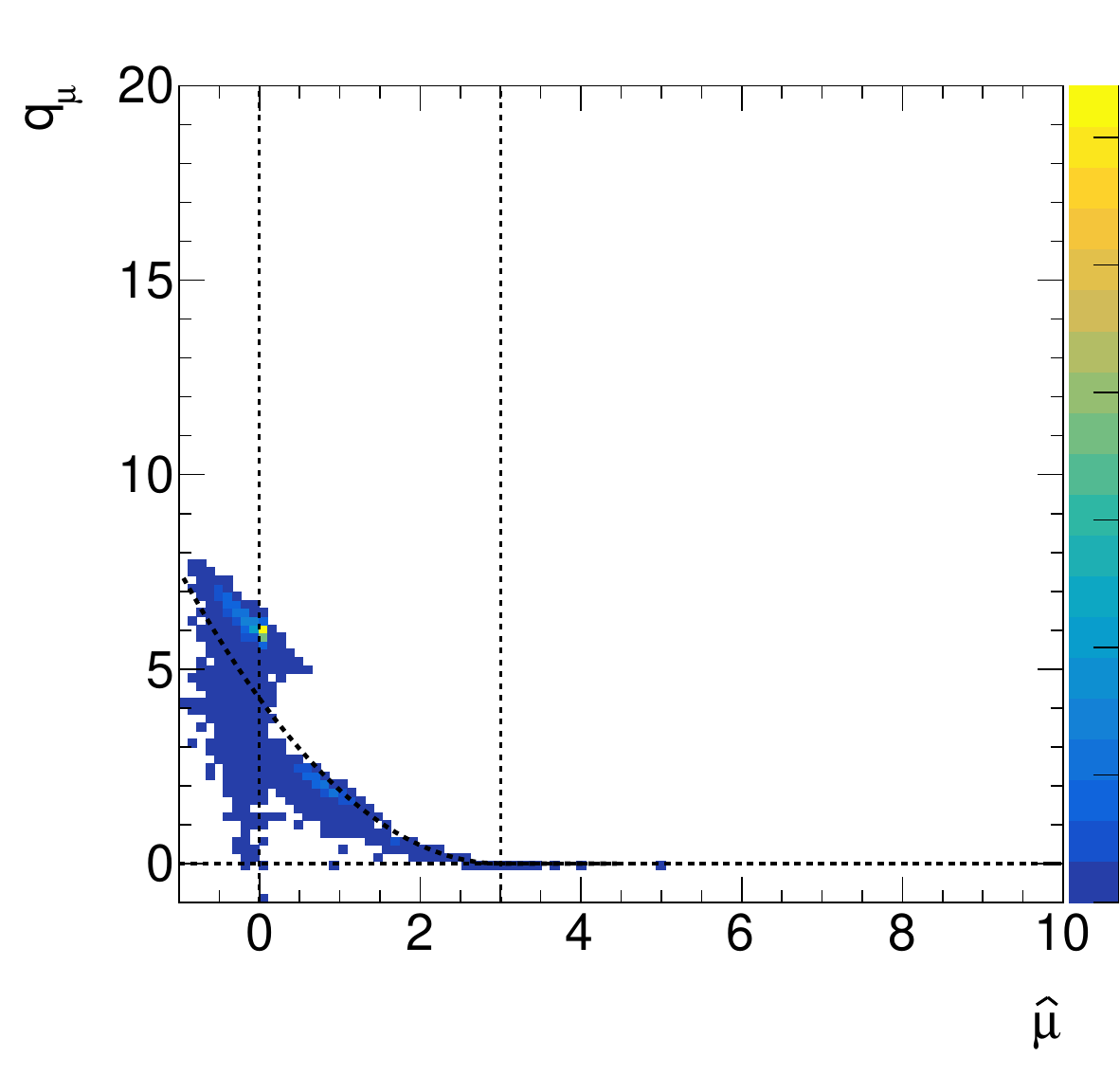}
     \includegraphics[width=0.45\textwidth]{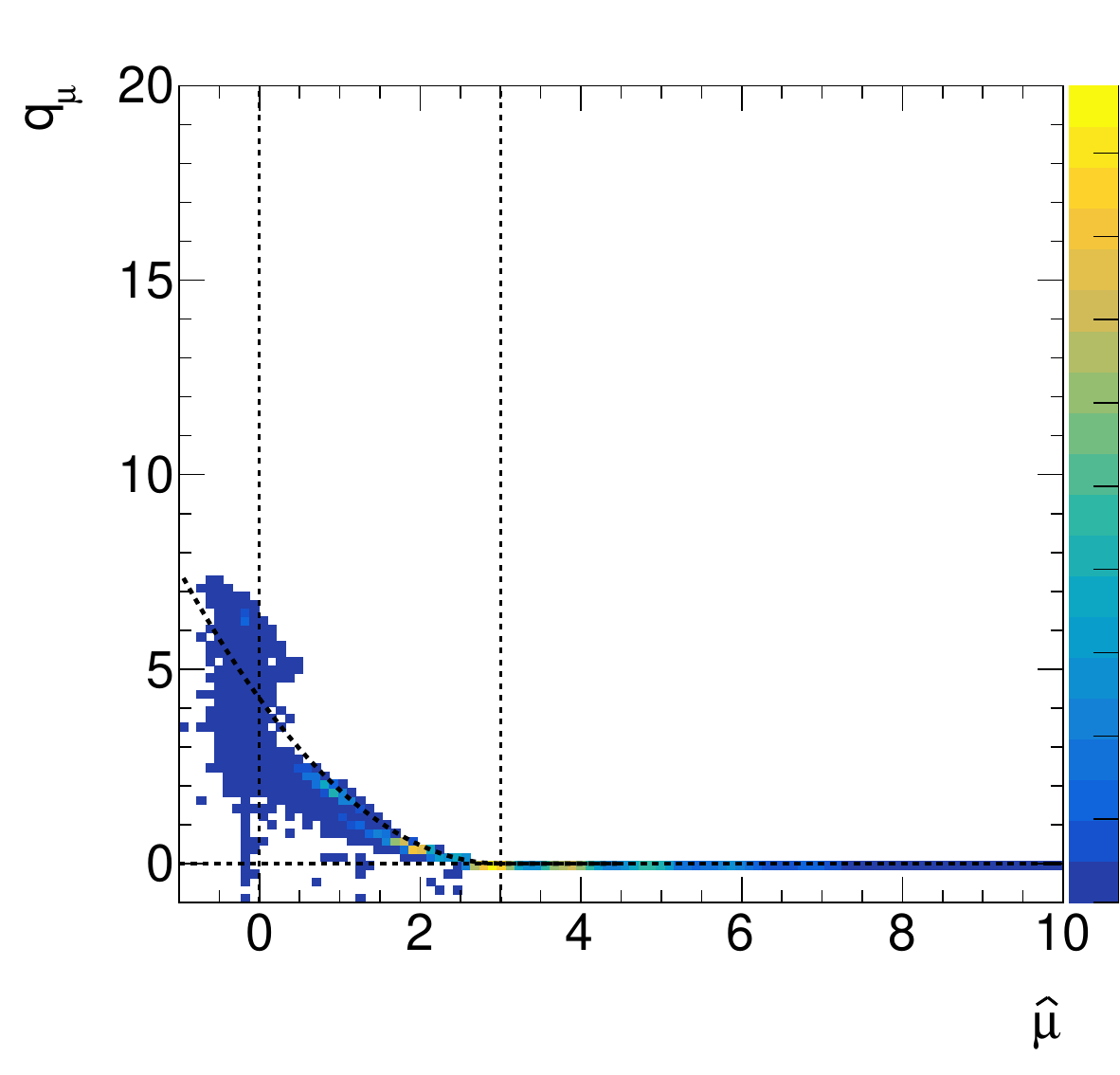}
     \caption{\label{fig:T3}
     The distribution of $q_{\mu}:\hatmu$ in Ex.~0 of Sec.~\ref{sec:example} for the hypothesized signal strength $\mu_H=0$ (L) and $\mu_H=3$ (R). The bold dashed curves represent the asymptotic relation from Wald's approximation. 
     }
 \end{figure}

\begin{figure}[htbp]
     \centering
     \includegraphics[width=0.45\textwidth]{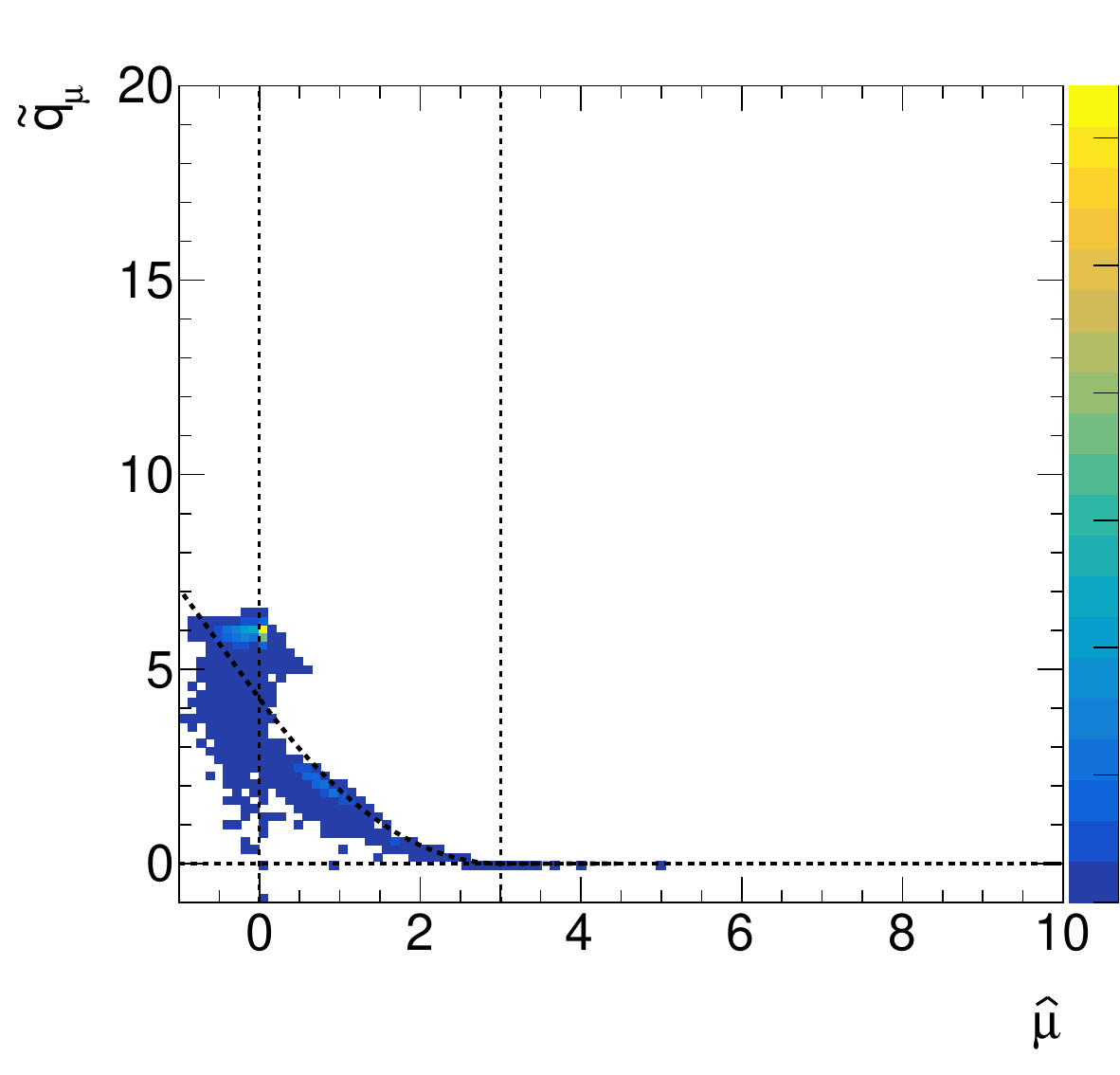}
     \includegraphics[width=0.45\textwidth]{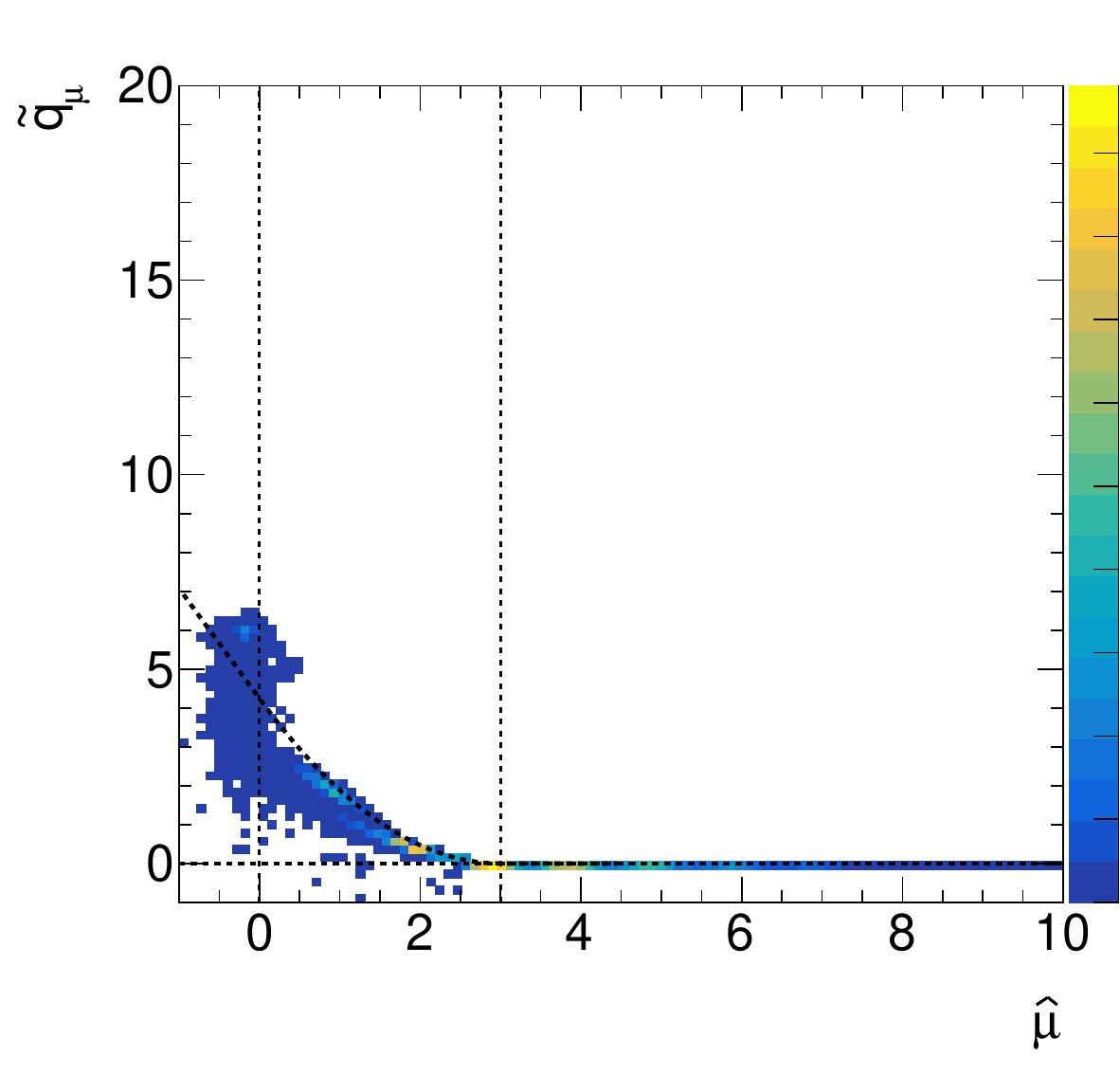}
     \caption{\label{fig:T6}
     The distribution of $\qtil_{\mu}:\hatmu$ in Ex.~0 of Sec.~\ref{sec:example} for the hypothesized signal strength $\mu_H=0$ (L) and $\mu_H=3$ (R). The bold dashed curves represent the asymptotic relation from Wald's approximation. 
     }
 \end{figure}

 \begin{table}
     \centering
     \caption{\label{tab:6t}
     Summary of the test statistics based on the likelihood ratio
     }
     \begin{tabular}{l l}
         \hline\hline
         Test statistic & Purpose \\
         \hline
         $t_0$ & establish the discovery of a signal \\
         $t_\mu$ & set a confidence interval of a signal at a given level \\
         $q_\mu$ & set an upper limit of a signal at a given level \\
         \hline
         $q_0$ & establish the discovery of a positive signal \\
         $\ttil_\mu$ & set a confidence interval of a positive signal at a given level \\
         $\qtil_\mu$ & set an upper limit of a positive signal at a given level \\
         \hline
         \hline
     \end{tabular}
 \end{table}
To gain an initial understanding, Fig.~\ref{fig:T1}-\ref{fig:T6} display the distributions of the six test statistics versus the signal strength, obtained from the toy MC simulations of an example (Ex.~0). These results, which will be analyzed in Sec.~\ref{sec:example}, are compared with the asymptotic relations predicted by Wald's theorem. 

For example, $q_\mu$ is recommended for setting an upper limit on $\mu$.
\begin{equation}\label{eq:qmu_def}
    q_\mu = \left\{\begin{matrix}
            0 & \hatmu > \mu \:,\\
            -2\ln\frac{\mL(\mu,\hat{\hat{\bm{\theta}}}(\mu))}{\mL(\hatmu,\hat{\bm{\theta}})} & \hatmu\leq\mu \:,\\
        \end{matrix}\right.
\end{equation}
If further considering the constraint $\mu>0$ (assuming that the signal contribution to the observed number of events is positive), the recommendation is $\qtil_\mu$.
\begin{equation}\label{eq:qmutilde_def}
    \qtil_\mu = \left\{\begin{matrix}
            0 & \hatmu > \mu \:,\\
            -2\ln\frac{\mL(\mu,\hat{\hat{\bm{\theta}}}(\mu))}{\mL(\hatmu,\hat{\bm{\theta}})} & \mu\geq \hatmu\geq 0 \:,\\
            -2\ln\frac{\mL(\mu,\hat{\hat{\bm{\theta}}}(\mu))}{\mL(0,\hat{\hat{\bm{\theta}}}(0))}    & \hatmu < 0 \: . 
        \end{matrix}\right.
\end{equation}
The recommended test statistic for rejecting the background-only hypothesis (namely, $\mu=0$) is $q_0$.
\begin{equation}\label{eq:q0_def}
    q_0 = \left\{\begin{matrix}
            -2\ln\frac{\mL(0,\hat{\hat{\bm{\theta}}}(0))}{\mL(\hatmu,\hat{\bm{\theta}})} & \hatmu\geq 0 \:,\\
            0    & \hatmu < 0 \: . 
        \end{matrix}\right.
\end{equation}

The asymptotic formulae in Ref.~\cite{asimov} to describe the probability distribution of these test statistics are based on Wald's theorem~\cite{Wald}. According to Wald's theorem, the logarithmic likelihood ratio, seen as a random variable, satisfies the following relation
\begin{equation}\label{eq:wald}
    t_\mu\equiv -2\ln\lambda(\mu) = \frac{(\hatmu-\mu)^2}{\sigma^2} + \bo(\frac{1}{\sqrt{N}}) \: ,
\end{equation}
where $\hatmu$ abides by a Gaussian distribution with mean $\mu_H$ and standard deviation $\sigma$; and $N$ represents the data sample size. 
The standard deviation $\sigma$ can be obtained from either the Fisher information matrix (second-order derivatives of the logarithmic likelihood function)~\cite{book_cowan,xia_constraint} or from Wald's
theorem (Eq.~\ref{eq:wald}) based on an
Asimov dataset (denoted by $\sigma$(Wald)). In the large sample limit, we can ignore the term $\bo(\frac{1}{\sqrt{N}})$ in
Eq.~\ref{eq:wald} (we call it ``Wald's approximation'' throughout this paper). This leads to the following asymptotic relation between the test statistics and $\hatmu$. 
\begin{equation}\label{eq:qmu_wald}
    q_\mu = \left\{\begin{matrix}
            0 & \hatmu > \mu \\
            \frac{(\hatmu-\mu)^2}{\sigma^2} & \hatmu\leq\mu\\
        \end{matrix}\right. \:,
\end{equation}
\begin{equation}\label{eq:qmutilde_wald}
    \qtil_\mu = \left\{\begin{matrix}
            0 & \hatmu > \mu \\
            \frac{(\hatmu-\mu)^2}{\sigma^2} & \mu\geq \hatmu\geq 0 \\
            \frac{\mu^2-2\mu\hatmu}{\sigma^2}    & \hatmu < 0 
        \end{matrix}\right. \:,
\end{equation}
and 
\begin{equation}
    q_0 = \left\{\begin{matrix}
            \frac{\hatmu^2}{\sigma^2} & \hatmu\geq 0 \\
            0    & \hatmu < 0  
        \end{matrix}\right.\:.
\end{equation}
Consequently, the probability distribution function (PDF) of these test statistics is obtained assuming that $\hatmu$ abides by a Gaussian distribution.

\section{New asymptotic formulae}\label{sec:new}
The classic asymptotic formulae perform well for sufficiently large sample sizes. 
To extend their applicability to cases with limited statistics, it is necessary to incorporate the leading-order correction term $\bo(\frac{1}{\sqrt{N}})$. However, this is case-dependent, requiring a dedicated analysis of the signal and background for each measurement.
A natural idea to address the challenge is to decompose the PDF of a test into two parts. 
One part is described by the asymptotic formulae (with appropriate corrections), the other part must be case-dependent and has to be estimated in a reasonable way.
This idea is inspired by the process of toy MC simulation. 
Imagine many people are doing the same measurement within the same experimental conditions independently.
Some experimenters would observe many events and believe it is fine to interpret their results using the classic asymptotic formulae. 
Some would observe few events and the classic formulae would be less reliable. 
The observed number of events follows a Poisson distribution with mean
$b+\mu s$ where $b$ and $s$ are the number of signal and background events and $\mu$ is the signal strength.

Let $T_\mu$ denote a test statistic, such as $q_\mu$ or $\qtil_\mu$. If the ``observed'' number of events exceeds a certain threshold, its contribution to the PDF of $T_\mu$ must be well approximated by the classic asymptotic formula. Otherwise, we explicitly model its contribution as accurately as possible. 
Fortunately, due to low event count, the computational cost of the latter case remains manageable. 
We will show that the PDF of $T_\mu$ in this regime is discrete, and its possible values can be reliably predicted. 
Therefore, letting $f(T_\mu|\mu_H)$ be the PDF of $T_\mu$ with hypothesized signal strength $\mu_H$, we have 
    \begin{eqnarray}
        f(T_\mu | \mu_H) = && \sum_{n=0}^{+\infty}f(T_\mu|n,\mu_H)P(n|b+\mu_Hs) \nonumber\\
         = && \sum_{n=0}^{\nsmall}f(T_\mu|n,\mu_H)P(n|b+\mu_Hs) + \sum_{n>\nsmall}f(T_\mu|n,\mu_H)P(n|b+\mu_Hs) \nonumber\\
         \approx && \sum_{n=0}^{\nsmall}f_{\ls}(T_\mu|n,\mu_H)P(n|b+\mu_Hs) + (1-\sum_{n=0}^{\nsmall}P(n|b+\mu_Hs))f_{\LS}(T_\mu|\nsmall,\mu_H) \:.
    \end{eqnarray}
    Here $P(n|\nu)$ is Poisson distribution function with mean $\nu$; $\nsmall$ is the boarder between large statistics (LS) and small statistics (SS), and has to be chosen appropriately. If the number of events is greater than $\nsmall$, the probability distribution of $T_\mu$ is described by a single function $f_{\LS}$.  $f_{\LS}$ is just the classic asymptotic formulae with a correction as explained in Sec.~\ref{sec:finalformulae}. For each possible number of events not greater than $\nsmall$, we obtain the probability distribution, $f_{\ls}$,
    based on simplified 6-bin distributions of the observables.  
    \begin{eqnarray}
        f_{\ls}(T_\mu|n,\mu_H) = && \sum_{k_0+k_1+k_2+k_3+k_4+k_5=n}\frac{n!}{k_0!k_1!\cdots k_5!}\Pi_{i=0}^5(\frac{b_i+\mu_Hs_i}{b+\mu_Hs})^{k_i}\nonumber \\
        &&\times f_{\text{binned}}(T_\mu|n_i=k_i,\text{for}\:i\in\{0,1,2,3,4,5\};\mu_H) \:,
    \end{eqnarray}
where $s_i$ ($b_i$) is the number of signal (background) events in the $i$-th bin; and $s$ ($b$) is the total number of signal (background) events. We will show the definition of $f_{\text{binned}}$ and explain how it is obtained in next section. But before presenting more details, here is another way to understand the new idea. Taking $q_\mu$ as example, its cumulative distribution function (CDF) is calculated below using an integral according to Wald's approximation.
\begin{eqnarray}
    F_{\text{Wald}}(q_\mu|\mu^\prime, \sigma) = &&\frac{1}{\sqrt{2\pi}\sigma}\int_{\mu-\frac{\sqrt{q_\mu}}{\sigma}}^{+\infty}e^{-\frac{1}{2}\frac{(x-\mu^\prime)^2}{\sigma^2}}dx\\
    =&&\Phi(\sqrt{q_\mu}-\frac{\mu-\mu^\prime}{\sigma})
\end{eqnarray}
Here $\Phi(x)\equiv \frac{1}{\sqrt{2\pi}}\int_{-\infty}^xe^{-x^{\prime2}/2}dx^\prime$. Alternatively, we can calculate it in two steps (using double integrals) below
\begin{eqnarray}
    F_{2-\text{step}}(q_\mu|\mu^\prime, \sigma_{\text{stat}}, \sigma_{\text{syst}}) = &&\frac{1}{\sqrt{2\pi}\sigma_{\text{stat}}}\int_{-\infty}^{+\infty}e^{-\frac{1}{2}\frac{(y-\mu^\prime)^2}{\sigma_{\text{stat}}^2}}\left[\frac{1}{\sqrt{2\pi}\sigma_{\text{syst}}}\int_{\mu-\frac{\sqrt{q_\mu}}{\sigma_{\text{stat}}/\sqrt{R}}}^{+\infty}e^{-\frac{1}{2}\frac{(x-y)^2}{\sigma_{\text{syst}}^2}}dx\right]dy \label{eq:2integrals}\nonumber\\
    =&&\Phi(\sqrt{q_\mu}\sqrt{\frac{\sigma_{\text{stat}}^2+\sigma_{\text{syst}}^2}{\sigma_{\text{stat}}^2}R}-\frac{\mu-\mu^\prime}{\sqrt{\sigma_{\text{stat}}^2+\sigma_{\text{syst}}^2}})
\end{eqnarray}
We get the same result if $\sigma=\sqrt{\sigma_{\text{stat}}^2+\sigma_{\text{syst}}^2}$ and $R=\frac{\sigma_{\text{stat}}^2}{\sigma^2}$. Here $\sigma_{\text{stat}}$ can be seen as the statistical uncertainty and $y$ is the signal strength with considering statistical uncertainty only. $\sigma_{\text{syst}}$ can be seen as the systematic uncertainty and $x$ is the final signal strength with the systematic effects included. 
The new idea works similarly. 
The first step is to randomize the number of events based on a binned model without nuisance parameters, and extract the information on signal strength and the test statistics. The second step is to consider the systematic effects by modeling the signal strength as a Gaussian-distributed variable with an appropriate width. This uncertainty is then propagated to the probability distribution of the test statistic.

It is important to note that the calculation above involves a convolution. When examining this convolution using the characteristic function method, we observe that the smearing effect induced by systematic uncertainties will suppress the higher-order variations in the original test statistic  distribution, such as the features arising from limited sample size. 
This provides an explanation for the empirical observation in Ref.~\cite{beyond} that the asymptotic formulae perform better in the presence of systematic effects. 

To get further insight into the new approach, we introduce the following integral.
\begin{equation}
    F_{\text{classic}}(T_\mu|\mu^\prime, \sigma^\prime, \sigma) = \frac{1}{\sqrt{2\pi}\sigma^\prime}\int_{\mu-\frac{\sqrt{T_\mu}}{\sigma}}^{+\infty}e^{-\frac{1}{2}\frac{(x-\mu^\prime)^2}{\sigma^{\prime2}}}dx 
\end{equation}
Then we have
\begin{eqnarray}
    && F_{\text{Wald}}(q_\mu|\mu,\sigma) = F_{\text{classic}}(q_\mu|\mu^\prime, \sigma, \sigma) \\
    && F_{2-\text{step}}(q_\mu|\mu,\sigma_0,\sigma_1) = \frac{1}{\sqrt{2\pi}\sigma_0}\int_{-\infty}^{+\infty}e^{-\frac{1}{2}\frac{(y-\mu^\prime)^2}{\sigma_0^2}}F_{\text{classic}}(q_\mu|\mu^\prime, \sigma_1, \sigma_0/\sqrt{R})dy \label{eq:2integrals1}
\end{eqnarray}
The key innovation of our approach involves replacing the Gaussian approximation in Eq.~\ref{eq:2integrals1}, which is valid in the large-sample limit, by a more accurate 6-bin model representation. Additionally, we assumes that the impact of the systematic effects is Gaussian-like. Thus we can see the structural similarity between our approach and the conventional method ($F_{\text{classic}}$ in the equations above). 

\subsection{A 2-bin model}\label{sec:2bin}
In this section, we employ a simplified 2-bin model to demonstrate the derivation of $f_{\text{binned}}$ and $f_{\ls}$ because we are able to obtain analytic expression of the parameter of interest, namely signal strength in most measurements.
Suppose the observable distribution is re-binned into just 2 bins. Let $b_i$, $s_i$ and $n_i$ denote the number of background, signal and observed events in the $i$-th bin ($i=0,1$). They are ordered with decreasing expected significances, namely, $Z_0>Z_1$, where $Z_i$ is defined as
\begin{equation}\label{eq:sig}
    Z_i = 2[(b_i+\mu_H s_i)\ln(1 + \frac{\mu_H s_i}{b_i}) -\mu_H s_i] \: .
\end{equation}
We further assume the binning is made to maximize the total expected significance, $Z_{\tot}\equiv\sum_i Z_i$. Under this condition, the purity in $0$-th bin is generally (and often substantially) greater than that in the $1$-st bin, namely, $s_0/b_0 > s_1/b_1$. 
Given the observed number of events, $n_i$, and ignoring other nuisance parameters and freely-floating parameters, the optimal signal strength estimator is obtained by maximizing the following binned likelihood function,
\begin{equation}\label{eq:binnedll}
    \mL(\mu) = \Pi_{i=0}^{\nbins} P(n_i | b_i + \mu s_i) \:,
\end{equation}
or equivalently the logarithmic likelihood function,
\begin{equation}
    \ln\mL(\mu) = \sum_{i=0}^{\nbins} n_i \ln(b_i + \mu s_i) - (b_i + \mu s_i) \:, 
\end{equation}
where $\nbins$ is the number of bins. In most of the cases, the best estimator, $\hatmu$, is found such that $\partial\ln\mL/\partial\mu=0$, namely,
\begin{equation}
    \sum_{i=0}^N \frac{n_is_i}{b_i+\mu s_i} - s_i = 0 \: .
\end{equation}
For $\nbins=2$, the equation can be solved easily,
\begin{eqnarray}
    && A = 2s_0 s_1 \\
    && B = s_0b_1 + s_1b_0 - \frac{n_0+n_1}{s_0+s_1}s_0s_1 \\
    && C = b_0b_1 - \frac{n_0s_0b_1 + n_1s_1b_0}{s_0+s_1} \\
    && \hatmu(n_0,n_1) = \frac{-B+\sqrt{B^2-4AC}}{2A} \: .
\end{eqnarray}
$\hatmu$ as a function of $n_0$ and $n_1$ has the following feature,
\begin{equation}
    -\frac{b_0}{s_0} = \hatmu(0,n) < \hatmu(1,n-1) < \hatmu(2,n-2) < \cdots < \hatmu(n, 0) = - \frac{b_0}{s_0} + \frac{n}{s_0+s_1} \: .  
\end{equation}
Especially, if $s_0/b_0>>s_1/b_1$, we have
\begin{equation}
     \hatmu(k,n-k) \approx  -\frac{b_0}{s_0} + \frac{k}{s_0+s_1} \:.
\end{equation}
Based on the solutions above, we have three observations.
\begin{itemize}
    \item For $n$ number of events, there will be $n+1$ possible values of $\hatmu$.
    \item The possible $\hatmu$ values are approximately equal-distance distributed. 
    \item For the value $\hatmu(k,n-k)$, its probability is proportional to $\frac{n!}{k!(n-k)!}$.  
\end{itemize}
The observations will be confirmed in the toy MC simulations in Sec.~\ref{sec:example}.

Without any nuisance parameter or free parameter, the distribution of $\hatmu$ is discrete and the distribution of $T_\mu$ is also discrete. Taking $\qtil_\mu$ and $q_\mu$ as example, we have
\begin{equation}\label{eq:qmubinned}
    T_\mu^{\binned}(\hatmu) =\left\{ 
    \begin{matrix}
        -2 [\sum_{i=0}^{\nbins} n_i\ln\frac{b_i + \mu s_i}{b_i+\hatmu s_i} - (\mu -\hatmu)s_i] \: , & \hatmu\geq 0\:\text{or using }q_\mu\\
         -2 [\sum_{i=0}^{\nbins} n_i\ln\frac{b_i + \mu s_i}{b_i} - \mu s_i] \: , &\hatmu < 0\\
    \end{matrix}
    \right..
\end{equation}
To summarize, the PDF of $T_\mu$ is then
\begin{equation}
    f_{\ls}(T_\mu|n,\mu_H) = \sum_{k_0+k_1=n} \frac{n!}{k_0!k_1!}(\frac{b_0+\mu_Hs_0}{b+\mu_Hs})^{k_0}(\frac{b_1+\mu_Hs_1}{b+\mu_Hs})^{k_1}f_{\binned}(T_\mu|n_0=k_0,n_1=k_1) 
\end{equation}
with
\begin{equation}\label{eq:fbinned_statonly}
    f_{\binned}(T_\mu|n_0=k_0,n_1=k_1) = \delta(T_\mu-T_\mu^{\binned}(\hatmu(n_0,n_1)))\:, 
\end{equation}
where $\delta(x)$ is Dirac $\delta$ function.

\subsection{A 6-bin model}\label{sec:6bin}
In the previous section, we have demonstrated the key ideas for extending the classic formulae using a simplified 2-bin observable distributions. We emphasize that better performance is generally achieved with more bins. 
Considering that 5 may be arguably a safe threshold between small-statistics and large-statistics regimes, and 
also taking into account the computational efficiency, it seems appropriate to use 5 bins. 
However, there are cases where we expect to see a large number of events but very few events in the signal-sensitive region and thus we still suffer from the effect of limited sample size.
Therefore, it is necessary to deal with the large-statistics part whose contribution to the signal detection is negligible. We put this part in the 6-th bin and hence propose a 6-bin model.  

In practice, here is the workflow to obtain the 6-bin model.
\begin{itemize}
    \item Merge the observable distributions in all signal regions into a fine-binning histogram for the signal and background component;
    \item Re-order the bins with the decreasing significance as defined in Eq.~\ref{eq:sig};
    \item Find the bin (denoted by $i_5$), the contribution of all the bins after which to the total significance is less than 0.1\%, which seems a safe threshold as conventionally we report results at 90~\% or 95~\% confidence level (C.L.). Define the signal and background yield summed over those bins as $s_5$ and $b_5$ (we use the index starting from 0).
    \item For the bins before $i_5$, we categorize them into 5 bins and the binning is determined by maximizing the significance. 
\end{itemize}
We should then update the summation in $f_{\ls}$ from the 2-bin model to the 6-bin model.
\begin{eqnarray}
    f_{\ls}(T_\mu|n,\mu_H) = &&\sum_{k_0+k_1+\cdots+k_5=n}\frac{n!}{k_0!k_1!\cdots k_5!}\Pi_{i=0}^5(\frac{b_i+\mu_Hs_i}{b+\mu_Hs})^{k_i}\nonumber \\
    &&\times f_{\binned}(T_\mu|n_i=k_i,\text{for}\:i\in\{0,1,2,3,4,5\};\mu_H) \label{eq:f_SS} \: .
\end{eqnarray}

\subsection{Systematic uncertainties}\label{sec:systs}
We have not yet accounted for any systematic uncertainty so far.
Systematic uncertainties are incorporated through nuisance parameters (NP), typically modeled as Gaussian-distributed variables with standard deviations representing their uncertainty magnitudes.
This means that we need to randomize both the number of observed events and the nuisance parameters at the same time in toy simulations. 
We assume the impact of the latter randomization (i.e. the impact of systematic uncertainties) makes $\hatmu$ abide by a Gaussian distribution, of which the mean is unchanged and obtained by maximizing the likelihood function in Eq.~\ref{eq:binnedll} and the standard deviation, denoted by $\sigma(\hatmu)$, is estimated as
\begin{equation}\label{eq:sigma_form}
    \sigma(\hatmu) = \sqrt{\sigma_0^2 + (\kappa\hatmu)^2} \: .
\end{equation}
The motivation for this form is explained in Appendix~\ref{app:sigma}. Here $\sigma_0$ and $k$ are determined using Asimov datasets, namely,
\begin{eqnarray}
    && \sigma_0 = \sqrt{\sigma_A^2(\mu=0)-\sigma_A^2(\mu=0,\text{stat. only})}\\
    && \kappa = \sqrt{\sigma_A^2(\mu=\mu_H)-\sigma_A^2(\mu=\mu_H,\text{stat. only}) - \sigma_0^2}/\mu_H \label{eq:k}\:,
\end{eqnarray}
where $\sigma_A(\mu)$ is the uncertainty of $\hatmu$ from fitting to an Asimov dataset with signal strength $\mu$; $\sigma_A(\mu,\text{stat. only})$ is the uncertainty with fixing all other nuisance parameters. Basically, $\sigma(\hatmu)$ has two contributions. One does not depend upon $\hatmu$ and the other does.

One the other hand, the calculation of $T_\mu^{\binned}(\hatmu)$ in Eq.~\ref{eq:qmubinned} is not correct as it is derived from the likelihood function in Eq.~\ref{eq:binnedll} without including any systematic uncertainty or freely-floating parameters.
This is overcome by applying a scale factor $R_\mu$ for calibration purpose. $R_\mu$ is actually the ratio of $T_\mu$ obtained from a background-only Asimov dataset to that calculated from the simplified likelihood function in Eq.~\ref{eq:binnedll}. 
Taking $q_\mu$ and $\qtil_\mu$ as example, it is
\begin{eqnarray}
    && R_\mu \equiv \frac{q_\mu^A(\mu_H=0)}{q_\mu^{\binned}(\mu_H=0)}\:,\label{eq:Rmu}\\
    && q_\mu^{\binned}(\mu_H=0) = -2\sum_{i=0}^{\nbins}b_i\ln\frac{b_i+\mu s_i}{b_i}-\mu s_i \: ,
\end{eqnarray}
where $q_\mu^A(\mu_H=0)$ is the expected value of $q_\mu$ in the background-only hypothesis. In view of the Wald approximation in Eq.~\ref{eq:wald}. $R_\mu$ can be seen as the ratio, $\sigma_{\text{stat.}}^2/\sigma_{\text{full}}^2$, where $\sigma_{\text{stat.}}$ is the signal strength uncertainty from the simple binned model without including any systematic uncertainty while $\sigma_{\text{full}}$ is that from the full measurement. 
This correction is already seen before in Eq.~\ref{eq:2integrals}. 
The distribution of $T_\mu$ is not discrete as in Eq.~\ref{eq:fbinned_statonly} any more. 
For simplicity, we introduce the following PDF.
\begin{equation}
    f_{\classic}(T_\mu|\mu^\prime, \sigma^\prime, \sigma) = \Phi(\frac{\mup-\mu}{\sigmap})\delta(T_\mu) + \left\{
        \begin{matrix}
            \frac{\sigma}{2\sqrt{T_\mu}}\frac{1}{\sqrt{2\pi}\sigmap}e^{-\frac{(\mu-\sigma\sqrt{T_\mu}-\mup)^2}{2\sigma^{\prime2}}} \:, & T_\mu \leq \frac{\mu^2}{\sigma^2}\:\text{or }T_\mu=q_\mu \\
            \frac{\sigma^2}{2\mu}\frac{1}{\sqrt{2\pi}\sigmap}e^{-\frac{(\frac{\mu^2-\sigma^2T_\mu}{2\mu}-\mup)^2}{2\sigma^{\prime2}}} \:, & T_\mu > \frac{\mu^2}{\sigma^2} \: .
        \end{matrix}\right.
\end{equation}
Then $f_{\binned}$ in Eq.~\ref{eq:fbinned_statonly} becomes
\begin{eqnarray}
    && f_{\binned}(T_\mu|n_0=k_0,n_1=k_1) = f_{\classic}(T_\mu|\hatmu(k_0,k_1),\sigma(\hatmu),\sigma_{\wald}^{\binned}) \:, \\
    && \sigma_{\wald}^{\binned} = \left\{
        \begin{array}{l}
            \frac{|\hatmu-\mu|}{\sqrt{R_\mu T_\mu^{\binned}(\hatmu)}} \:, \hatmu > 0 \:\text{or using }q_\mu\\
            \sqrt{\frac{-2\mu\hatmu+\mu^2}{R_\mu T_\mu^{\binned}(\hatmu)}} \:, \hatmu \leq 0 
        \end{array}
    \right..
\end{eqnarray}

It should be noted that the case of observing 0 events and the case with the optimal $\hatmu$ attained at its lowest bound will be studied explicitly in Appendix~\ref{app:zero_events}, respectively. The lowest bound is the smallest number to make the yield non-negative in all bins. The probability of these cases may be significant in searching for new physics with very low background. 
The simplified binned model is able to predict the center value of $T_\mu$ well, but fail to describe its width due to systematic uncertainties. 
Necessary corrections to $\sigma(\hatmu)$ will be introduced because the first-order derivation of the logarithmic likelihood does not vanish at $\hatmu$.

\subsection{The final formulae}\label{sec:finalformulae}
In this section, we explain a correction to the asymptotic function in the LS part, recommend the choice of $\nsmall$  and present the final formulae.

Firstly, the SS part with the number of events not greater than $\nsmall$ is considered in the 6-bin model. Generally, the expectation value of $\hatmu$ will not be $\mu_H$ any more. To recover the right expectation value, $\mu_H$ in LS part has to be modified to be
\begin{eqnarray}
    &&\mu_H^{\LS}(\nsmall,\mu_H) = \nonumber\\
    &&\frac{\mu_H - \sum_{n\leq\nsmall}P(n|b+\mu_Hs)\sum_{k_0+\cdots+k_5=n}\frac{n!}{k_0!\cdots k_5!}\Pi_{i=0}^5\left(\frac{b_i+\mu_Hs_i}{b+\mu_Hs}\right)^{k_i}\hatmu(k_0,\cdots,k_5)}{1-\sum_{n\leq \nsmall}P(n|b+\mu_Hs)} \: . \label{eq:muH_LS}
\end{eqnarray}

Secondly, we recommend the following choice of $\nsmall$
\begin{equation}
    \nsmall = \min\{b+\mu s-1,10\} \: ,
\end{equation}
with the modification of fixing $b_5$ at 10 and scaling $s_5$ to $s_5\times 10/b_5$ if $b_5>10$ ($b_5$ is the number of background events in the region whose contribution to the signal detection is negligible). 
We choose $\nsmall$ to be around $b+\mu s-1$ because we want a conservative improvement and do not expect the updated part to be more than 50~\%. However, the computation consumption is significant if $b+\mu s -1$ is too big and hence $\nsmall$ is capped at 10.
Although the definition of $\nsmall$ is unserious, the performance of the new formulae is robust against varying $\nsmall$ as we will see in Sec.~\ref{sec:nsmall}. 
It should be noted that we propose to choose 6 bins and cap $b_5$ and $\nsmall$ at 10 because of the computation cost. This can be loosened and better performance is expected. 

Eventually, we summarize the complete set of formulae below used in the new method for convenience. 
    \begin{eqnarray}
        && f(T_\mu | \mu_H) = \sum_{n=0}^{\nsmall}f_{\ls}(T_\mu|n,\mu_H)P(n|b+\mu_Hs) + (1-\sum_{n=0}^{\nsmall}P(n|b+\mu_Hs))f_{\LS}(T_\mu|\nsmall,\mu_H)  \label{eq:key0}\nonumber \\
        && \\
        && f_{\ls}(T_\mu|n,\mu_H) = \sum_{k_0+k_1+\cdots+k_5=n}\frac{n!}{k_0!k_1!\cdots k_5!}\Pi_{i=0}^5(\frac{b_i+\mu_Hs_i}{b+\mu_Hs})^{k_i}\nonumber\\
        &&\quad\quad\quad\quad\quad\quad\quad\times f_{\text{binned}}(T_\mu|n_i=k_i,\text{for}\:i\in\{0,1,\cdots,5\};\mu_H)\label{eq:key1}\\
        && f_{\LS}(T_\mu|\nsmall,\mu_H) = f_{\classic}(T_\mu|\mu_H^{\LS}(\nsmall,\mu_H), \sigma_{\wald}, \sigma_{\wald}) \\
        && f_{\binned}(T_\mu|n_i\:\text{for}\:i\in\{0,1,\cdots,5\};\mu_H) = f_{\classic}(T_\mu|\hatmu(n_0,n_1,\cdots,n_5),\sigma(\hatmu),\sigma_{\wald}^{\binned})\: . 
\end{eqnarray}
Here (taking $q_\mu$ and $\qtil_\mu$ as example)
    \begin{eqnarray}
        && \sigma_{\wald} = \frac{\mu}{\sqrt{q_\mu^A(\mu_H=0)}}\\
        && \sigma(\hatmu) = \left\{
            \begin{array}{ll}
                \kappa\frac{\mu-c\hatmu}{2}\:, &\text{if}\: k_0=\cdots=k_4=0\\
                \sqrt{\sigma_0^2+(\kappa\hatmu)^2+(\kappa\frac{\mu-c\hatmu}{2})^2}\:, &\text{else if}\: \hatmu=-b_0/s_0\\
                \sqrt{\sigma_0^2+(\kappa\hatmu)^2}\:, &\text{otherwise}
        \end{array}
    \right.\\
    && \sigma_{\wald}^{\binned} = \left\{
        \begin{array}{ll}
            \frac{|\hatmu-\mu|}{\sqrt{R_\mu T_\mu^{\binned}(\hatmu)}} \:,  & \hatmu > 0 \:\text{or using }q_\mu\\
            \sqrt{\frac{-2\mu\hatmu+\mu^2}{R_\mu T_\mu^{\binned}(\hatmu)}} \:, & \hatmu \leq 0 \\
        \end{array}
    \right.\\
        &&T_\mu^{\binned}(\hatmu) =\left\{ 
        \begin{array}{ll}
            -2 [\sum_{i=0}^5 n_i\ln\frac{b_i + \mu s_i}{b_i+\hatmu s_i} - (\mu -\hatmu)s_i] \: , & \hatmu>0\text{ or using }q_\mu\\
            -2 [\sum_{i=0}^5 n_i\ln\frac{b_i + \mu s_i}{b_i} - \mu s_i] \: , &\hatmu \leq 0\\
    \end{array}
    \right.\\
        && \nsmall = \min\{b+\mu s-1, 10\} \:,
\end{eqnarray}
where $c=1$ for $q_\mu$ and 2 for $\qtil_\mu$ (more details in Appendix~\ref{app:zero_events}); $R_\mu$ is defined in Eq.~\ref{eq:Rmu}; $\mu_H^{\LS}(\nsmall,\mu_H)$ is defined in Eq.~\ref{eq:muH_LS}; $\hatmu(k_1,\cdots,k_5)$ is obtained by maximizing the binned likelihood function in Eq.~\ref{eq:binnedll}.

\section{Two examples}\label{sec:example}
\begin{figure}[htbp]
     \centering
     \includegraphics[width=0.45\textwidth]{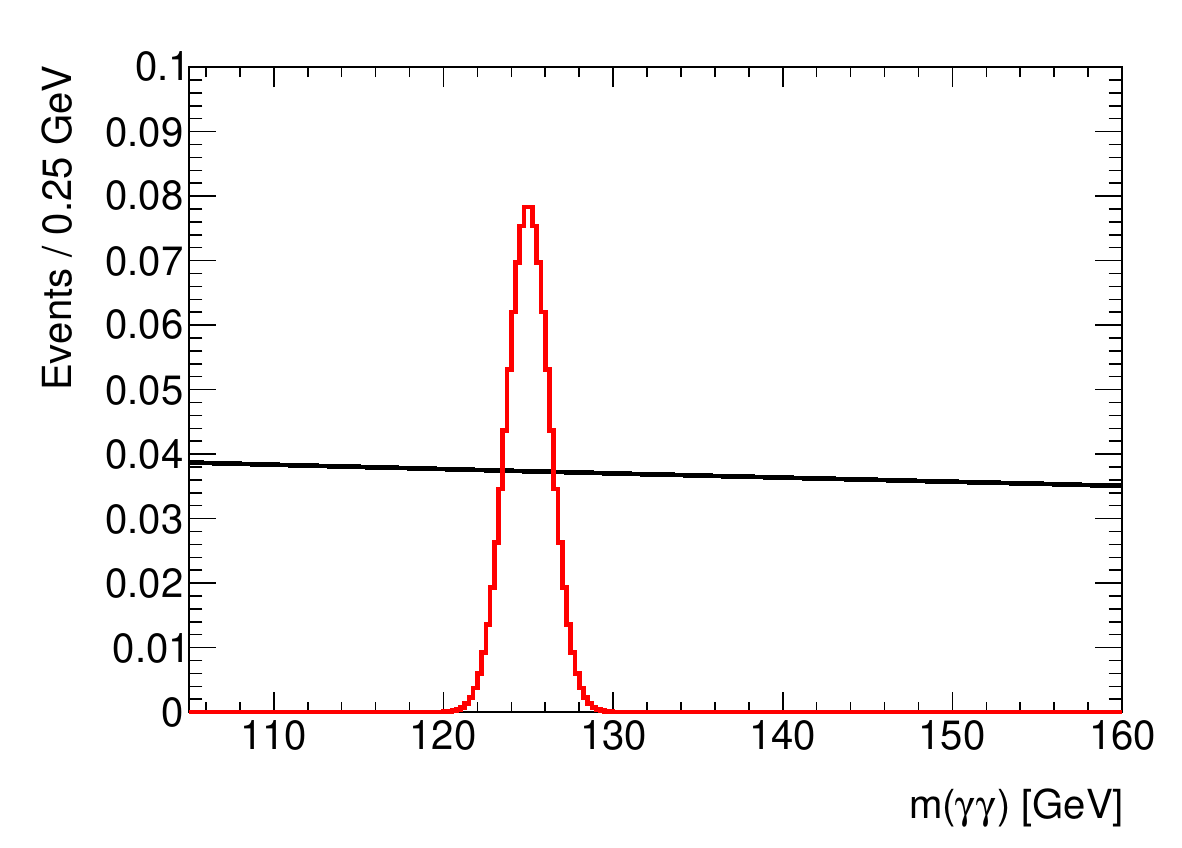}
     \includegraphics[width=0.45\textwidth]{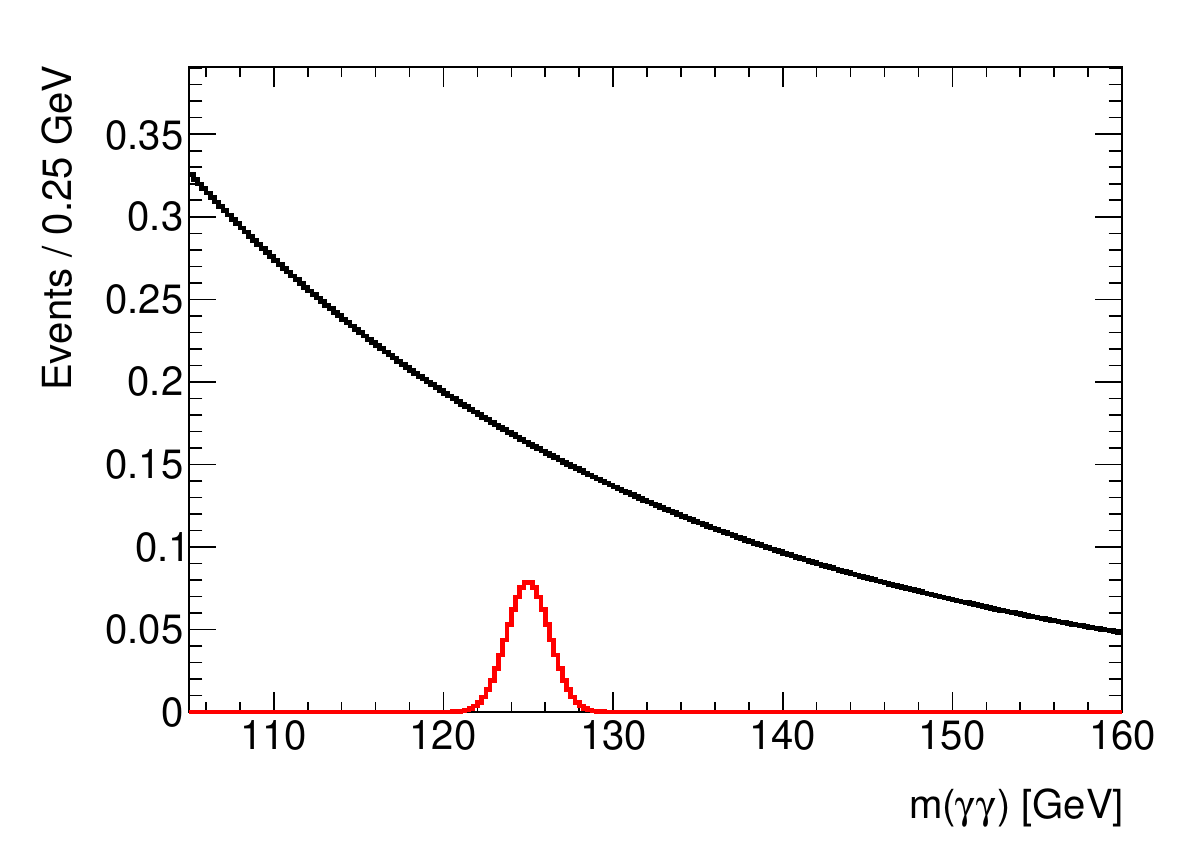}
     \caption{\label{fig:ex_myy}
     The distribution of $m(\gamma\gamma)$ in Ex.~0 (L) and Ex.~1 (R). The red histograms represent the signal and the black histograms represent the background.
     }
 \end{figure}

In this section, we apply the new formulae to two toy examples to compare the performance of the classic and new asymptotic formulae.  The applications to two real physics analyses are documented in Appendix~\ref{app:real_phys_app}. The two examples are denoted by Ex.~0 and Ex.~1 with increasing sample size. The physics behind the examples is to measure Higgs production cross section using the $H\to\gamma\gamma$ mode. The signal strength is obtained by fitting to the $\gamma\gamma$ invariant mass spectrum. Table~\ref{tab:ex_yields} summarizes the expected signal and background yields in the mass region $123<m(\gamma\gamma)<127$~GeV. 
The expected background and signal yields are low in both examples. 
The signal shape is simulated by a Gaussian distribution and the background shape is simulated by an exponential distribution. They are shown in Fig.~\ref{fig:ex_myy}. According to the strategy in Sec.~\ref{sec:6bin}, the 6-bin model is built and the expected yield in each bin in the background-only hypothesis is shown in Table~\ref{tab:ex_6bin}. It should be emphasized that these numbers vary under different hypotheses. 

  \begin{table}
      \centering
      \caption{\label{tab:ex_yields}
          Summary of the yields expected in the mass region  $123<m(\gamma\gamma)<127$~GeV in the two examples.
      }
      \begin{tabular}{l| l l}
          \hline\hline
          Yield & signal& background \\
          \hline
          Ex.~0 & 0.91 & 0.64 \\
          Ex.~1 & 0.91 & 2.79 \\
          \hline
          \hline
      \end{tabular}
  \end{table}

  \begin{table}
      \centering
      \caption{\label{tab:ex_6bin}
      The signal and background yields in the 6-bin model.
      }
      \begin{tabular}{l| l| l l l l l l}
          \hline\hline
           & Bin & 0 & 1 & 2 & 3 & 4 & 5 \\
          \hline
          Ex.~0 & sig. & 0.571 & 0.194 & 0.122 & 0.066 & 0.038 & 0.010 \\
                & bkg. & 0.146 & 0.070 & 0.068 & 0.067 & 0.098 & 2.98 \\
                \hline
          Ex.~1 & sig. & 0.571 & 0.194 & 0.122 & 0.066 & 0.038 & 0.010 \\
                & bkg. & 0.328 & 0.163 & 0.162 & 0.161 & 0.240 & 7.514 \\
          \hline
          \hline
      \end{tabular}
  \end{table}
 
 We further consider three systematic uncertainty sources. They are the luminosity measurement, our knowledge on Higgs mass and the spurious signal (due to background modeling), the last of which directly affects the expected signal yield and is usually dominant in real analyses~\cite{Hyy2022}. The uncertainty sizes are summarized in Table~\ref{tab:sysunc}. 
 In addition, three ``observed'' data samples with increasing injected signal strength are also prepared for each example. The injected signal strengths are negative (-0.15 in Ex.~0 and -0.5 in Ex.~1), +0.5 and +2, respectively. 
 For the MC method, we produce 40000 toys for each signal-plus-background hypothesis and 10000 toys for each background-only hypothesis, which estimate the CLs at around $5\%$ (corresponding to the $95\%$ C.L.) with a precision of about $2\%$.

\begin{figure}[htbp]
    \centering
    \includegraphics[width=0.45\textwidth]{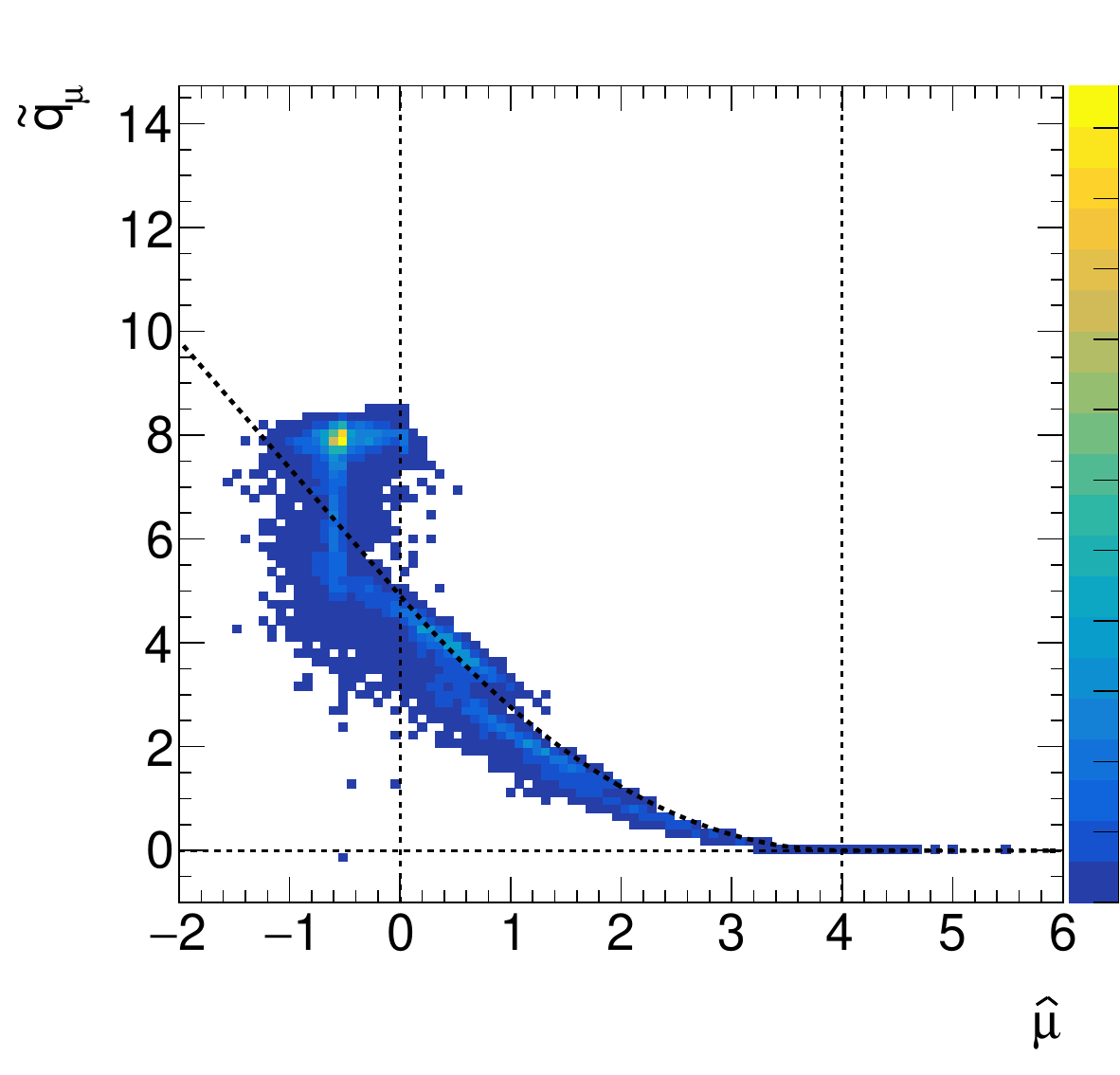}
    \includegraphics[width=0.45\textwidth]{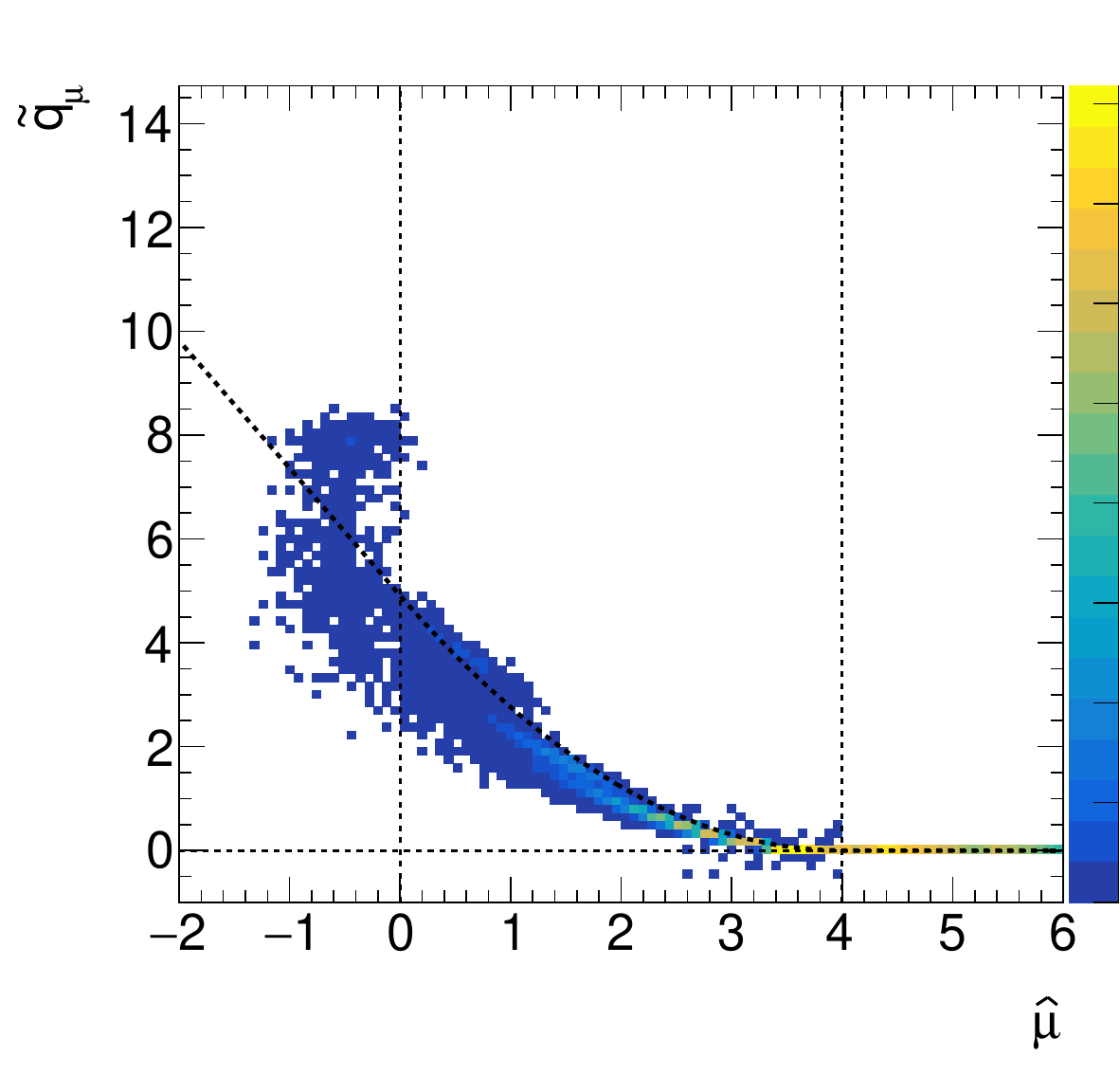}
    \caption{\label{fig:qmutilde_muhat_ex1}
    The scattering plot of $\qtil_\mu:\hatmu$ with $\mu=4$ in Ex.~1 from the toy experiments under the hypothesis$\muH=0$ (L) and $\muH=\mu=4$ (R). The bold dashed curve shows the asymptotic relation according to Wald's approximation.
    }
\end{figure}

  \begin{table}
      \centering
      \caption{\label{tab:sysunc}
      Summary of systematic uncertainties.
      }
      \begin{tabular}{l l l l}
          \hline\hline
          & Luminosity & Higgs mass & Spurious signal\\
          \hline
          Uncertainty & $\pm2$\% & $\pm 0.2$ GeV &  $\pm 15\%$\\
          \hline
          \hline
      \end{tabular}
  \end{table}
First of all, we investigate asymptotic relation between $\qtil_\mu$ and $\hatmu$ in Eq.~\ref{eq:qmutilde_wald} using the toy simulations. Figure~\ref{fig:T6} and ~\ref{fig:qmutilde_muhat_ex1} are the scattering plot of $\qtil_\mu:\hatmu$ in Ex.~0 and Ex.~1. 
On the one hand, we can see that the asymptotic form still looks good even in these low-statistics cases. On the other hand, there are clear structures which reflect the discrete feature in the distribution of $\qtil_\mu$ or $\hatmu$.

\begin{figure}[htbp]
     \centering
     \includegraphics[width=0.45\textwidth]{ 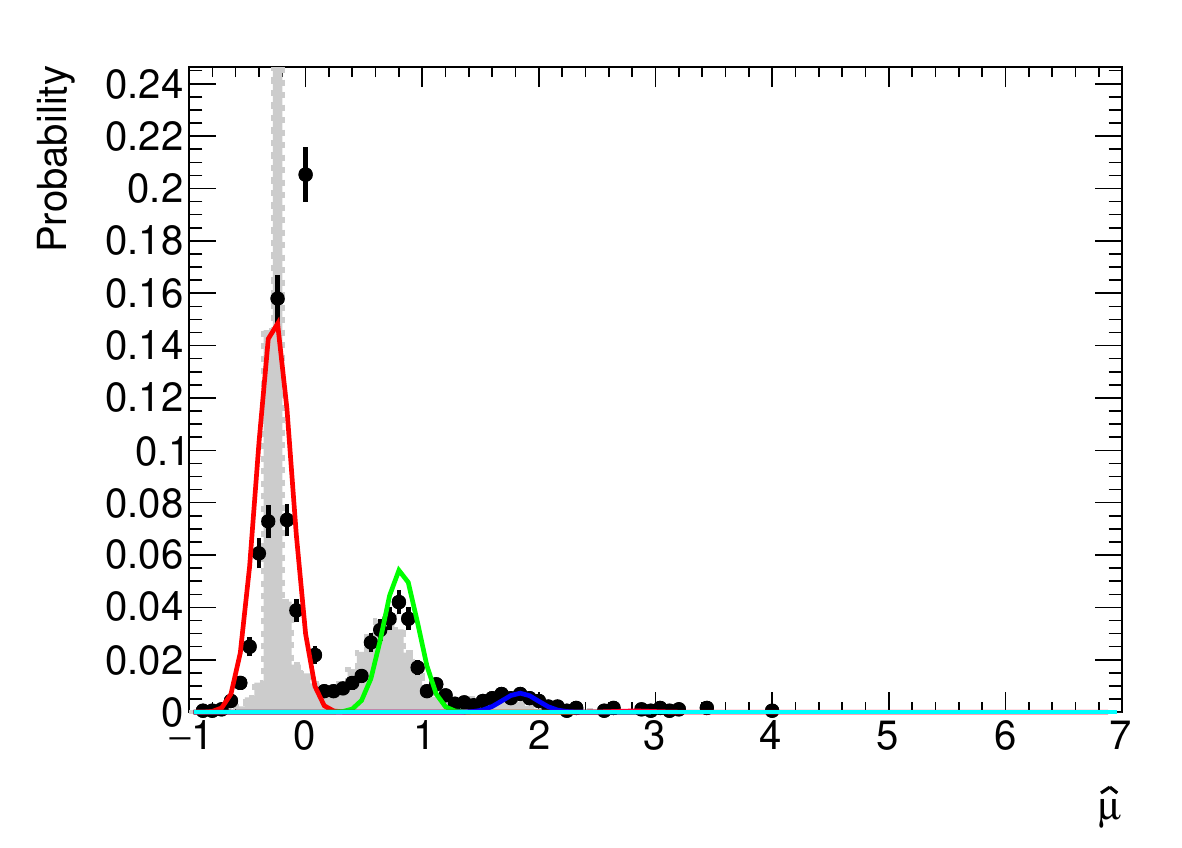}
     \includegraphics[width=0.45\textwidth]{ 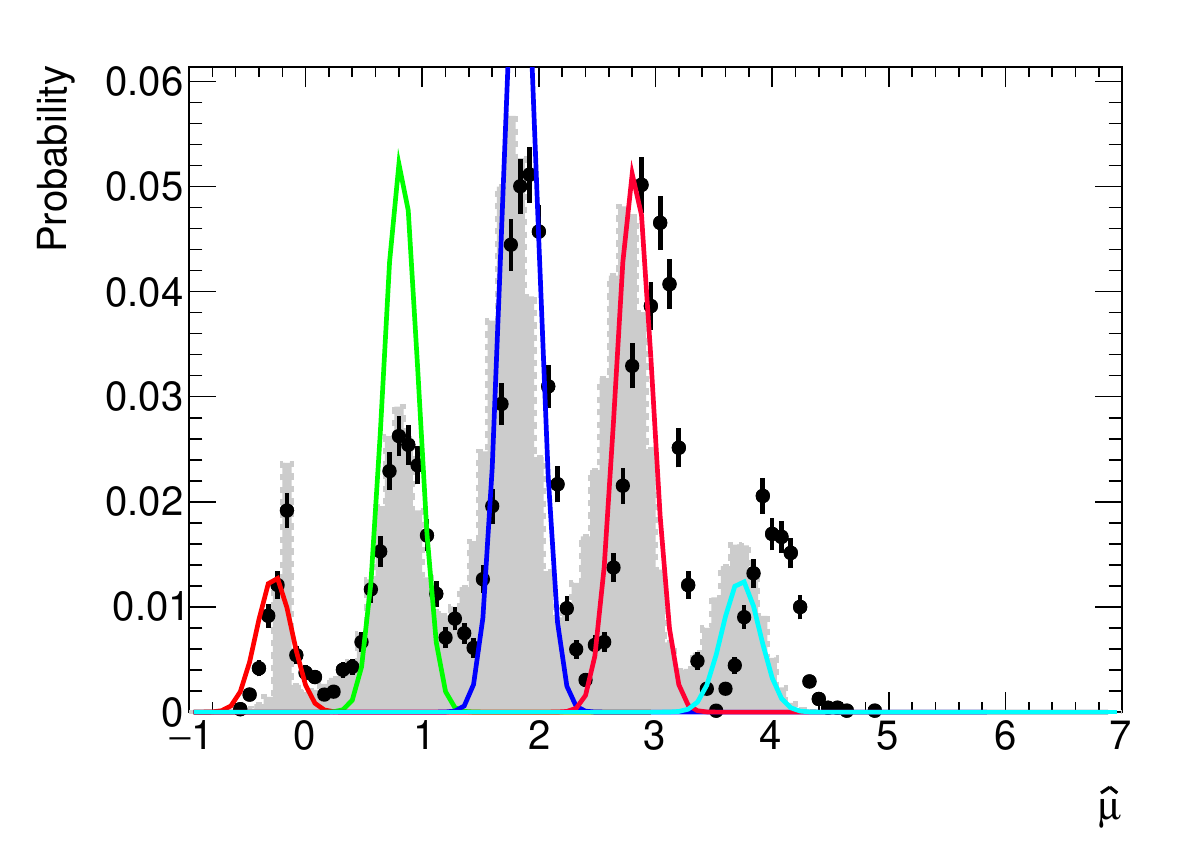}
     \caption{\label{fig:illustration_muhat}
     The distribution of $\hatmu$ in Ex.~0 under the hypothesis $\muH=0$ (L) and $\muH=\mu=3$ (R) for the number of total events being 4. The black dots represent the toy experiments. The curves with different colors represent the solutions predicted in the 2-bin model. The gray histograms are the prediction from the 6-bin model. 
     }
\end{figure}

\begin{figure}[htbp]
     \centering
     \includegraphics[width=0.45\textwidth]{ 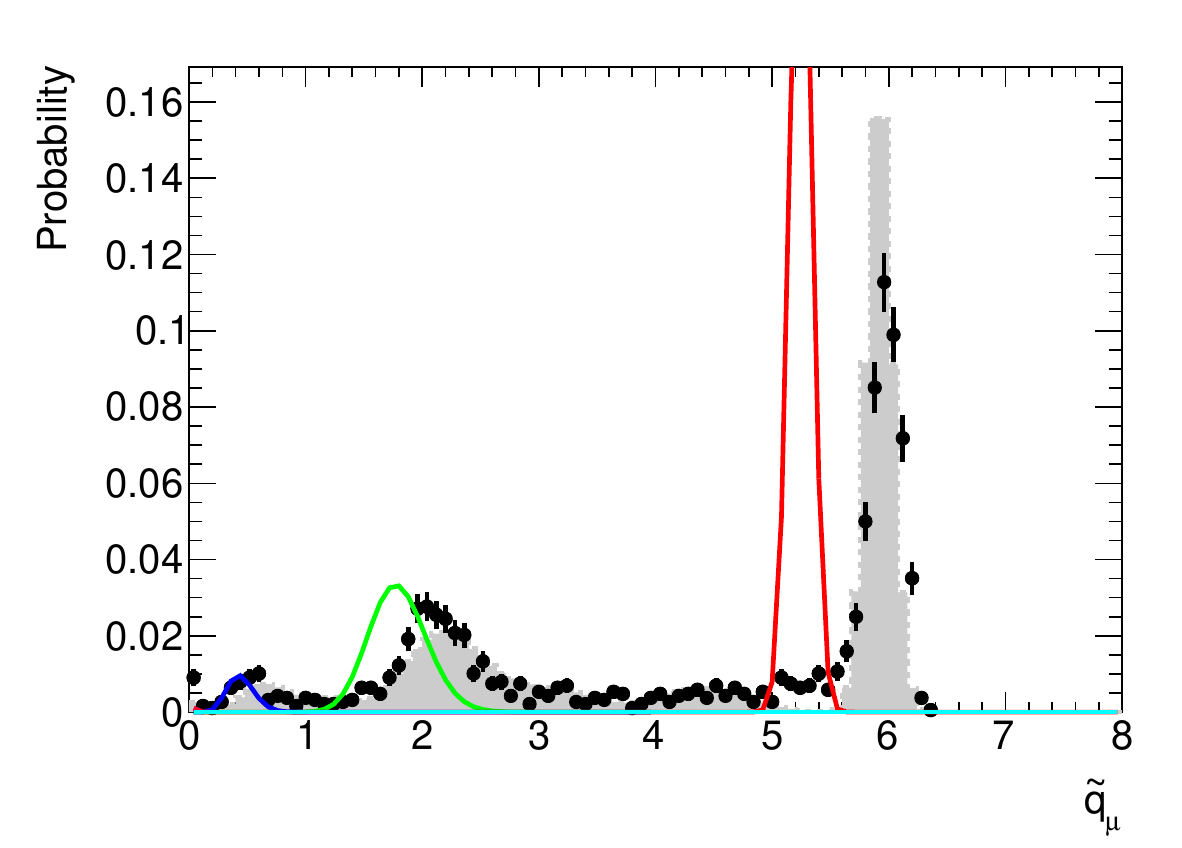}
     \includegraphics[width=0.45\textwidth]{ 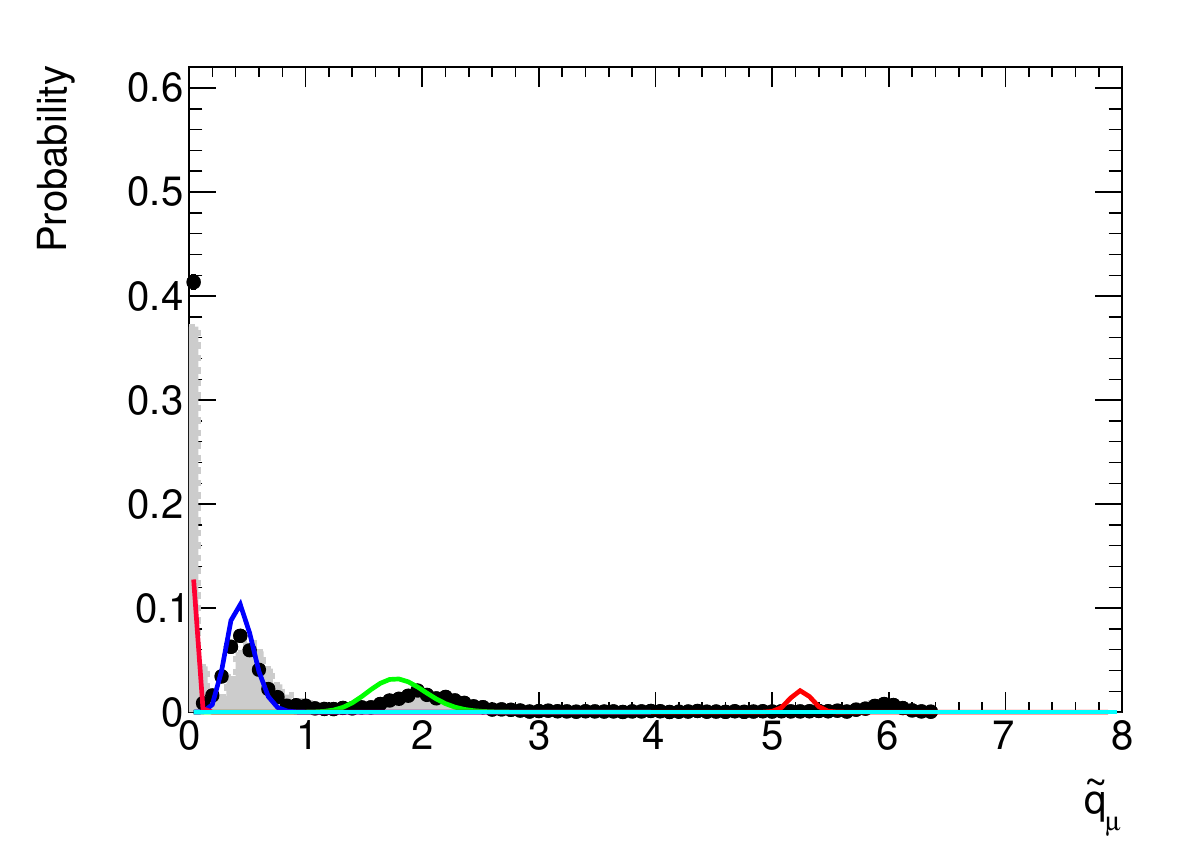}
     \caption{\label{fig:illustration_qmutilde}
     The distribution of $\qtil_\mu$ in Ex.~0 under the hypothesis $\muH=0$ (L) and $\muH=\mu=3$ (R) for the number of total events being 4. The black dots represent the toy experiments. The curves with different colors represent the solutions predicted in the 2-bin model. The gray histograms are the prediction from the 6-bin model. 
     }
\end{figure}

Second, we examine the discrete features. For toy simulations with 4 observed events, the distribution of $\hatmu$ and $\qtil_\mu$ in Ex.~0 is shown and compared with the predictions from the 2-bin and 6-bin models in Fig.~\ref{fig:illustration_muhat} and Fig.~\ref{fig:illustration_qmutilde}, respectively. The $\hatmu$ distribution exhibits 5 equal-distant peaks (mostly visible in right plot of Fig.~\ref{fig:illustration_muhat}), which follow a roughly binomial distribution.
These results validate the three observations presented in Sec.~\ref{sec:2bin}.  
Moreover, the 6-bin model demonstrates better agreement with the toy simulation results compared to the 2-bin model.

\begin{figure}[htbp]
    \centering
     \includegraphics[width=0.45\textwidth]{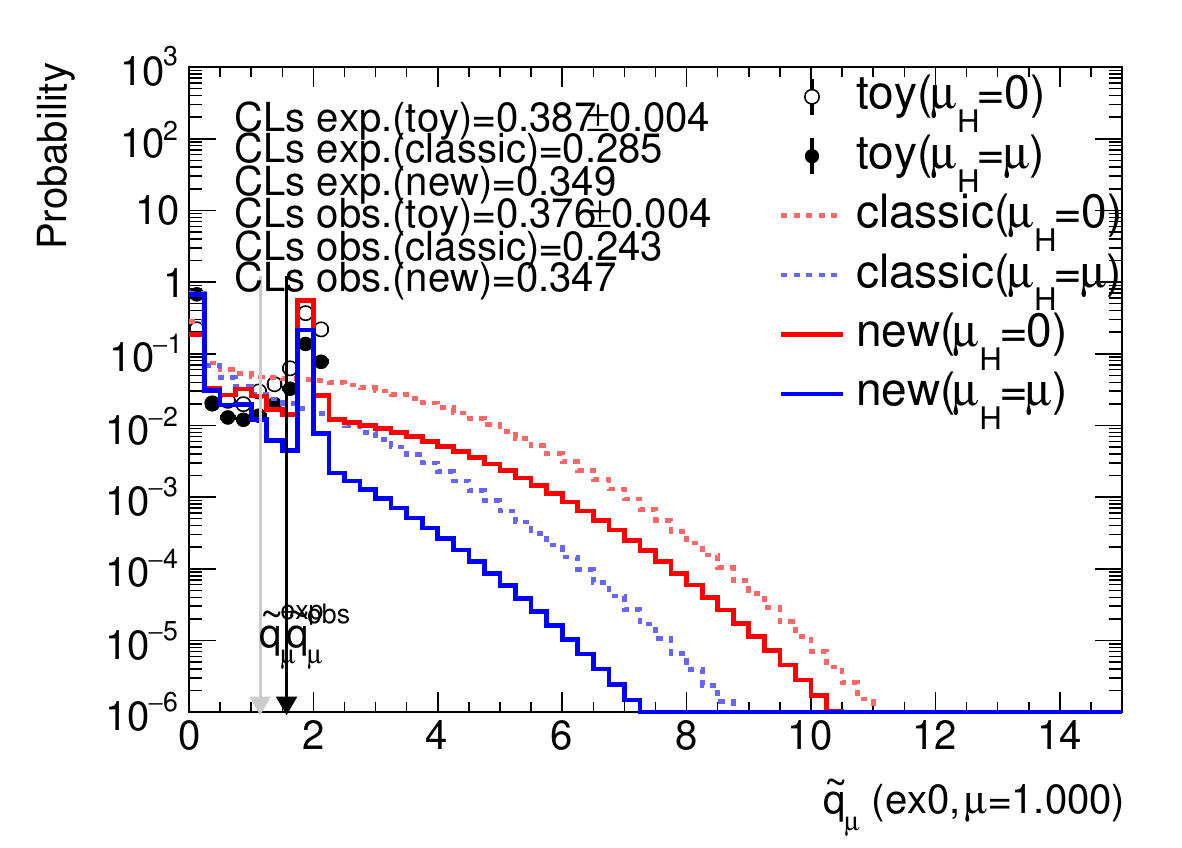}
     \includegraphics[width=0.45\textwidth]{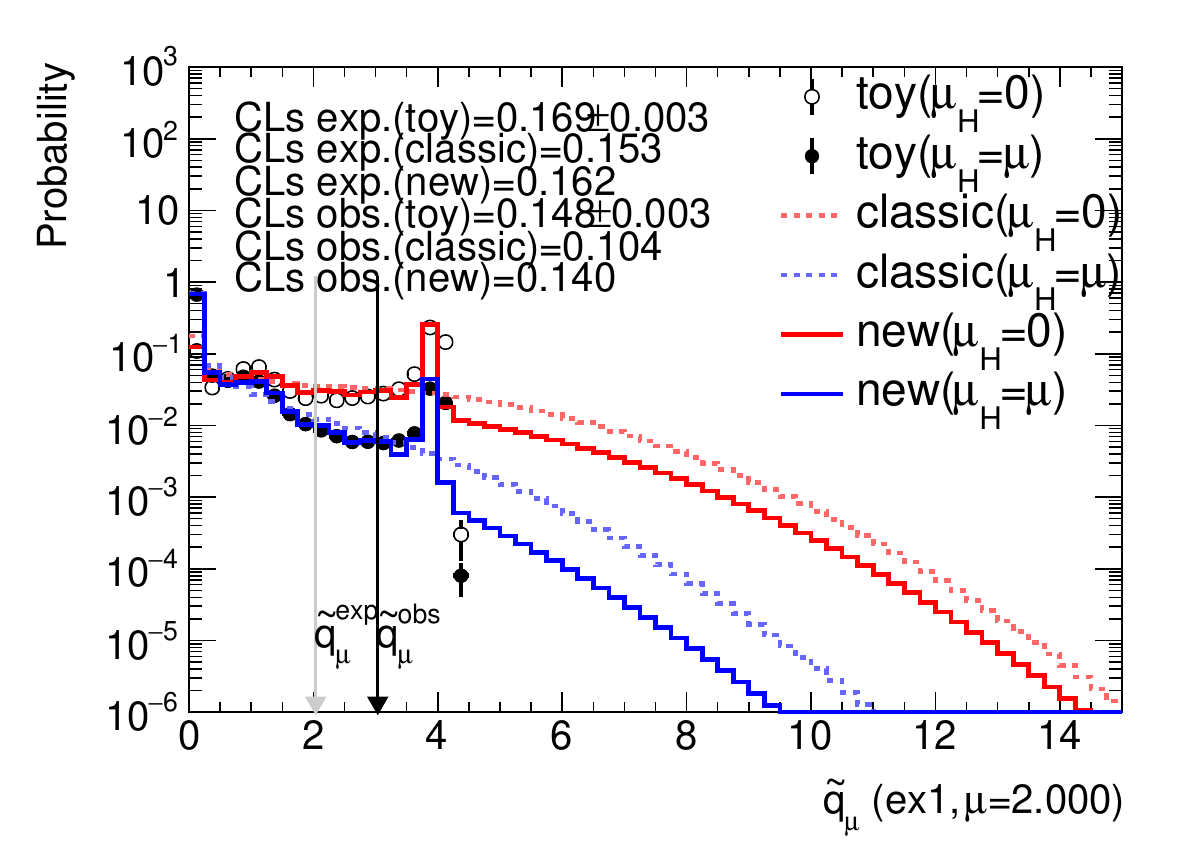}\\
     \includegraphics[width=0.45\textwidth]{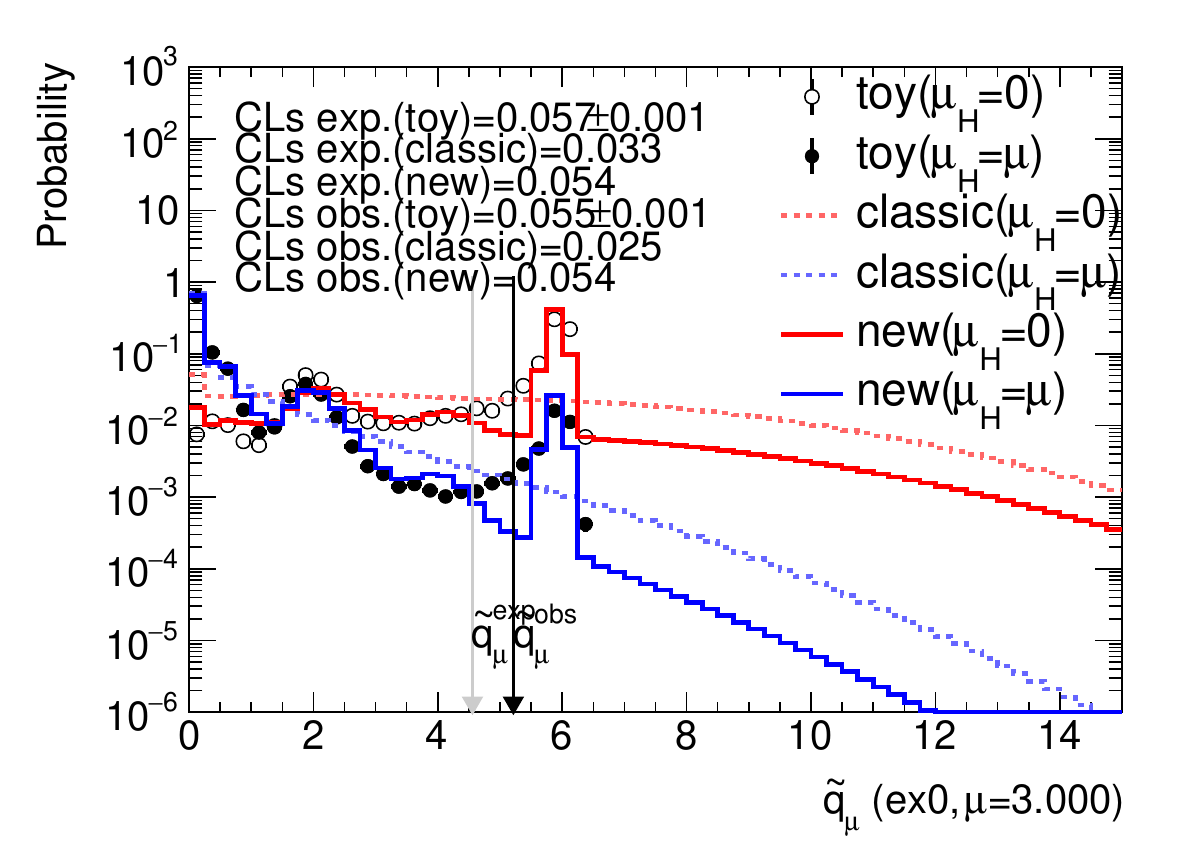}
     \includegraphics[width=0.45\textwidth]{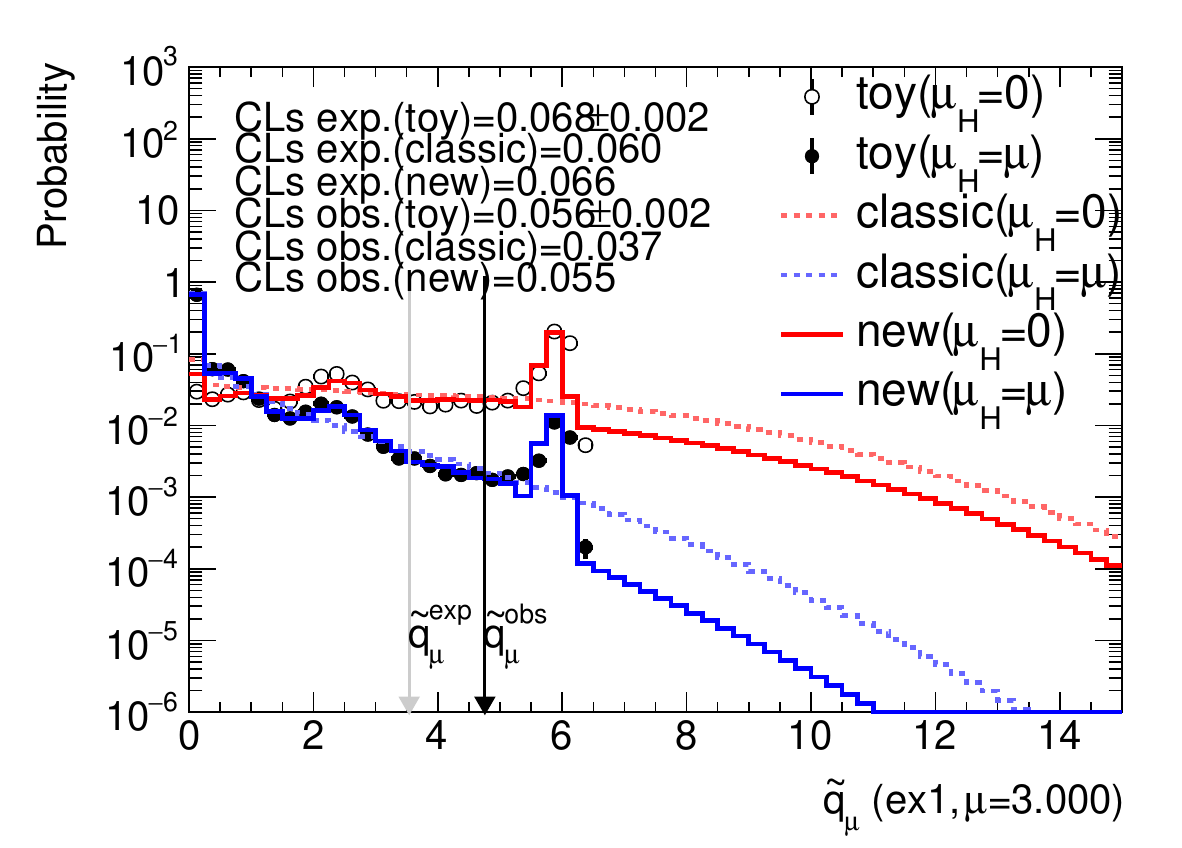}\\
     \includegraphics[width=0.45\textwidth]{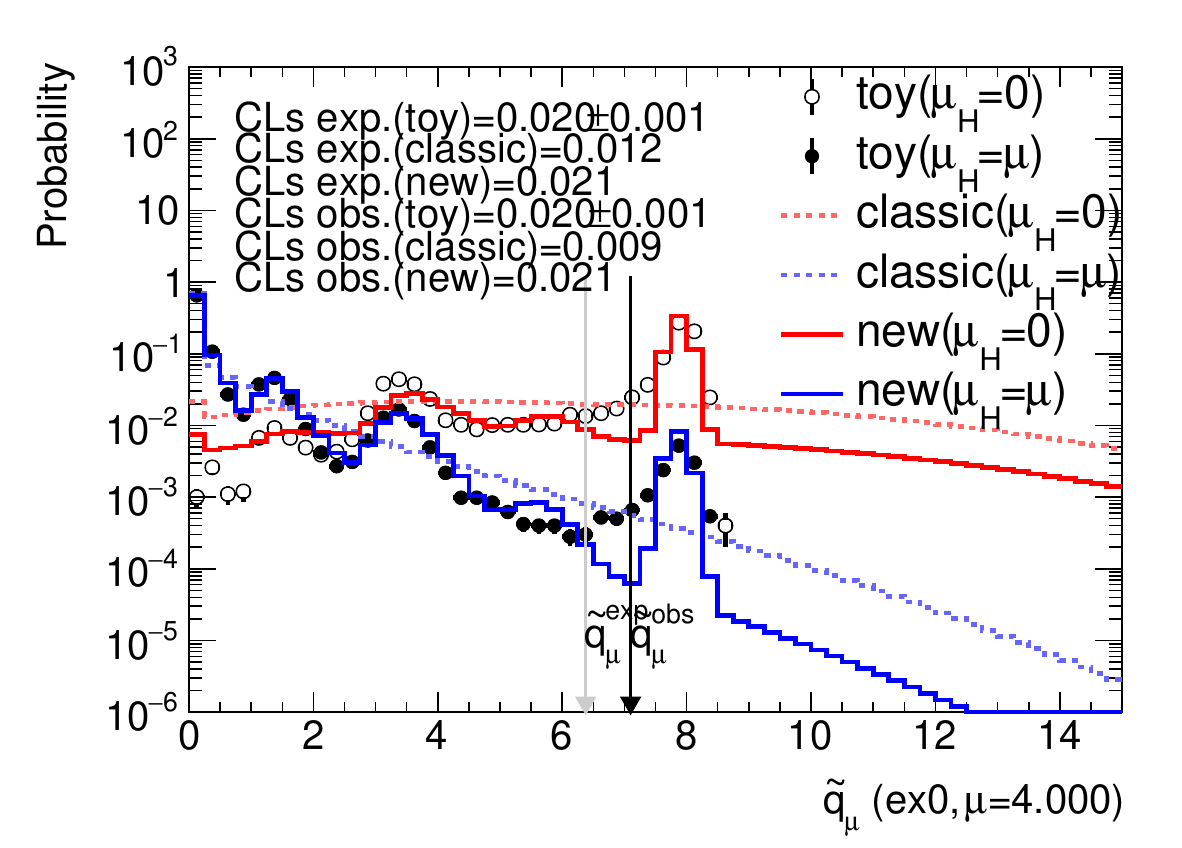}
     \includegraphics[width=0.45\textwidth]{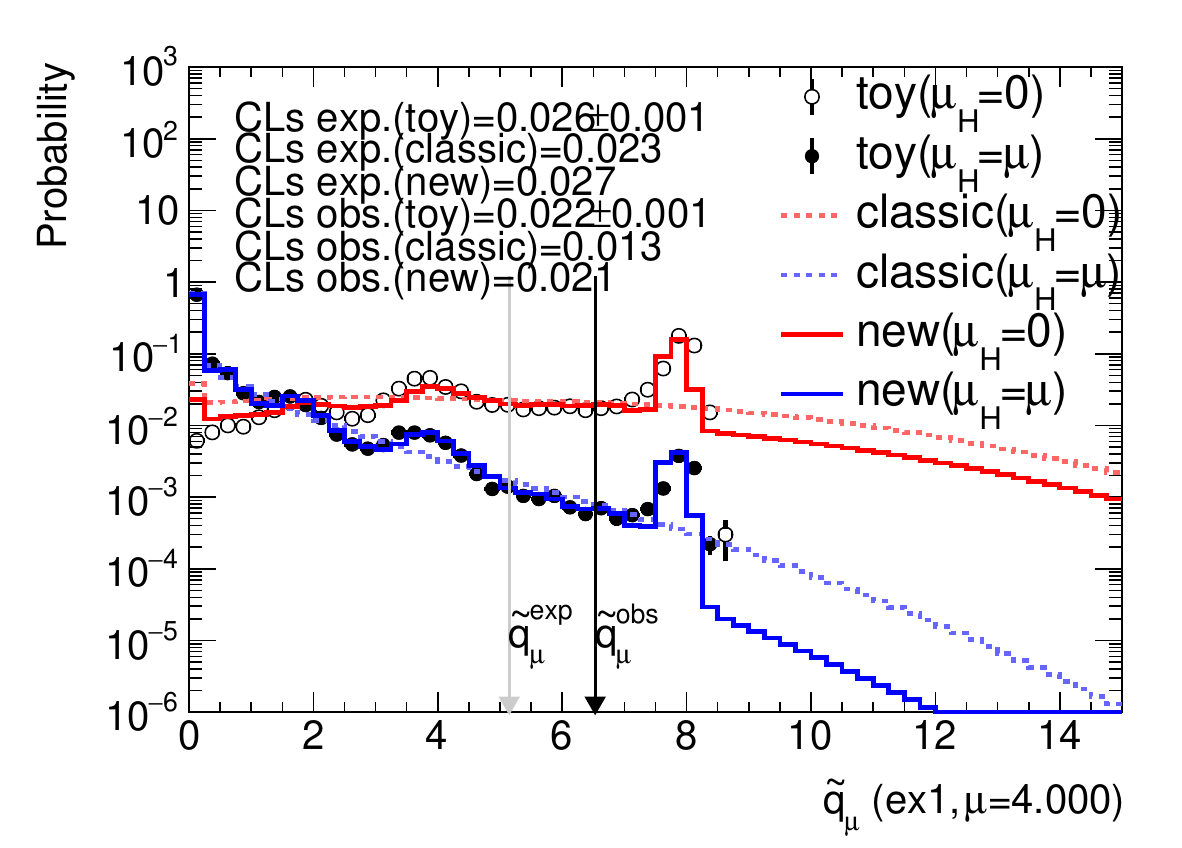}
     \caption{\label{fig:qmutilde_datanegative}
     The probability distributions of $\qtil_\mu$ in Ex.~0 (Left column) and Ex.~1 (Right column) for an ``observed'' dataset with a negative signal strength.
     The black dots and open circles represent the toy MC results. The blue/red solid histograms represent the new asymptotic formulae in this work while the blue/red dashed histograms represent the classic asymptotic  formulae from Wald's approximation. The black and gray arrows represent the observed and expected $\tilde{q}_\mu$, respectively.
     }
\end{figure}

\begin{figure}[htbp]
    \centering
     \includegraphics[width=0.45\textwidth]{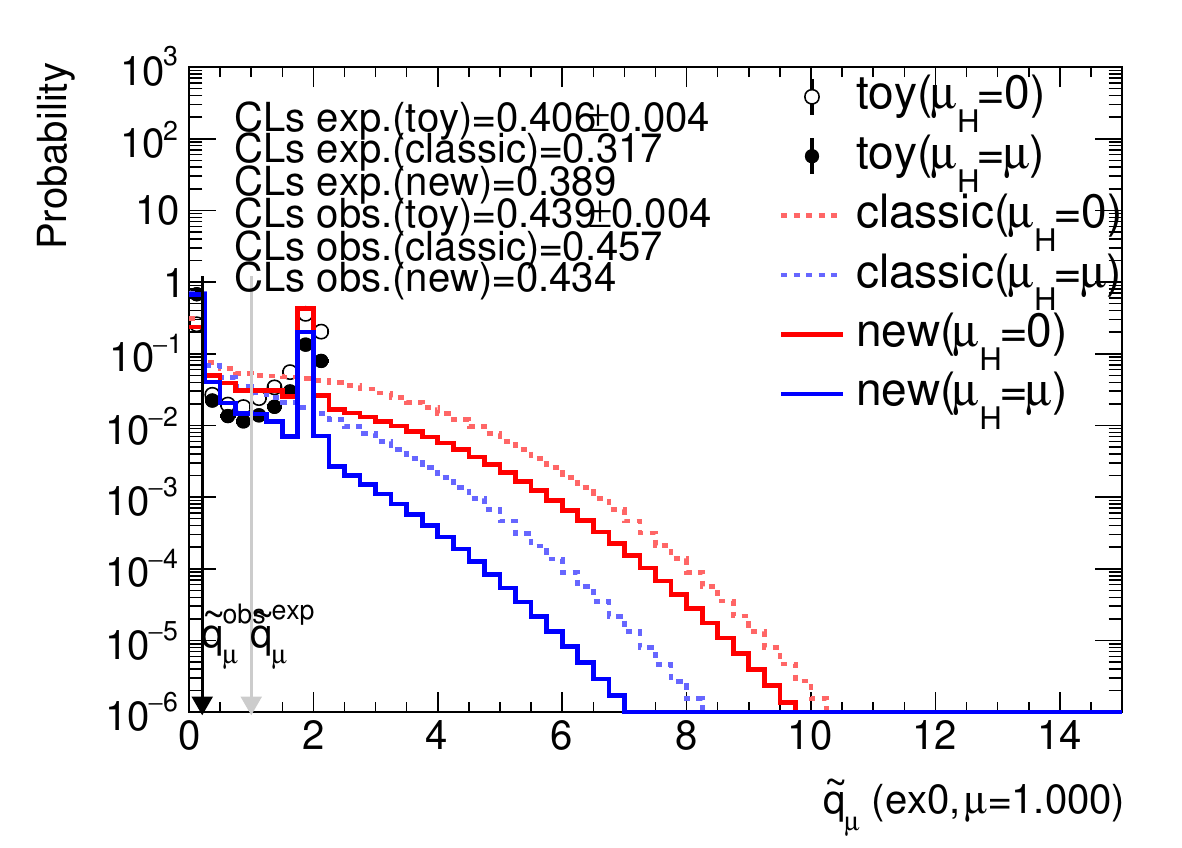}
     \includegraphics[width=0.45\textwidth]{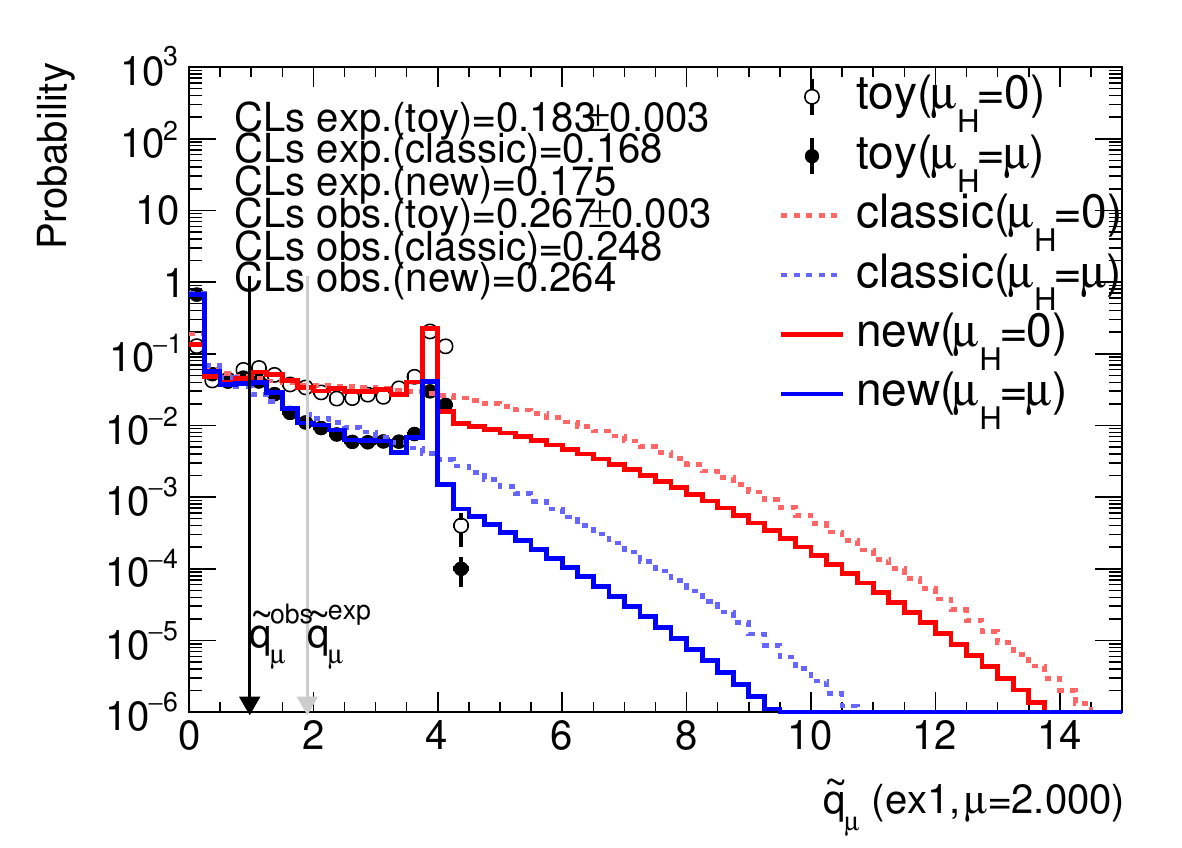}\\
     \includegraphics[width=0.45\textwidth]{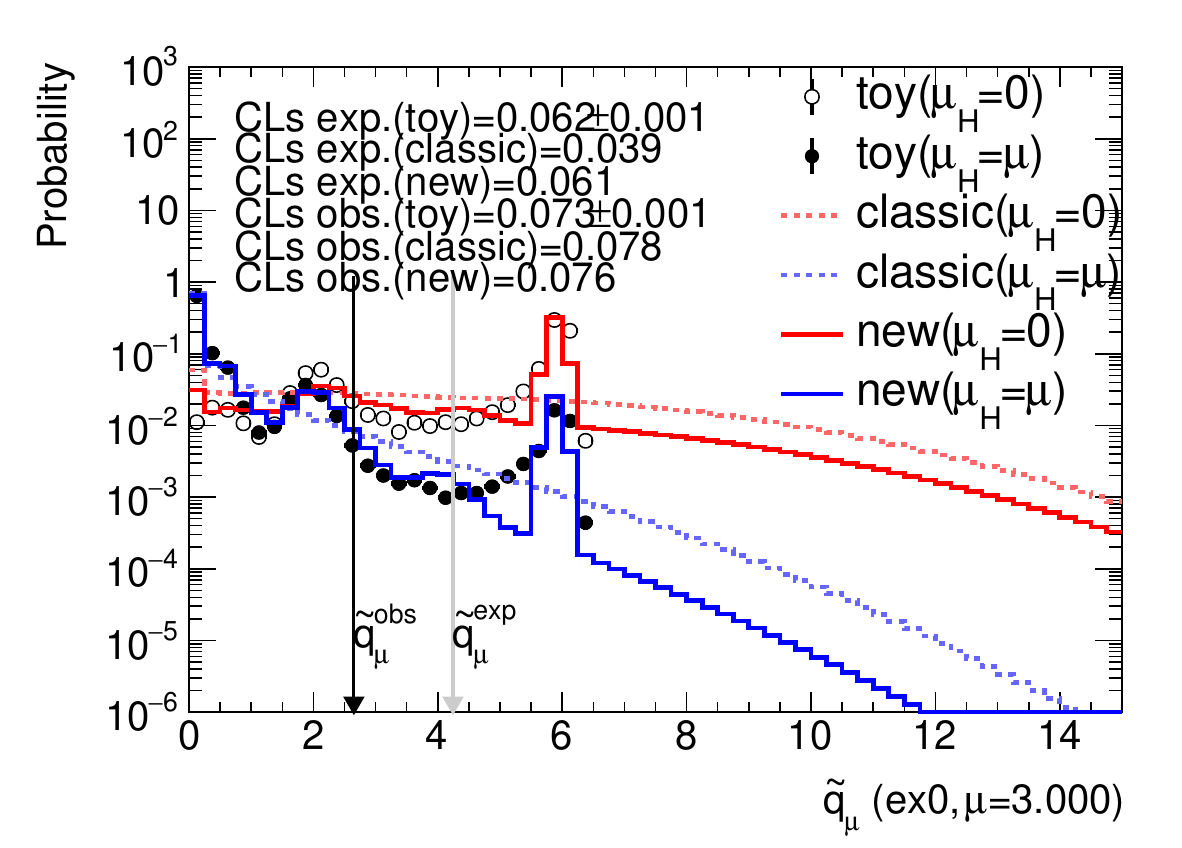}
     \includegraphics[width=0.45\textwidth]{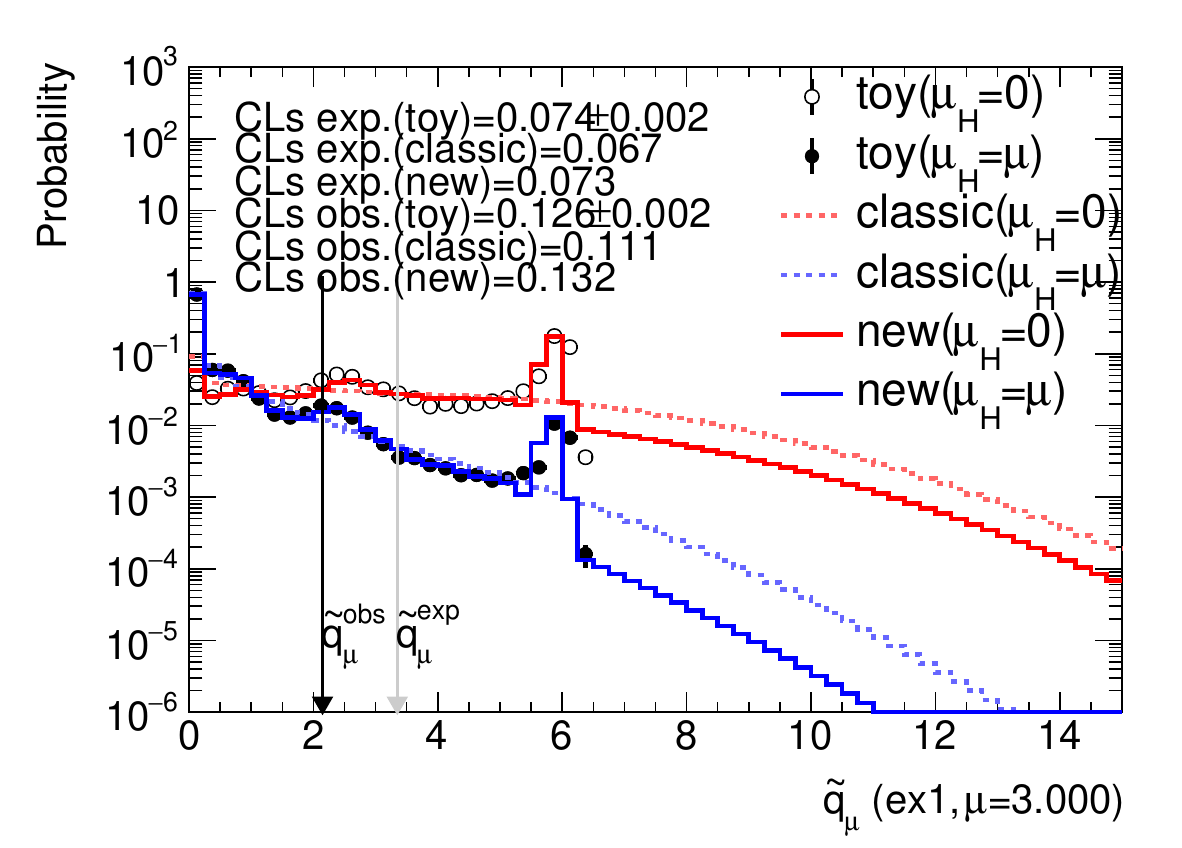}\\
     \includegraphics[width=0.45\textwidth]{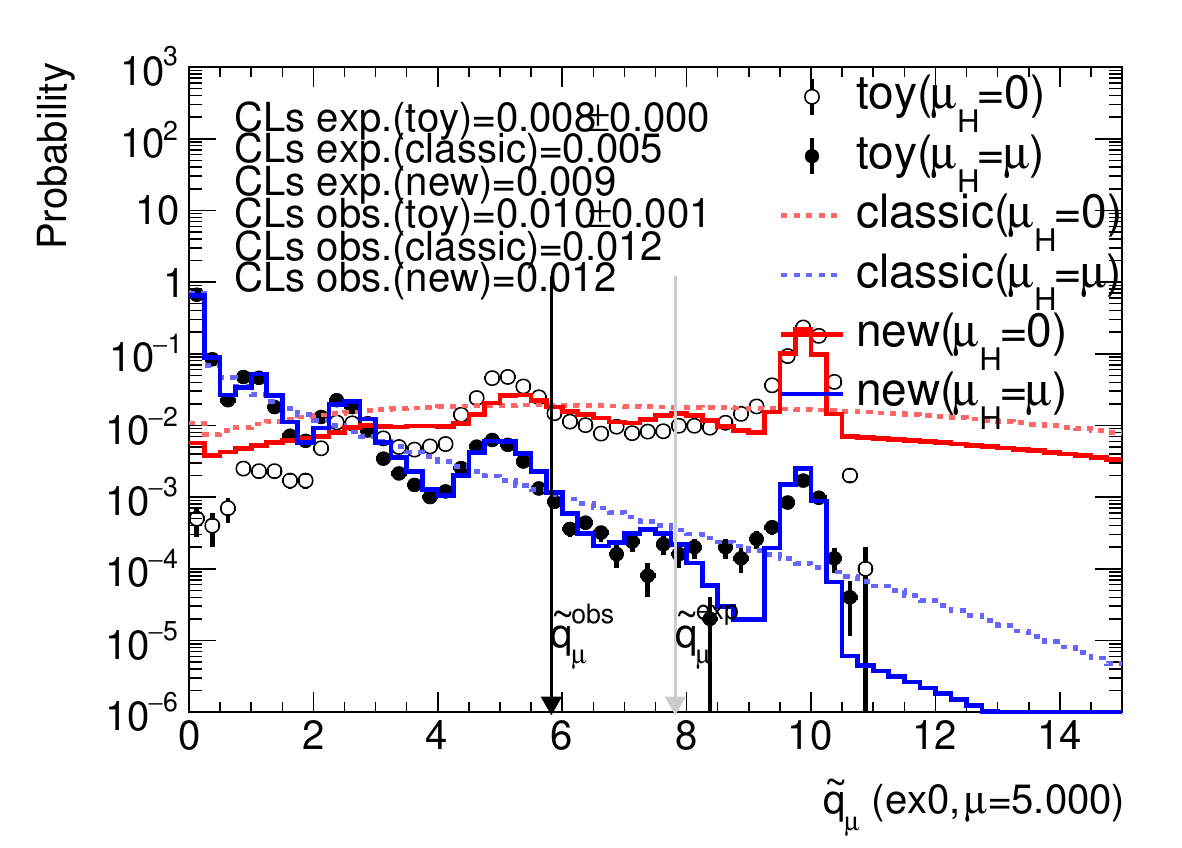}
     \includegraphics[width=0.45\textwidth]{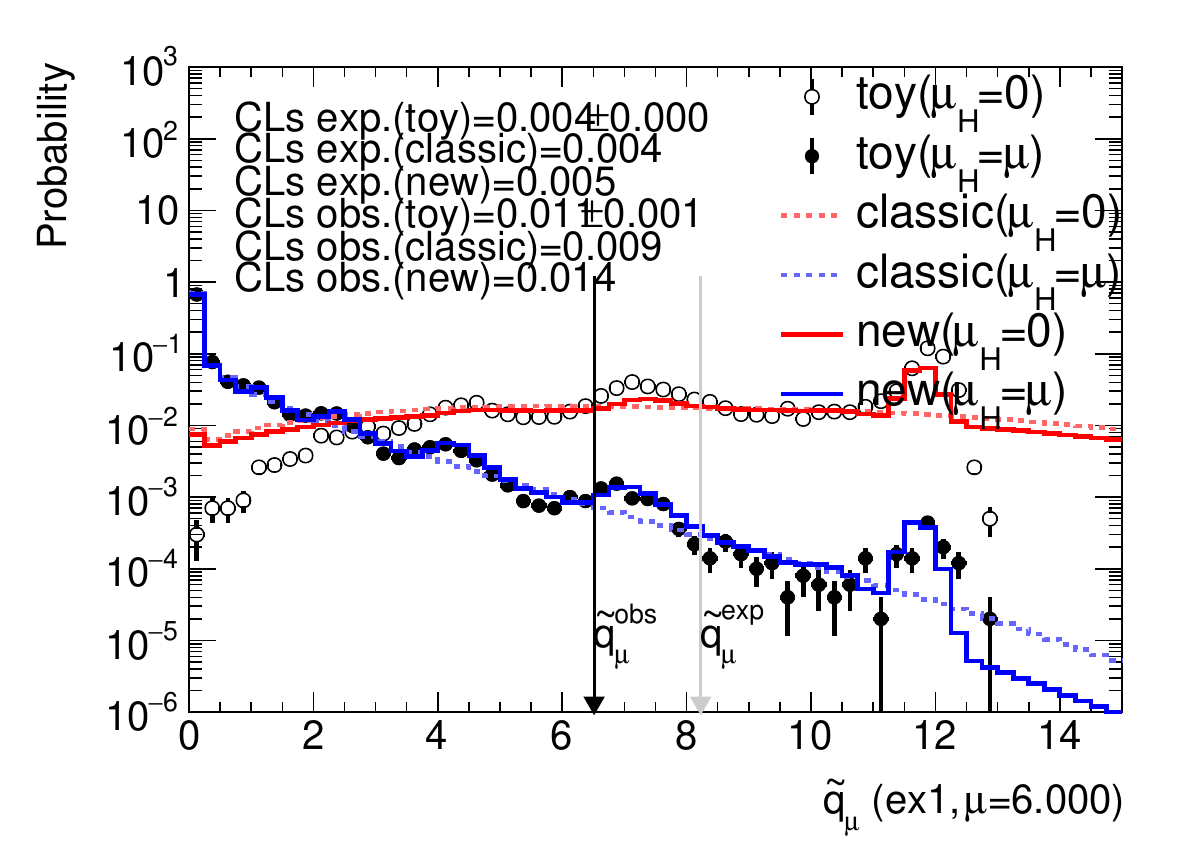}
     \caption{\label{fig:qmutilde_data0p5}
     The probability distributions of $\qtil_\mu$ in Ex.~0 (Left column) and Ex.~1 (Right column) for an ``observed'' dataset with signal strength equal to 0.5.
     The black dots and open circles represent the toy MC results. The blue/red solid histograms represent the new asymptotic formulae in this work while the blue/red dashed histograms represent the classic asymptotic  formulae from Wald's approximation. The black and gray arrows represent the observed and expected $\tilde{q}_\mu$, respectively.
     }
\end{figure}

\begin{figure}[htbp]
    \centering
     \includegraphics[width=0.45\textwidth]{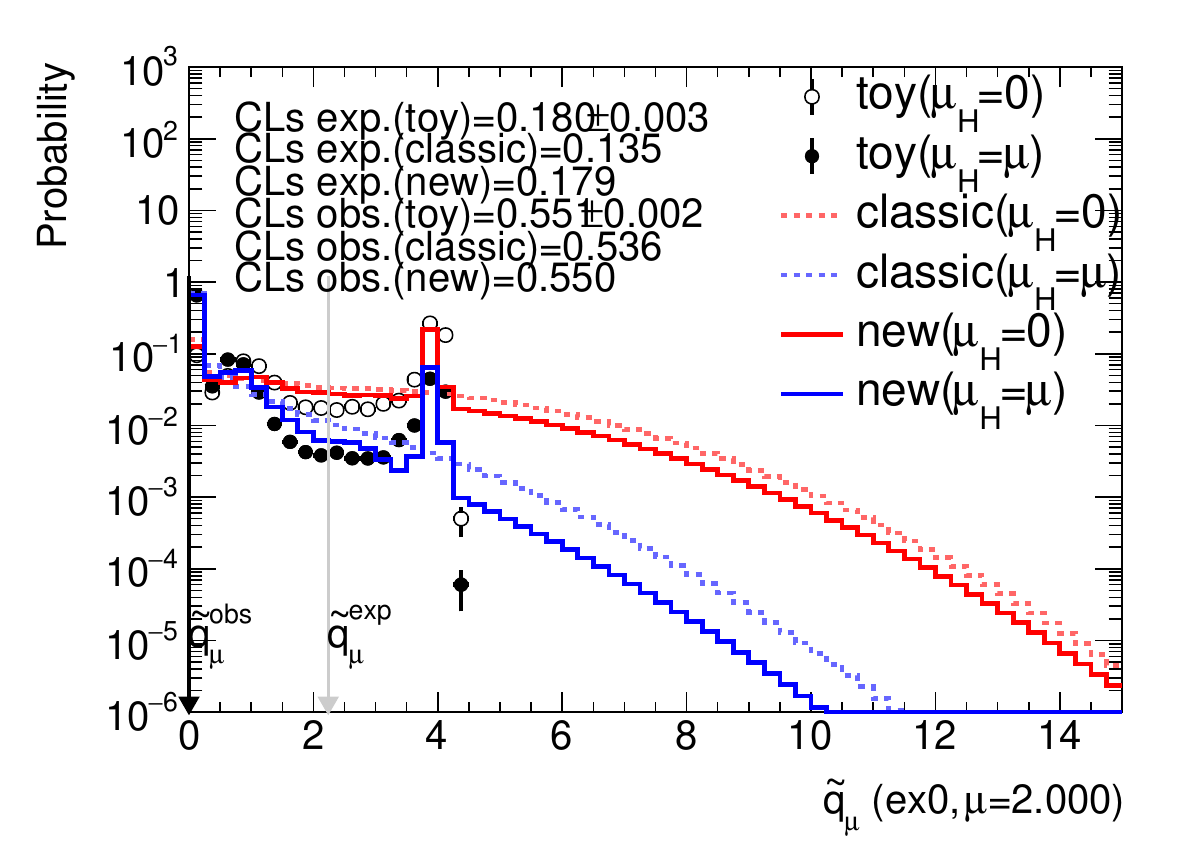}
     \includegraphics[width=0.45\textwidth]{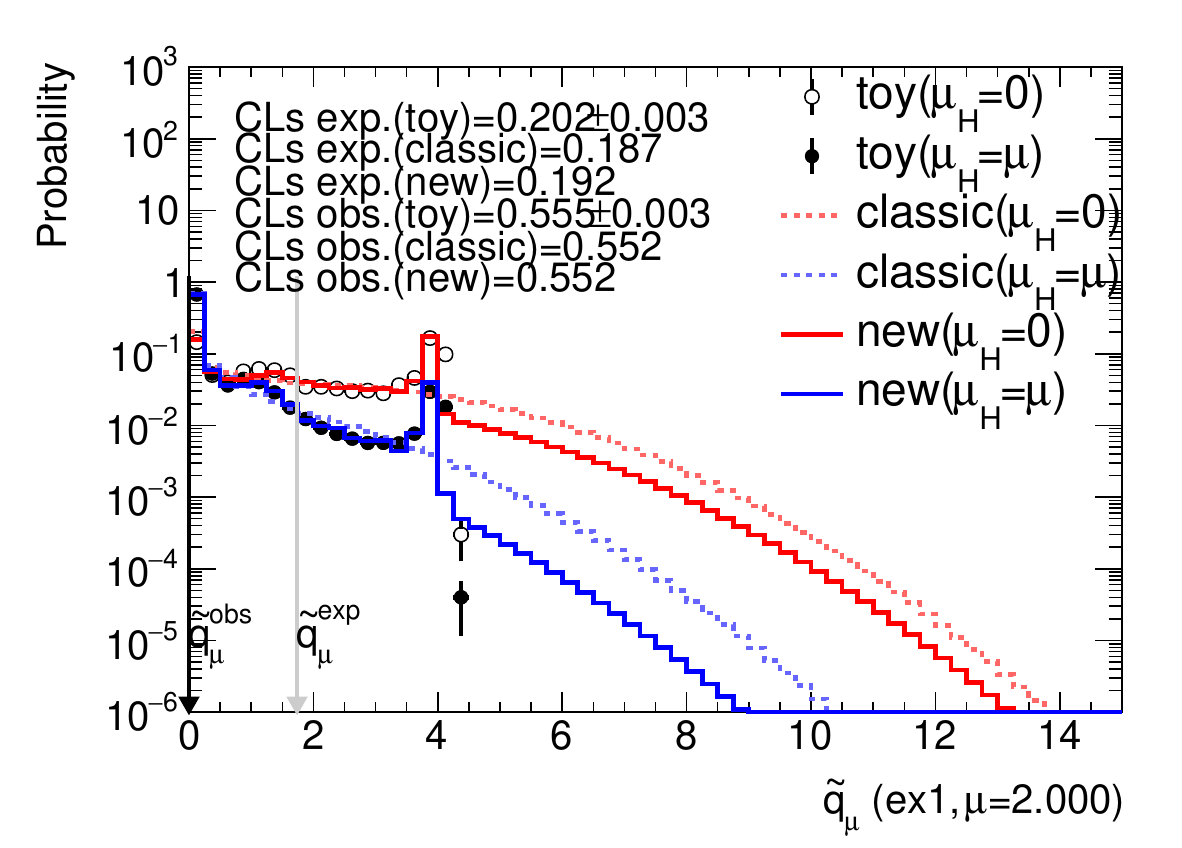}\\
     \includegraphics[width=0.45\textwidth]{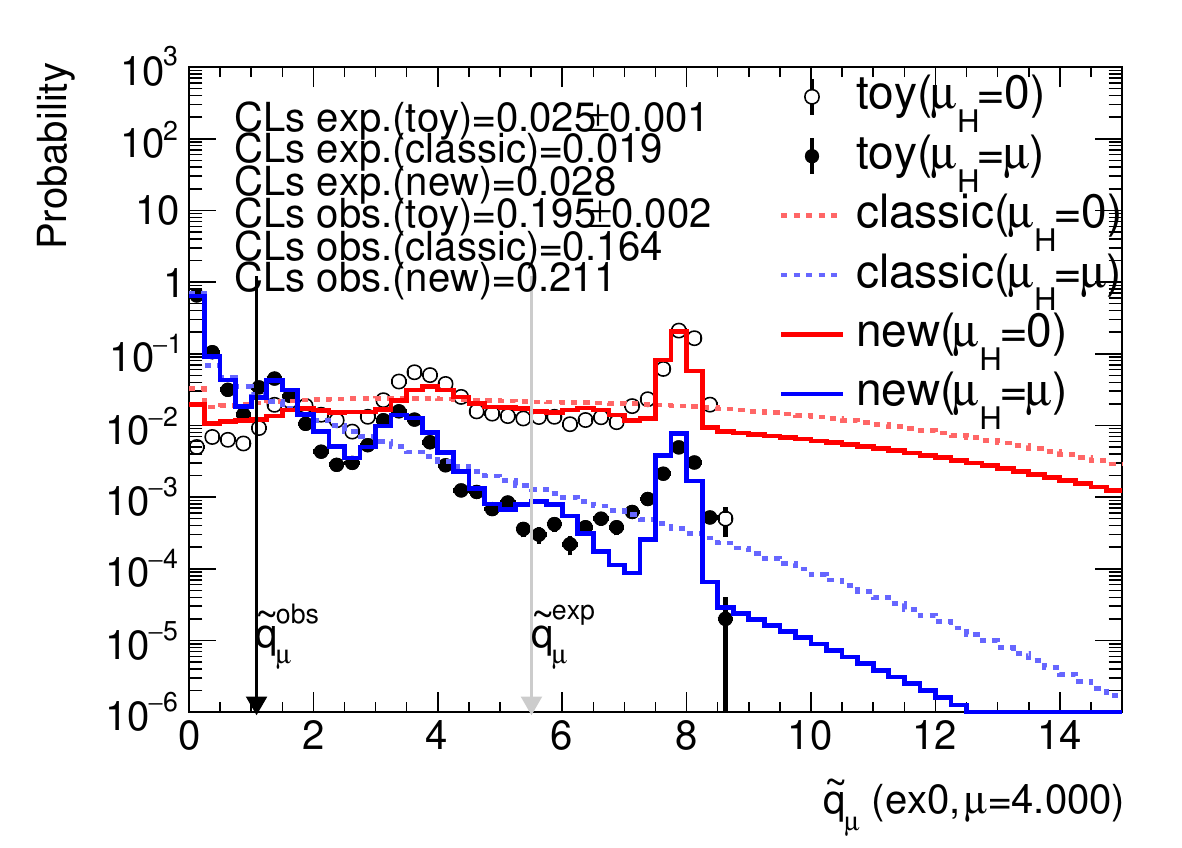}
     \includegraphics[width=0.45\textwidth]{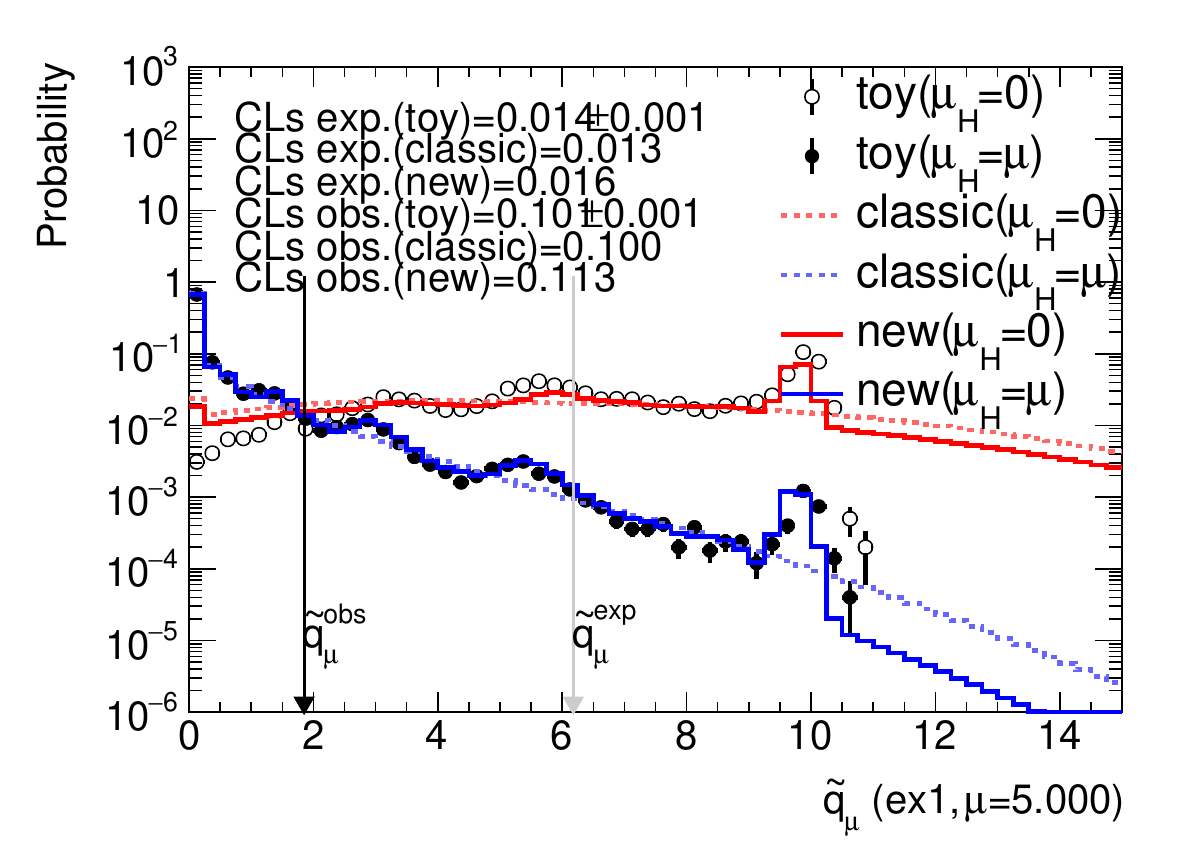}\\
     \includegraphics[width=0.45\textwidth]{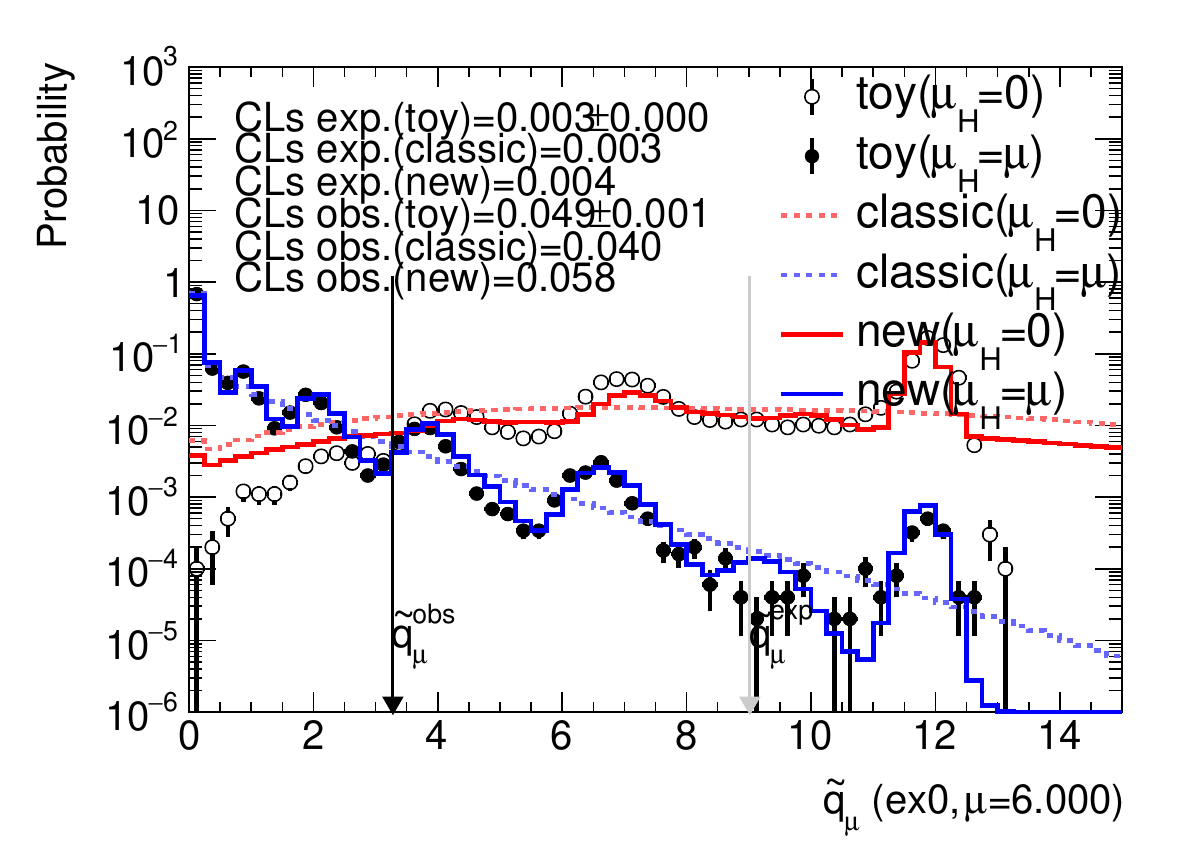}
     \includegraphics[width=0.45\textwidth]{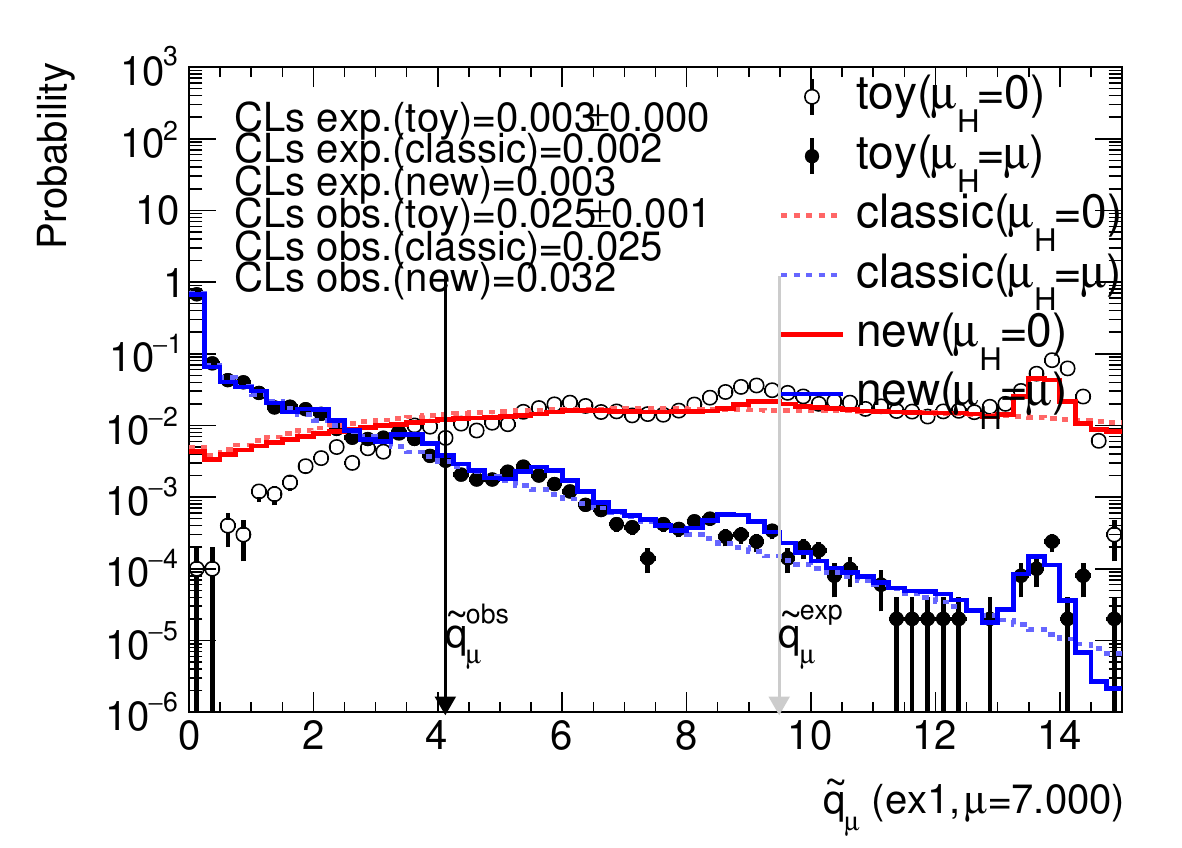}
     \caption{\label{fig:qmutilde_data2}
     The probability distributions of $\qtil_\mu$ in Ex.~0 (Left column) and Ex.~1 (Right column) for an ``observed'' dataset with signal strength equal to 2.
     The black dots and open circles represent the toy MC results. The blue  solid/dashed histograms represent the new asymptotic formulae in this work while the red solid/dashed histograms represent the classic asymptotic  formulae from Wald's approximation. The black and gray arrows represent the observed and expected $\tilde{q}_\mu$, respectively.
     }
\end{figure}

\begin{figure}[htbp]
     \centering
     \includegraphics[width=0.45\textwidth]{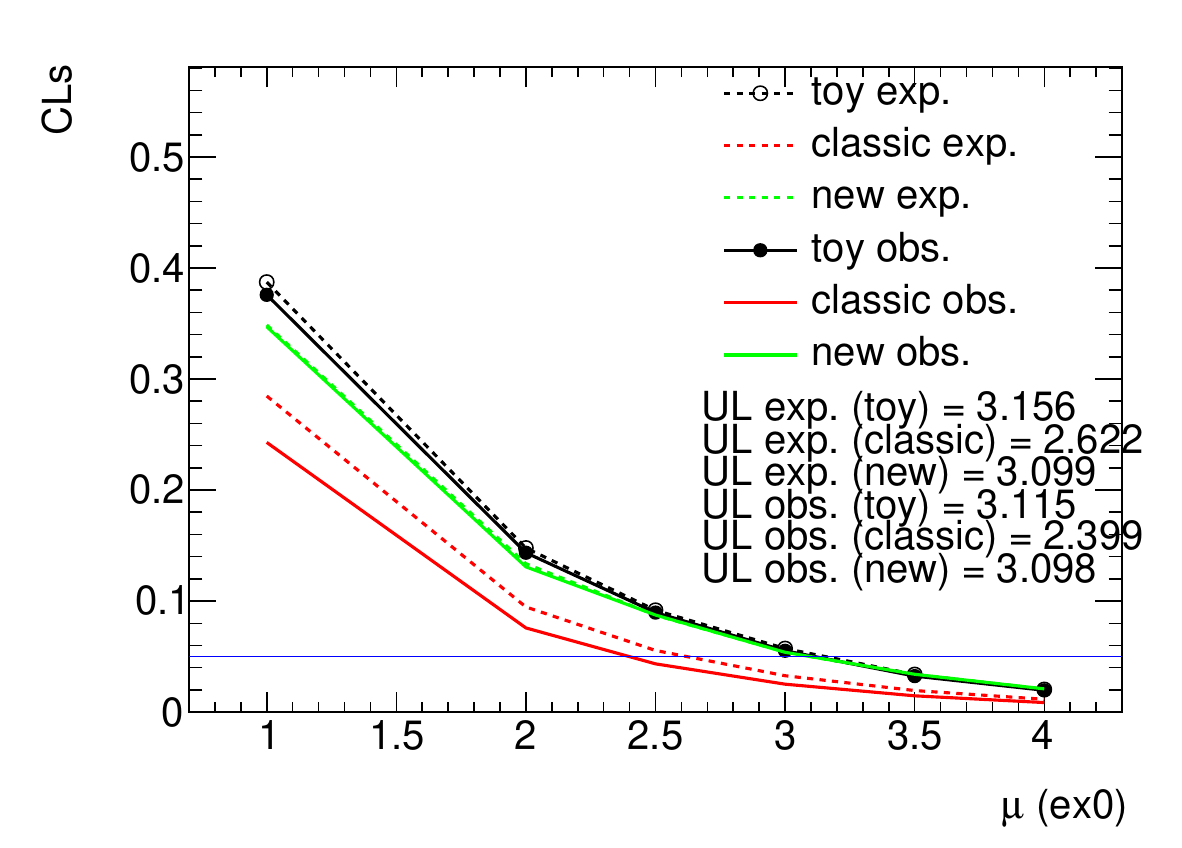}
     \includegraphics[width=0.45\textwidth]{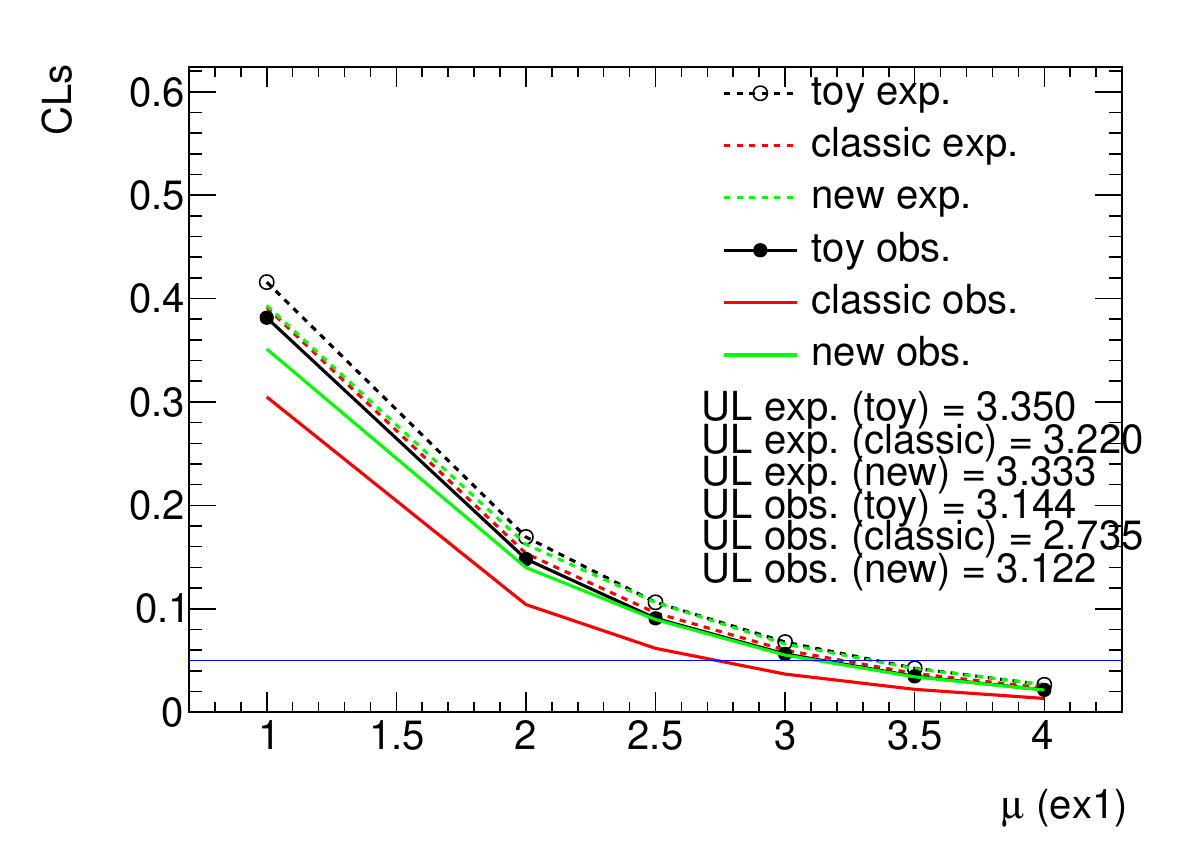}\\
     \includegraphics[width=0.45\textwidth]{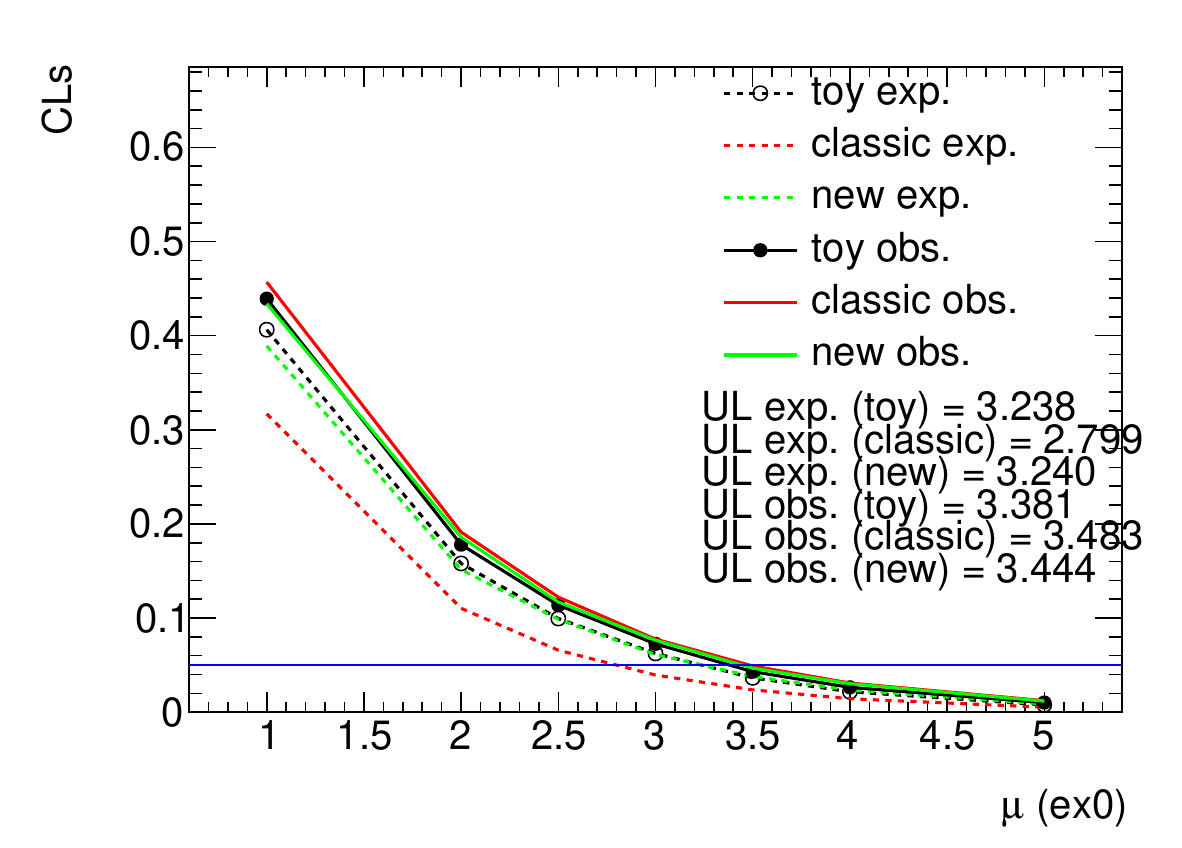}
     \includegraphics[width=0.45\textwidth]{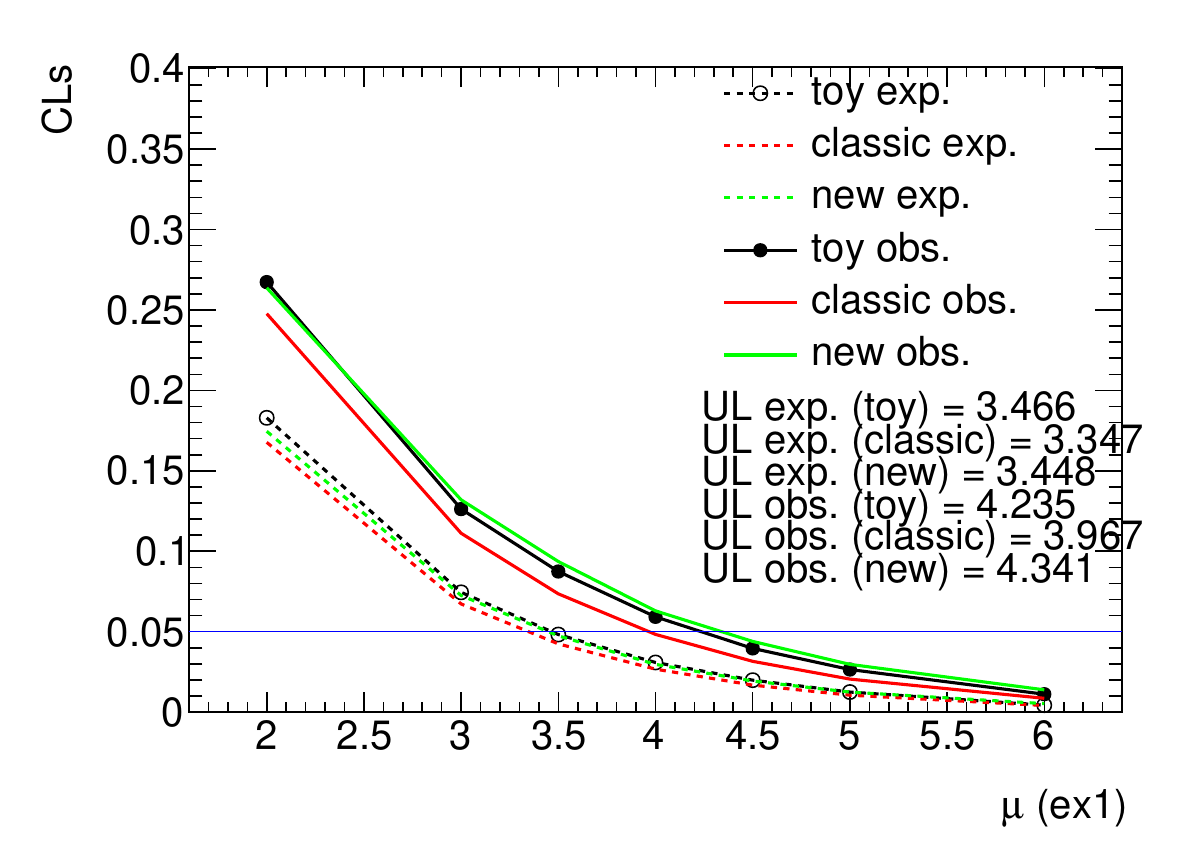}\\
     \includegraphics[width=0.45\textwidth]{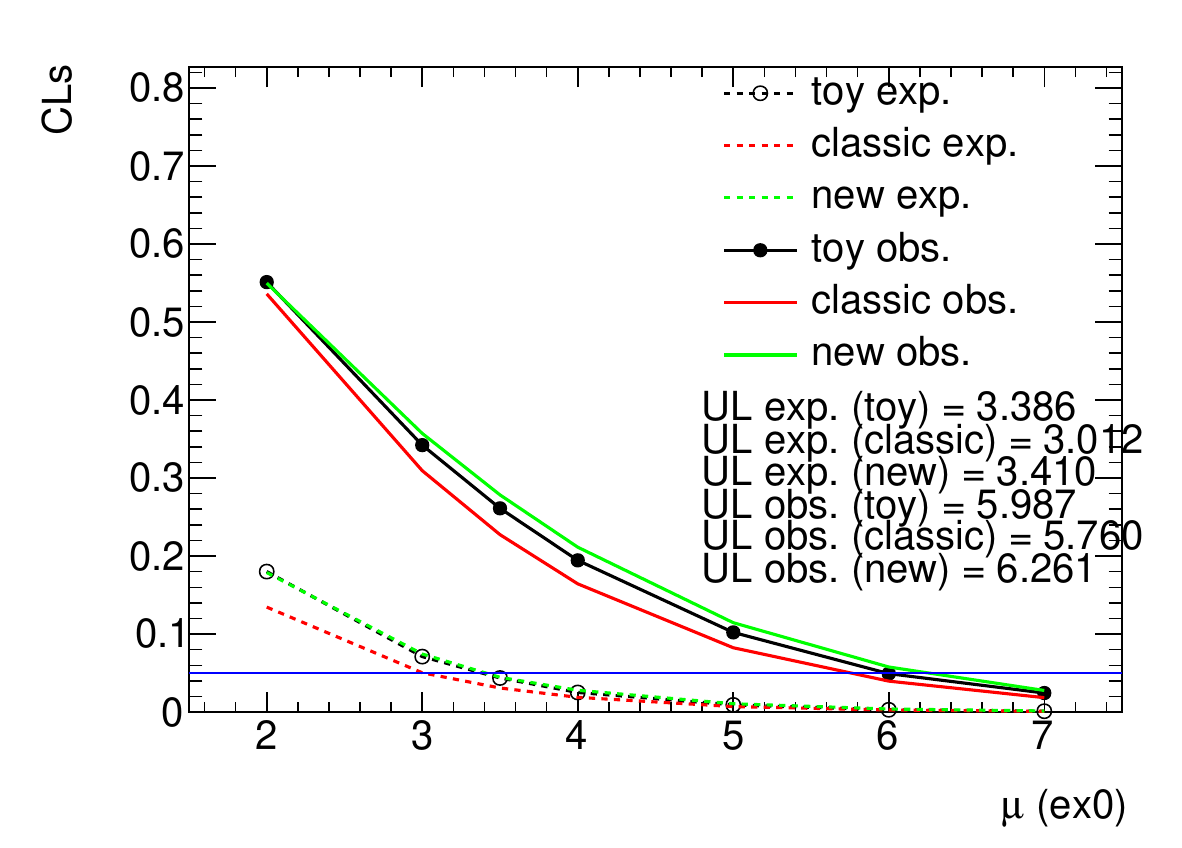}
     \includegraphics[width=0.45\textwidth]{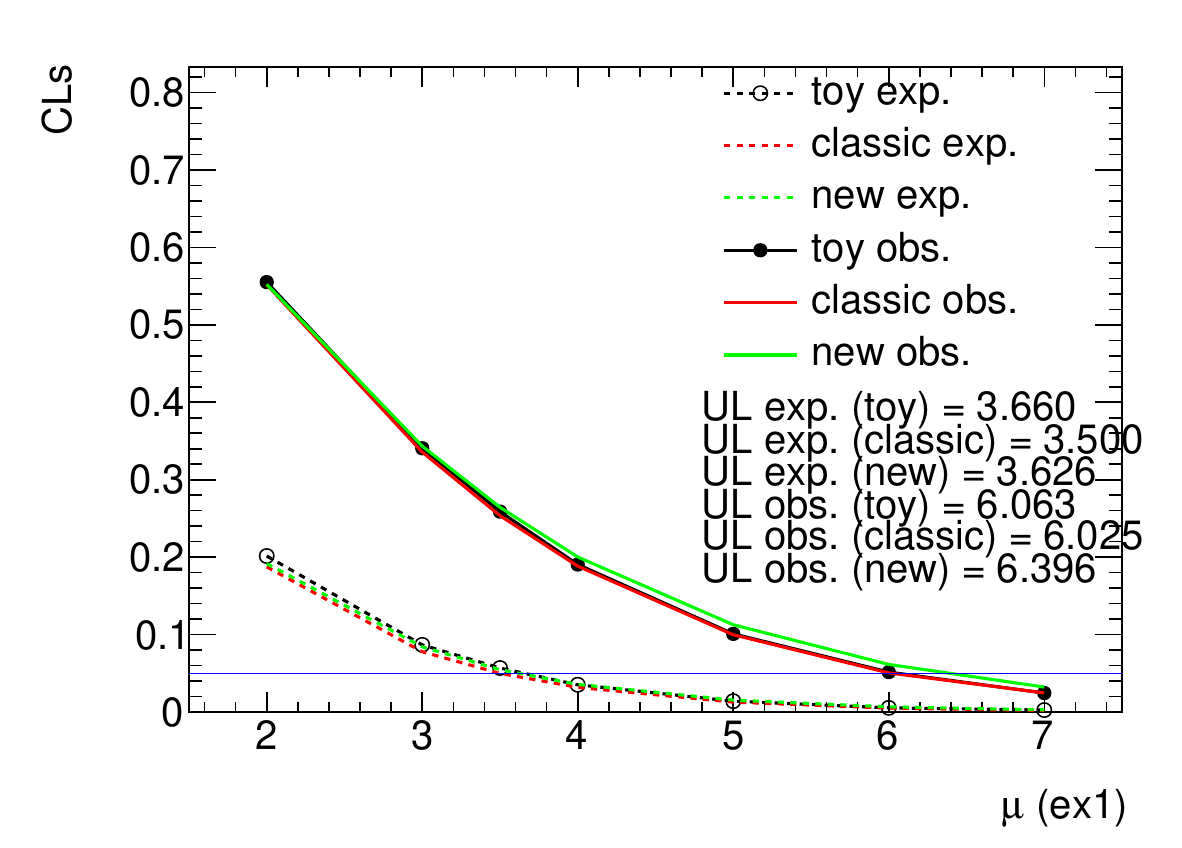}
     \caption{\label{fig:CLs}
     CLs as a function of $\mu$ in Ex.~0 (L) and Ex.~1 (R) using the test statistic $\qtil_\mu$. From top to bottom, they represent different observed datasets. The black curves with markers show the toy MC results. The red and green curves are the predictions from the classic and new asymptotic formulae, respectively.  
     }
 \end{figure}

Finally, Fig.~\ref{fig:qmutilde_datanegative}~\ref{fig:qmutilde_data0p5} and ~\ref{fig:qmutilde_data2} present the distributions of $\qtil_\mu$ from the toy simulations in Ex.~0 and Ex.~1, comparing results for different test signal strengths and ``observed'' datasets with predictions from both the classic and new approaches. 
The new formalism successfully captures the discrete features arising from low statistics. In Fig.~\ref{fig:CLs}, we show CLs~\cite{CLs_Zech,CLs} as a function of $\mu$
and also upper limits at 95~\% confidence level for different observed datasets. 
A direct comparison of the limits in Fig.~\ref{fig:limits_nsmall} confirms that the new approach consistently outperforms the classic one. 

\subsection{The effect of the choice of $\nsmall$}\label{sec:nsmall}
\begin{figure}[htbp]
    \centering
     \includegraphics[width=0.45\textwidth]{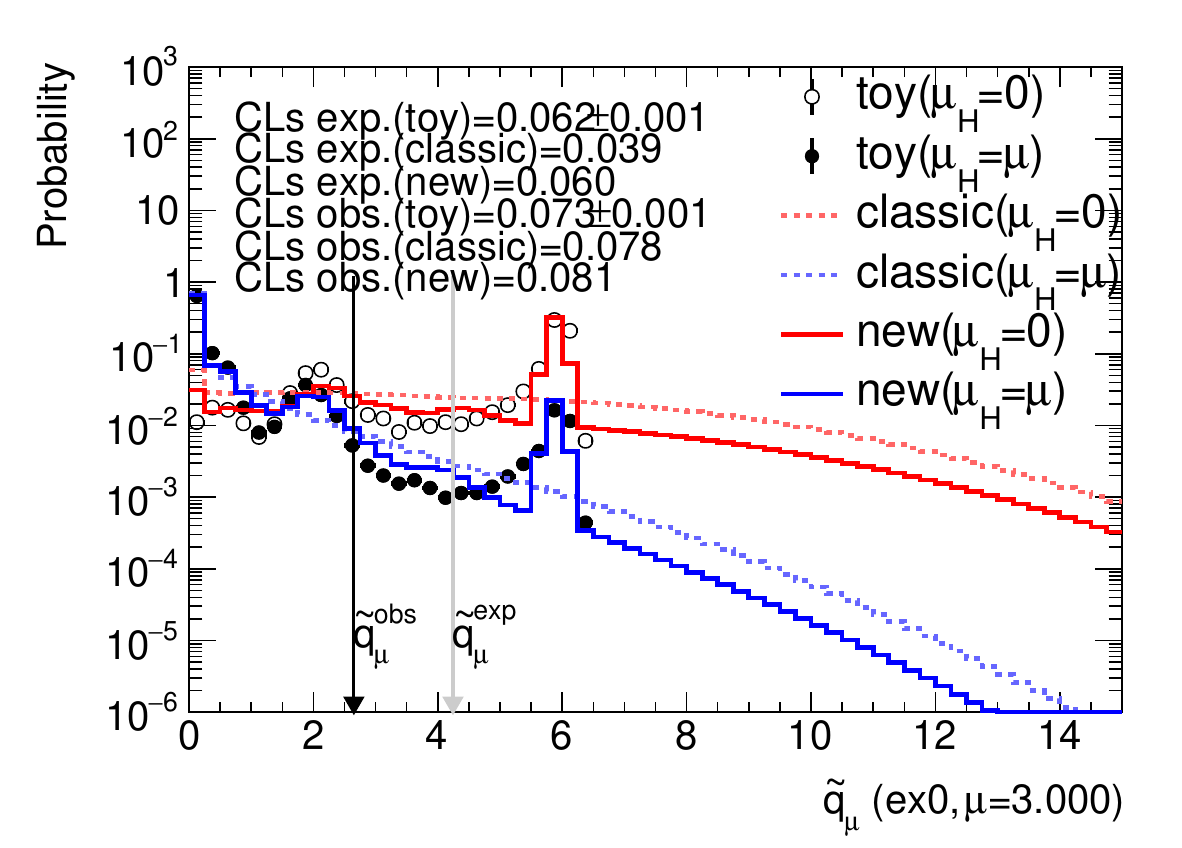}
     \includegraphics[width=0.45\textwidth]{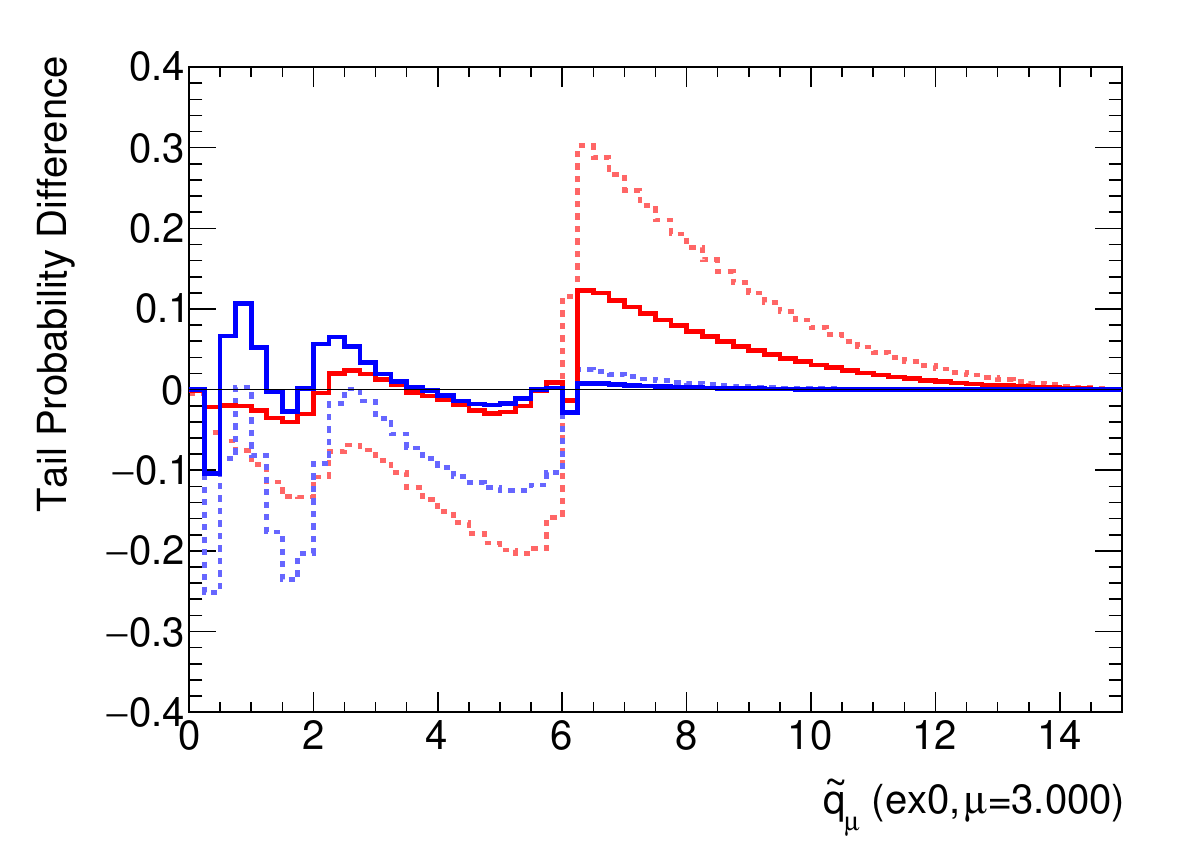}\\
     \includegraphics[width=0.45\textwidth]{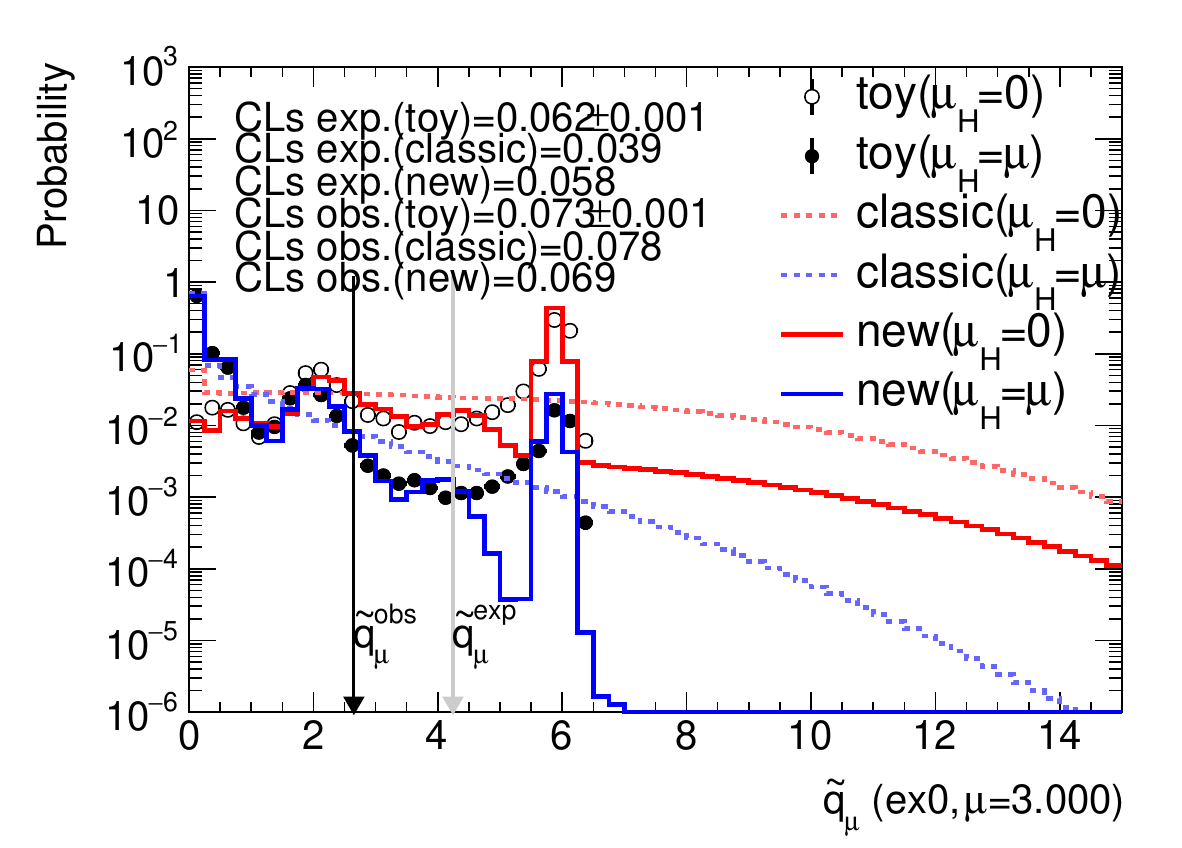}
     \includegraphics[width=0.45\textwidth]{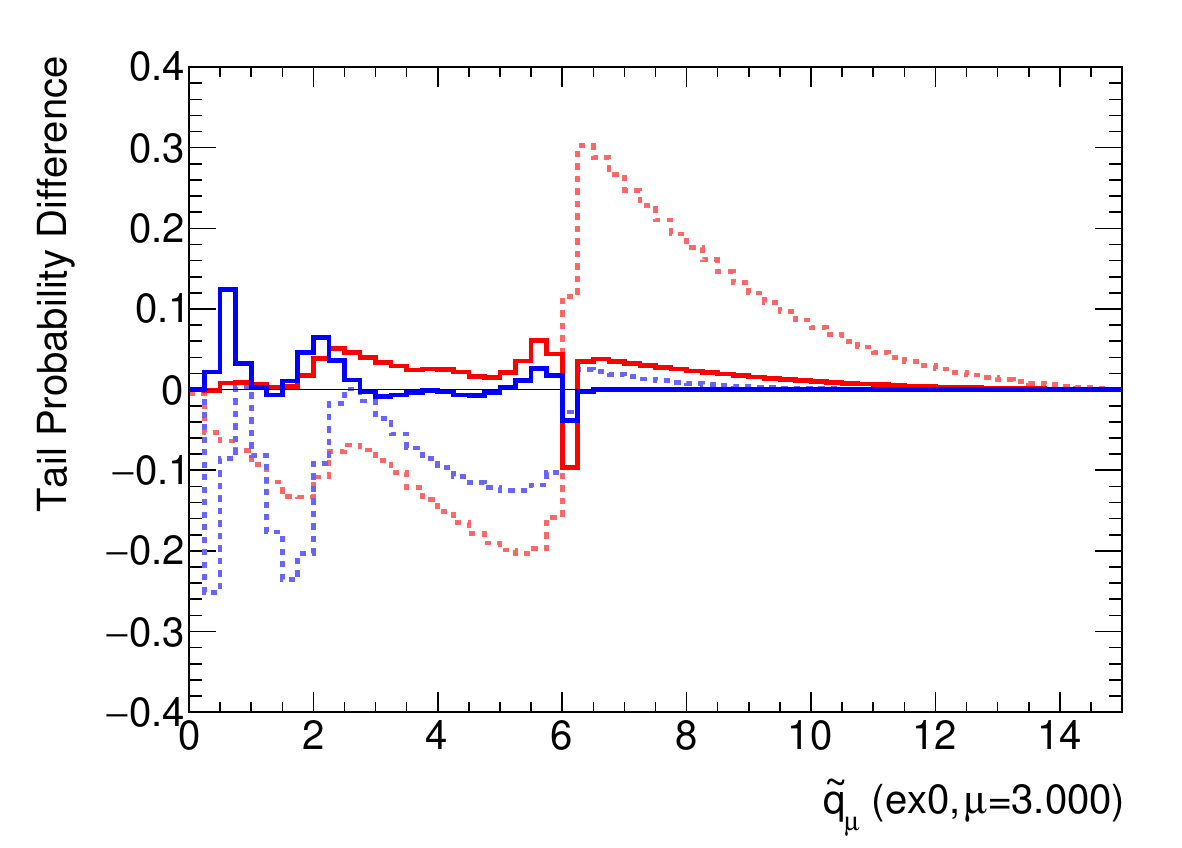}\\
     \includegraphics[width=0.45\textwidth]{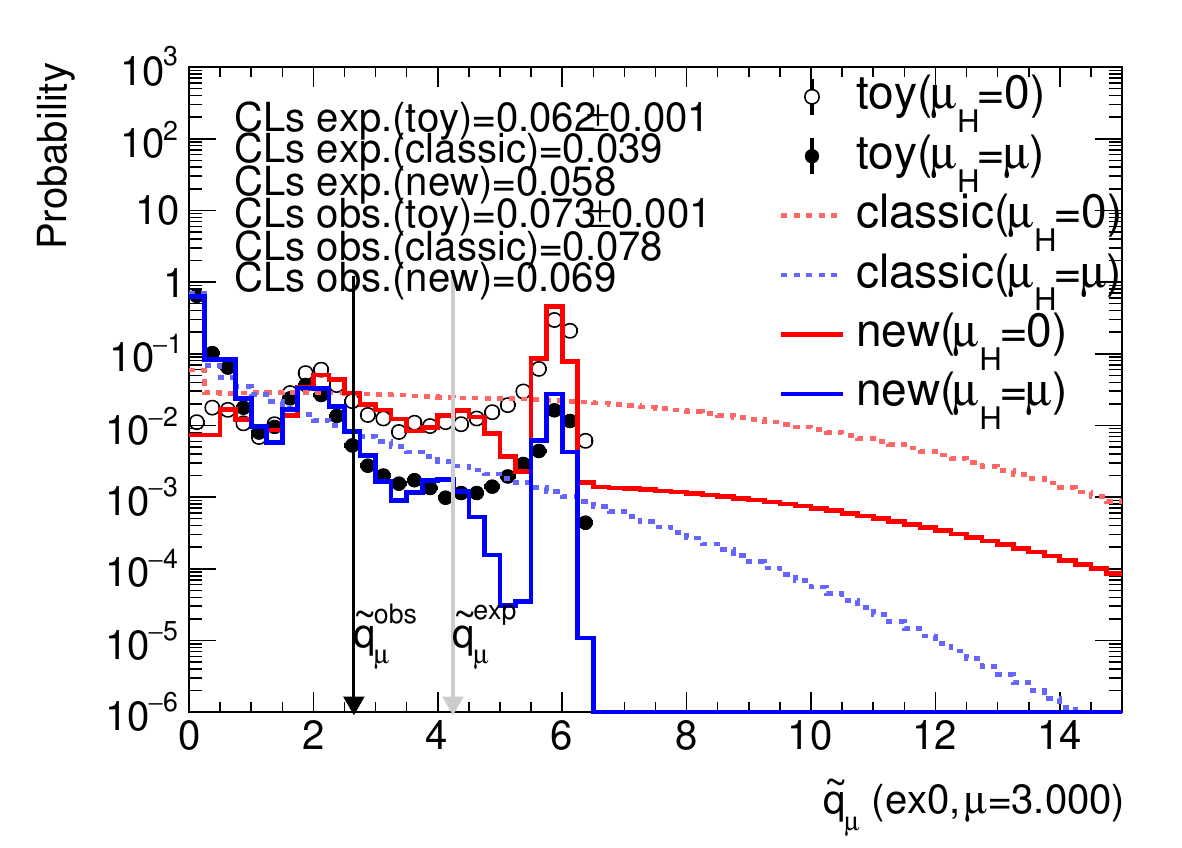}
     \includegraphics[width=0.45\textwidth]{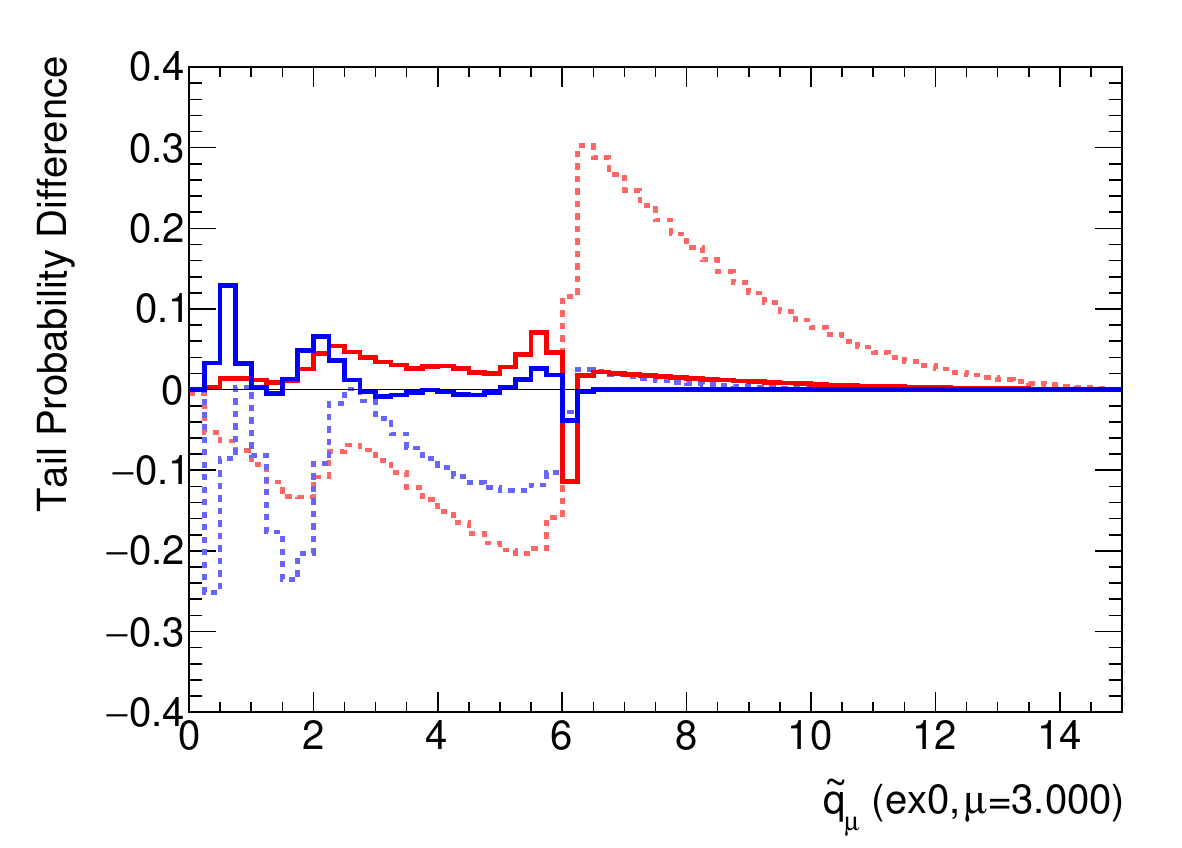}
     \caption{\label{fig:qmutilde_ex0_nsmall}
     Left column: the probability distributions of $\qtil_\mu$ in Ex.~0. From top to bottom, $\nsmall$ is the nominal value -1, +3 and +5, respectively.
     Right column: the relative difference of the tail probability compared to the toy results (for better visibility, the relative difference for the hypothesis $\mu_H=\mu$ is scaled by a factor of 5). 
     The black dots and open circles represent the toy MC results. The blue/red solid histograms represent the new asymptotic formulae in this work while the blue/red dashed histograms represent the classic asymptotic  formulae from Wald's approximation. The black and gray arrows represent the observed and expected $\tilde{q}_\mu$, respectively. 
     }
\end{figure}

\begin{figure}[htbp]
    \centering
     \includegraphics[width=0.45\textwidth]{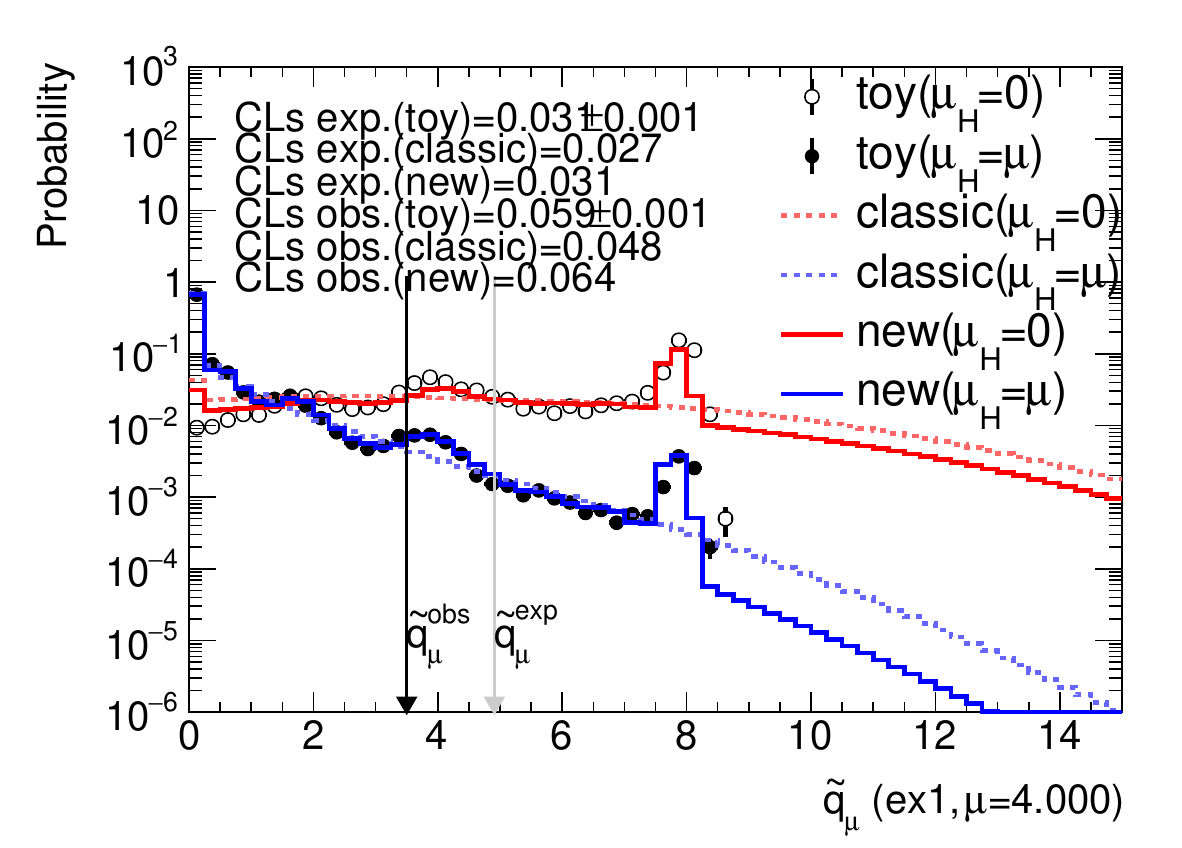}
     \includegraphics[width=0.45\textwidth]{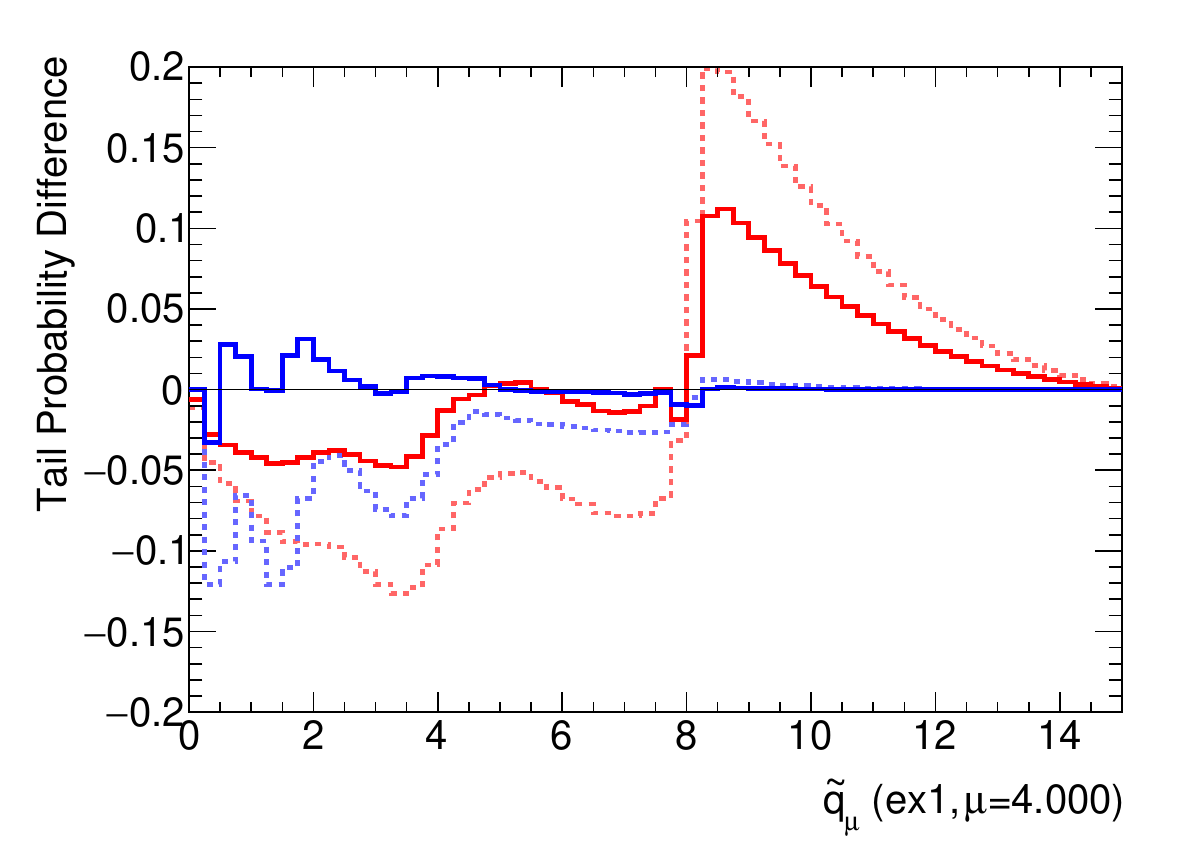}\\
     \includegraphics[width=0.45\textwidth]{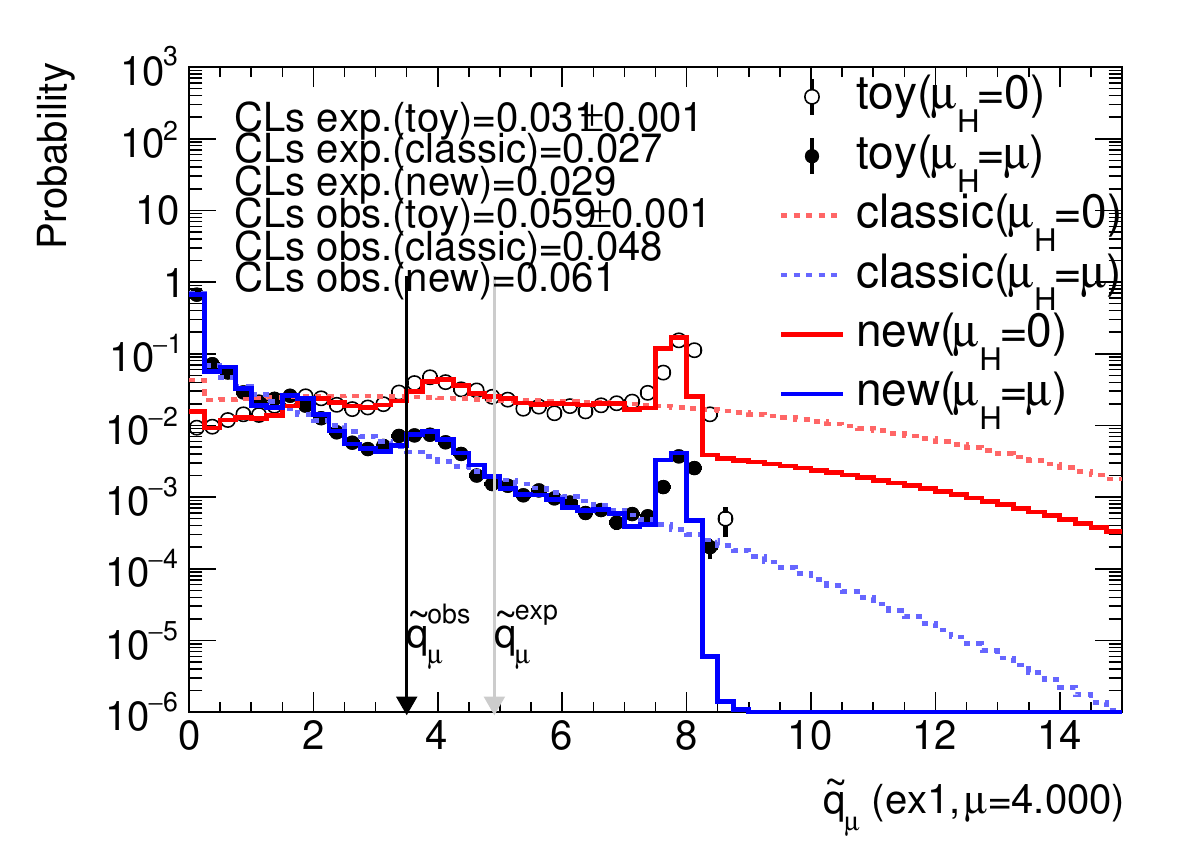}
     \includegraphics[width=0.45\textwidth]{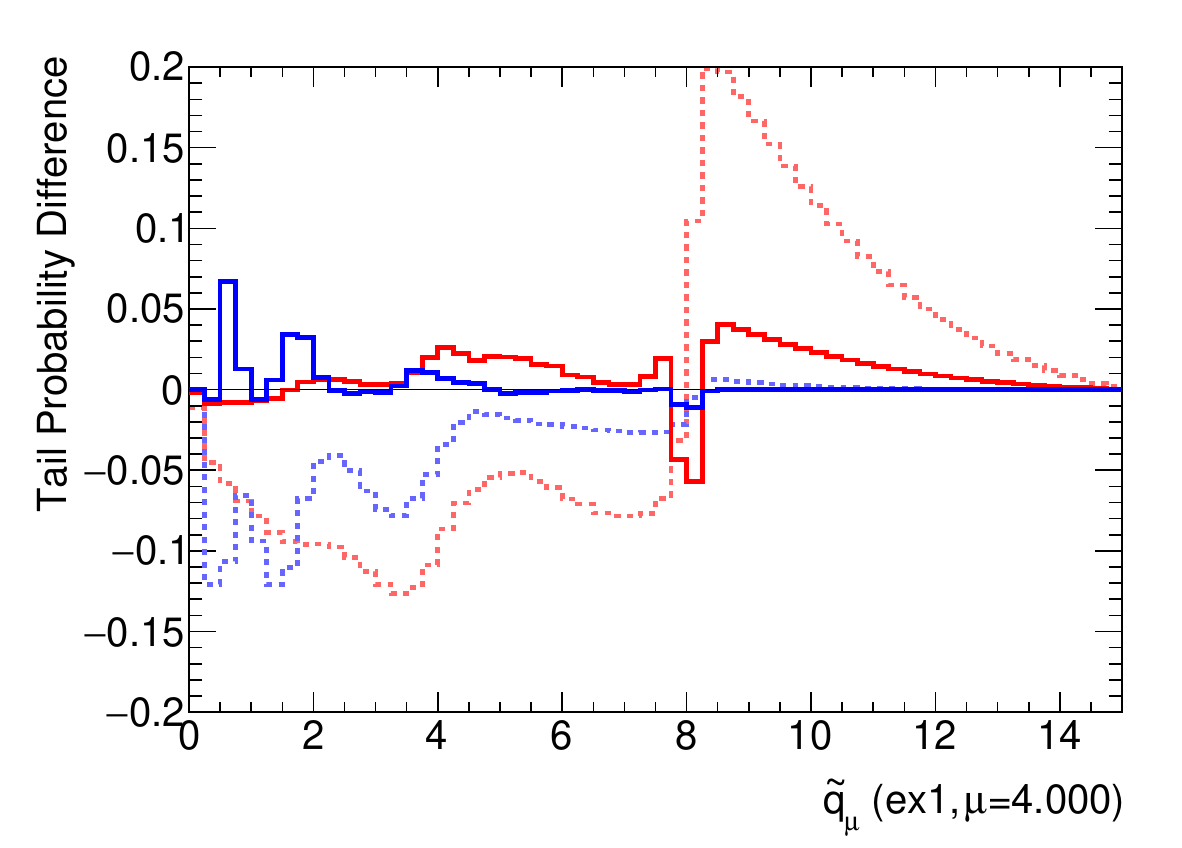}\\
     \includegraphics[width=0.45\textwidth]{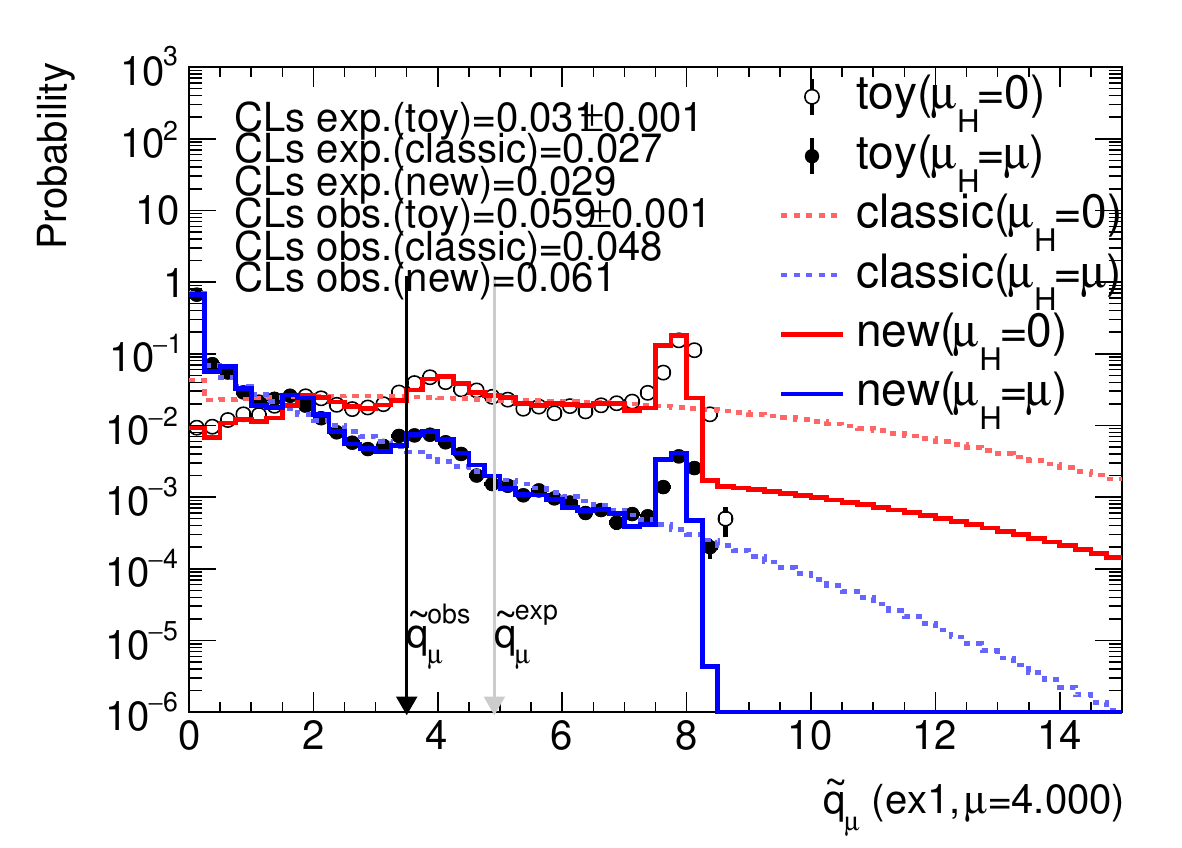}
     \includegraphics[width=0.45\textwidth]{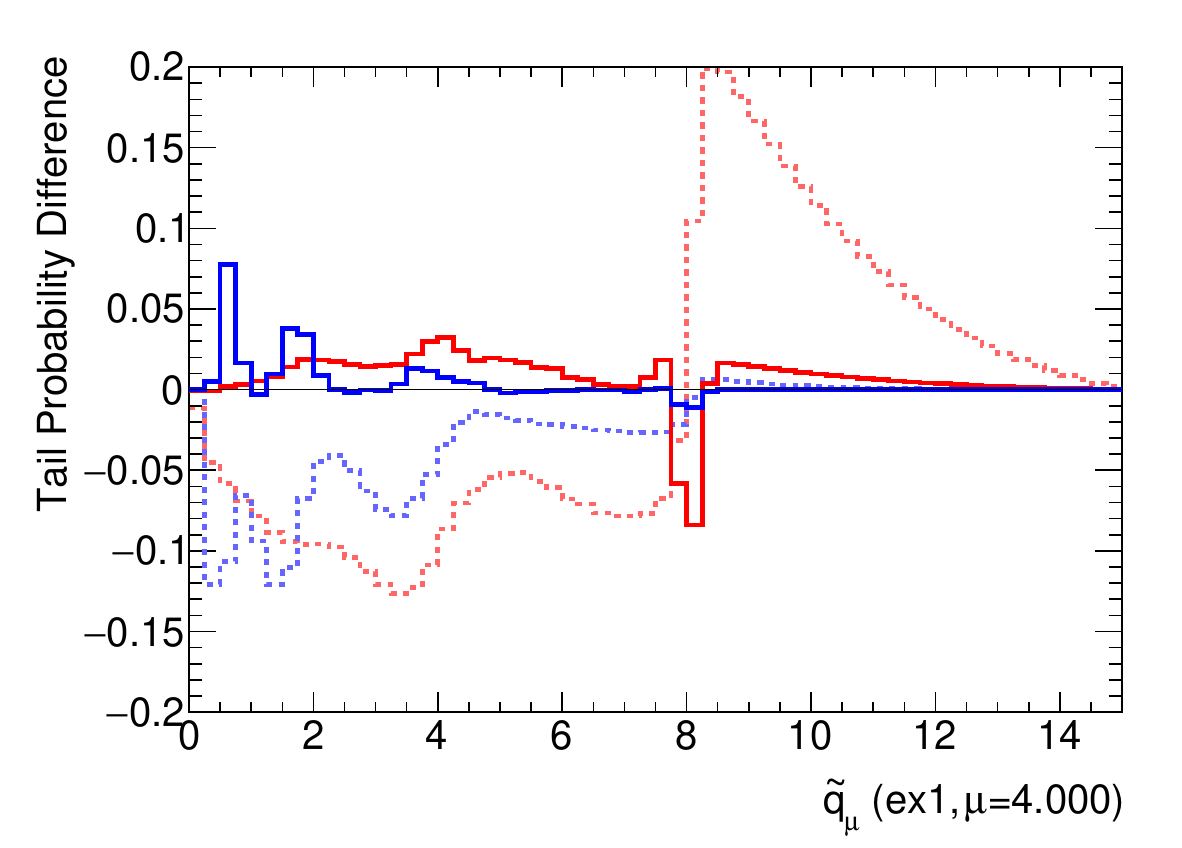}
     \caption{\label{fig:qmutilde_ex1_nsmall}
     Left column: the probability distributions of $\qtil_\mu$ in Ex.~1. From top to bottom, $\nsmall$ is the nominal value -1, +3 and +5, respectively.
     Right column: the relative difference of the tail probability compared to the toy results (for better visibility, the relative difference for the hypothesis $\mu_H=\mu$ is scaled by a factor of 5). 
     The black dots and open circles represent the toy MC results. The blue/red solid histograms represent the new asymptotic formulae in this work while the blue/red dashed histograms represent the classic asymptotic  formulae from Wald's approximation. The black and gray arrows represent the observed and expected $\tilde{q}_\mu$, respectively. 
     }
\end{figure}

\begin{figure}[htbp]
    \centering
     \includegraphics[width=0.45\textwidth]{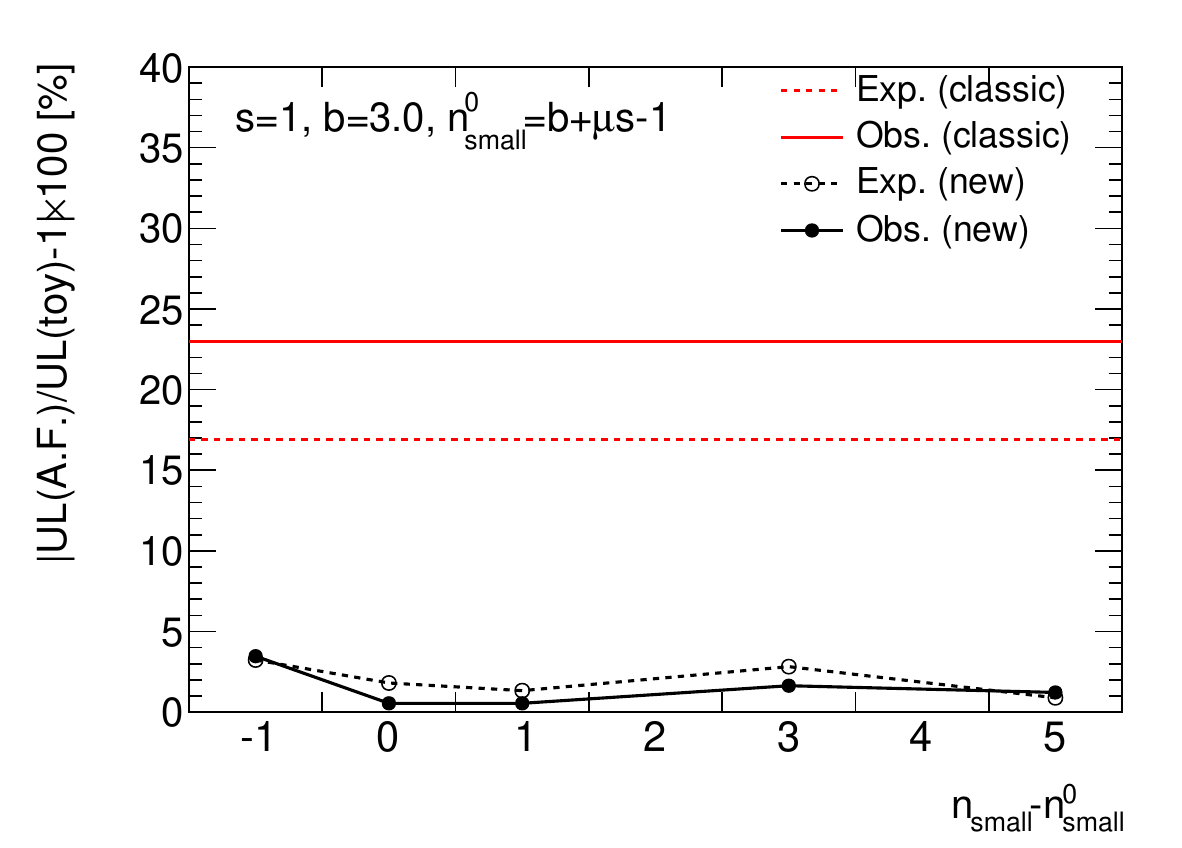}
     \includegraphics[width=0.45\textwidth]{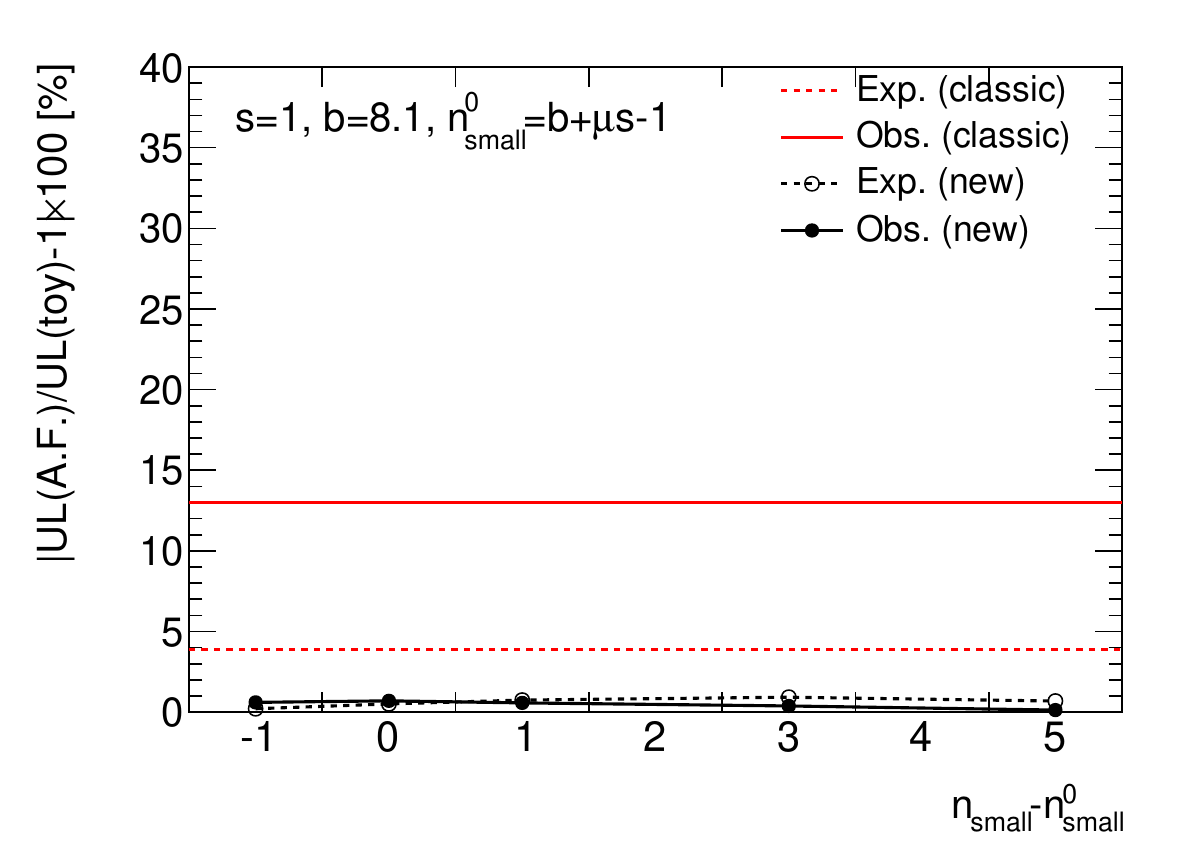}\\
     \includegraphics[width=0.45\textwidth]{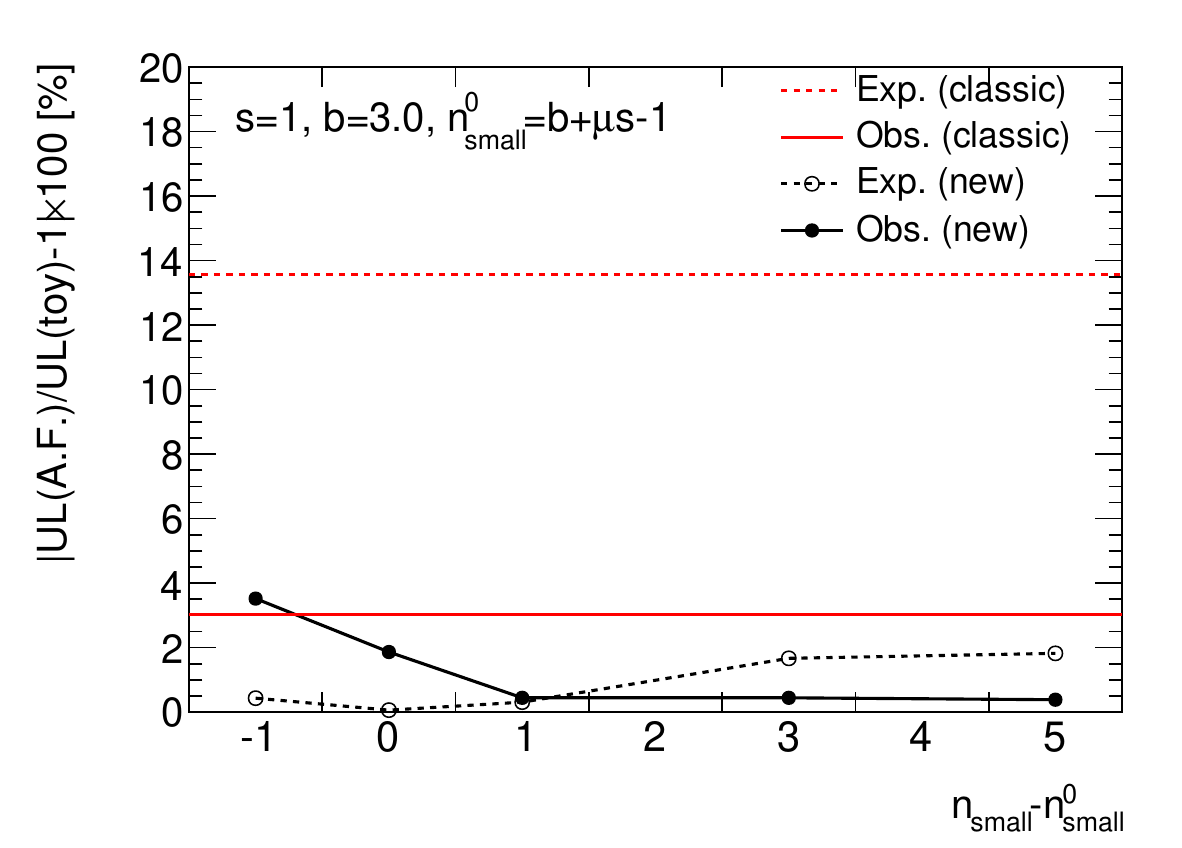}
     \includegraphics[width=0.45\textwidth]{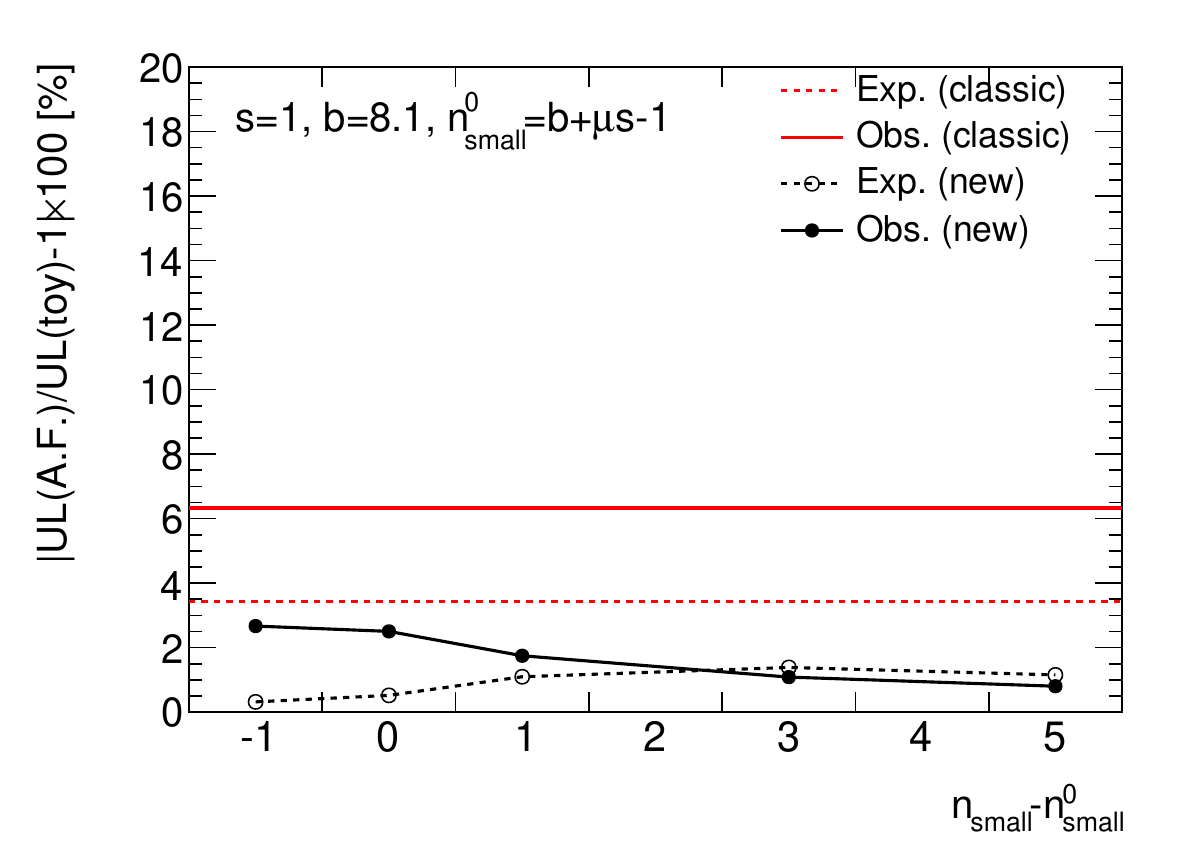}\\
     \includegraphics[width=0.45\textwidth]{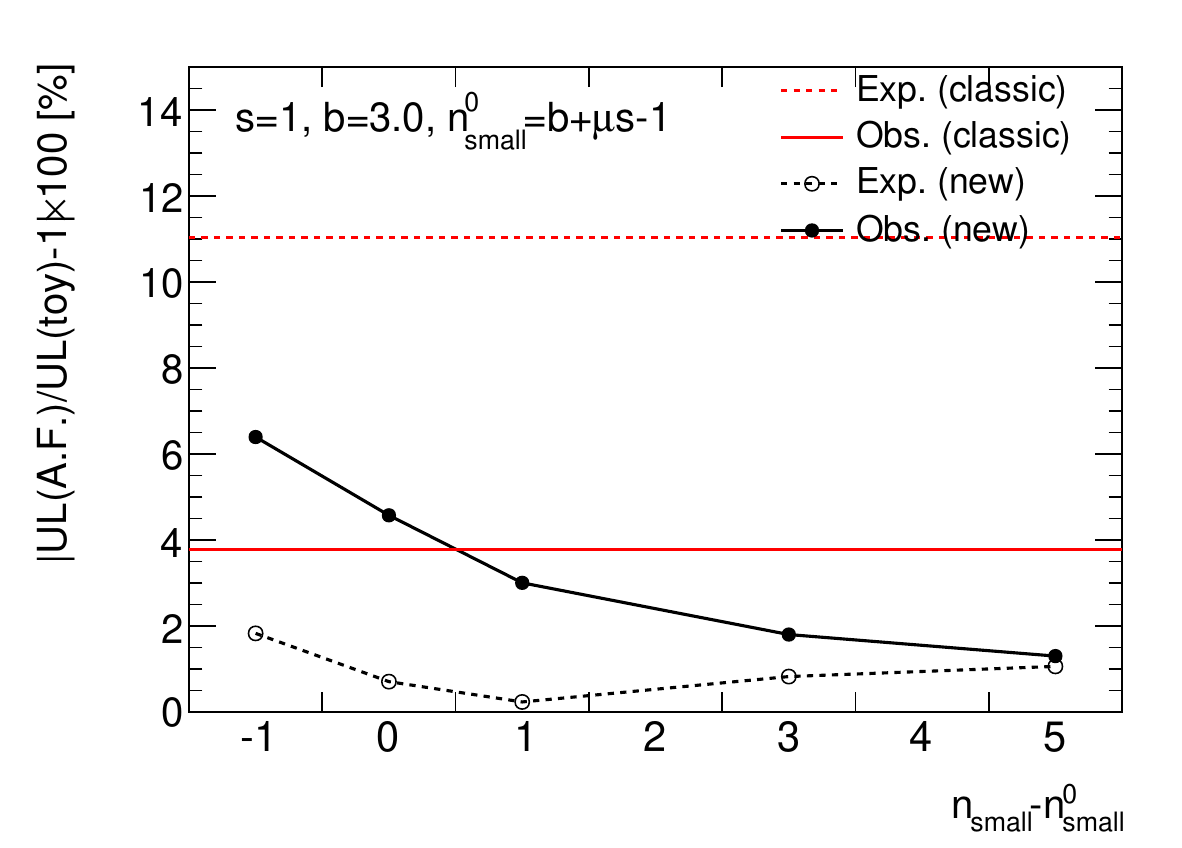}
     \includegraphics[width=0.45\textwidth]{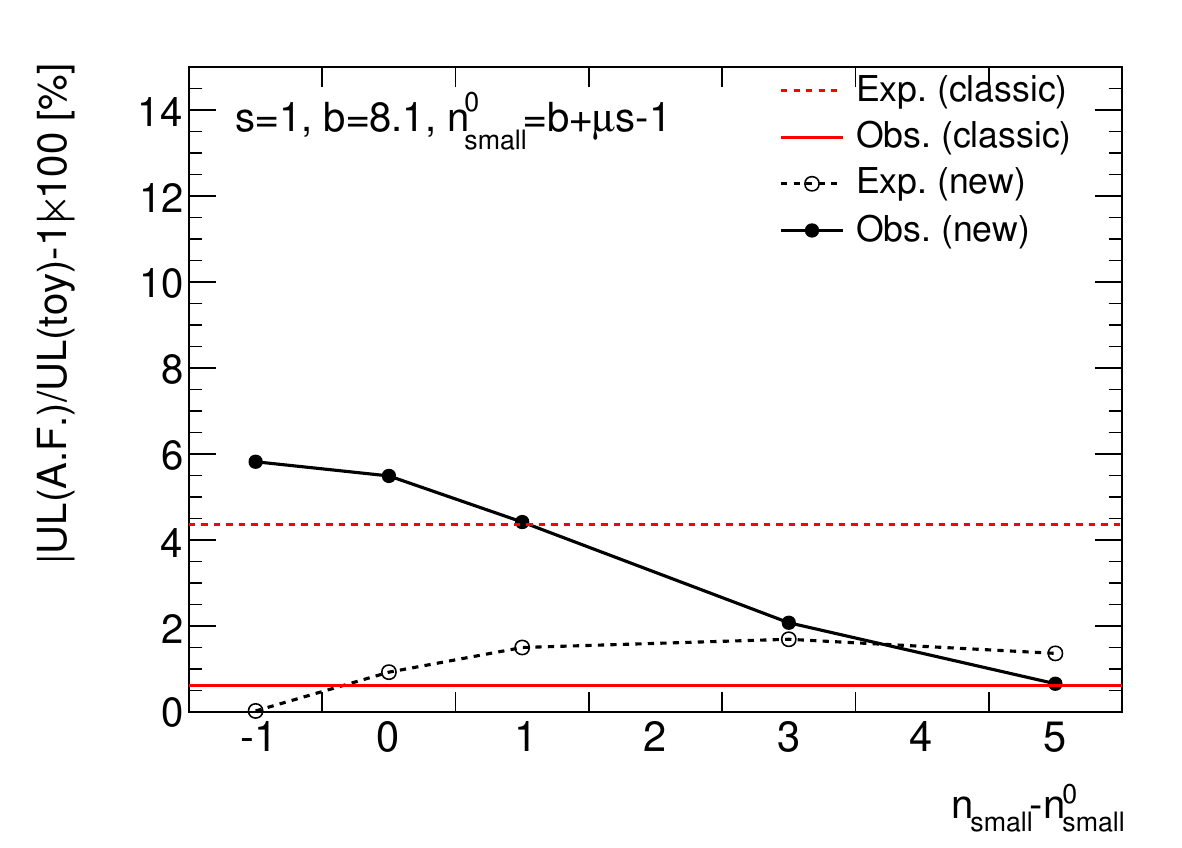}
     \caption{\label{fig:limits_nsmall}
     The relative difference of upper limit compared to the toy results as a function of the choice of $\nsmall$ in Ex.~0 (Left column) and Ex.~1 (Right column). From top to bottom, they correspond to an observed data with a negative signal strength, $\mu=0.5$, and $\mu=2$. The red horizontal lines represent the results using the classic formulae. 
     }
\end{figure}

In Sec.~\ref{sec:new}, $\nsmall$ for the test signal strength $\mu$ is proposed to be $b+\mu s-1$ (denoted by $\nsmall^0$) based on the expectation that the modified component should not exceed 50\% for conservation. To verify the robustness of this choice, we examine how different $\nsmall$ values affect the upper limits. Given the background yields of 0.64 (2.79) in Ex.~0 (Ex.~1) from Table~\ref{tab:ex_yields}, we vary $\nsmall$ across a range of -1 to +5. 
Figure~\ref{fig:qmutilde_ex0_nsmall} and~\ref{fig:qmutilde_ex1_nsmall} show some examples of $\qtil_\mu$ distribution for different $\nsmall$ choices. The upper limit variation relative to the toy results as a function of $\nsmall$ is summarized in Fig.~\ref{fig:limits_nsmall}. 
These results demonstrate that the upper limits derived from the new formalism remain stable across the tested $\nsmall$ range.

\subsection{The test statistic $q_\mu$}\label{sec:qmu}
\begin{figure}[htbp]
    \centering
     \includegraphics[width=0.45\textwidth]{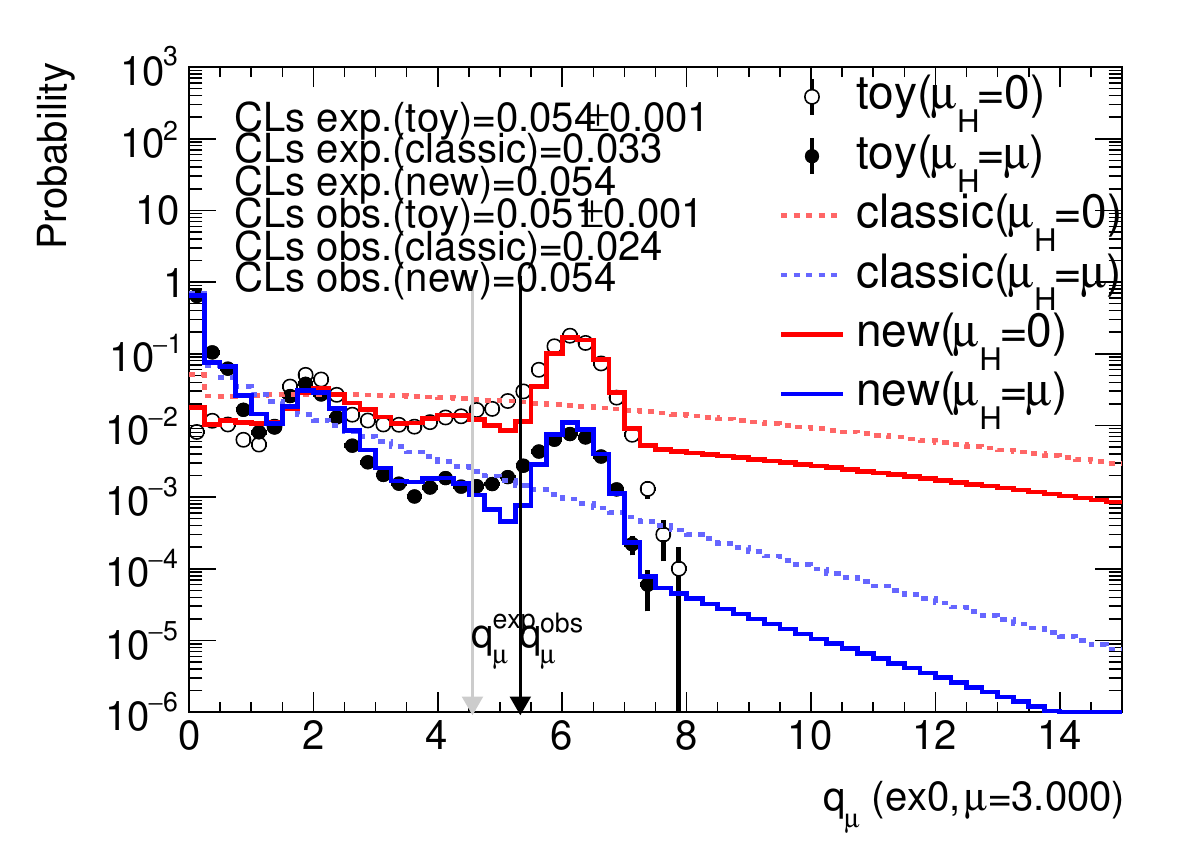}
     \includegraphics[width=0.45\textwidth]{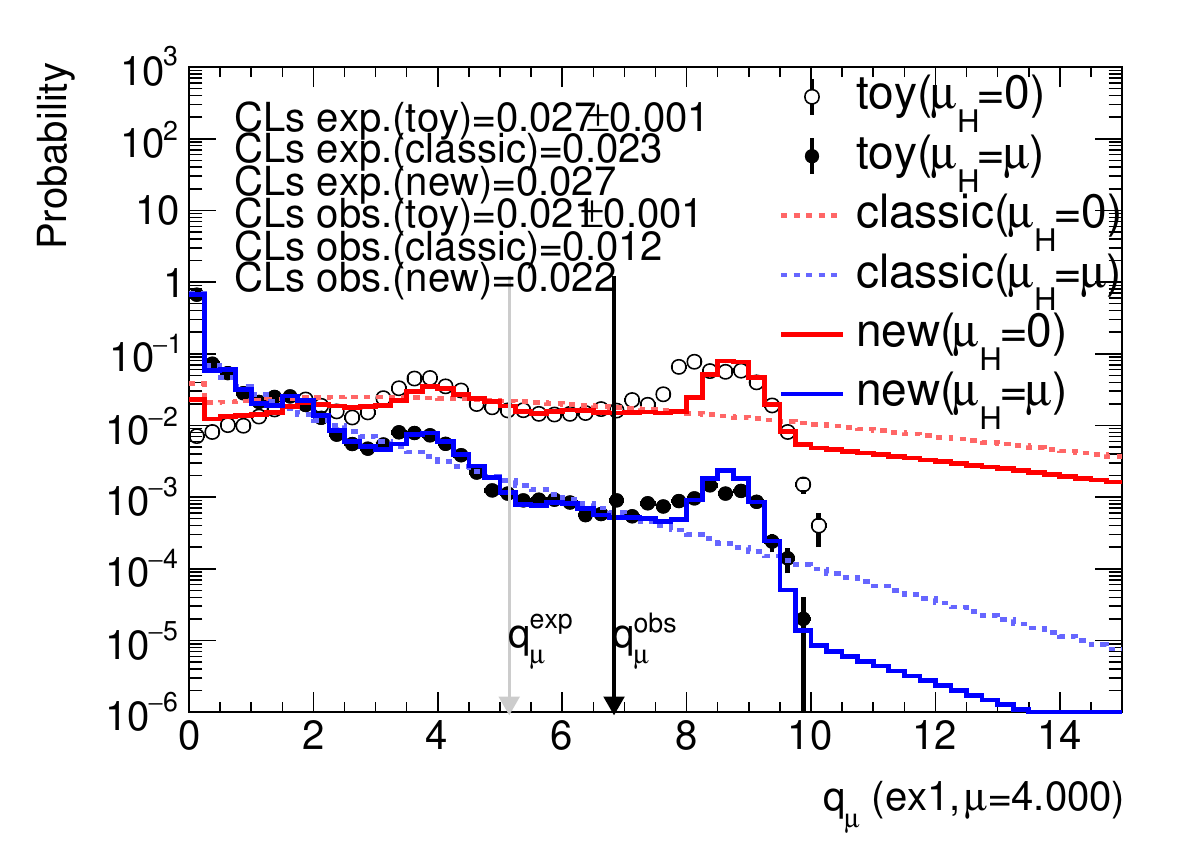}\\
     \includegraphics[width=0.45\textwidth]{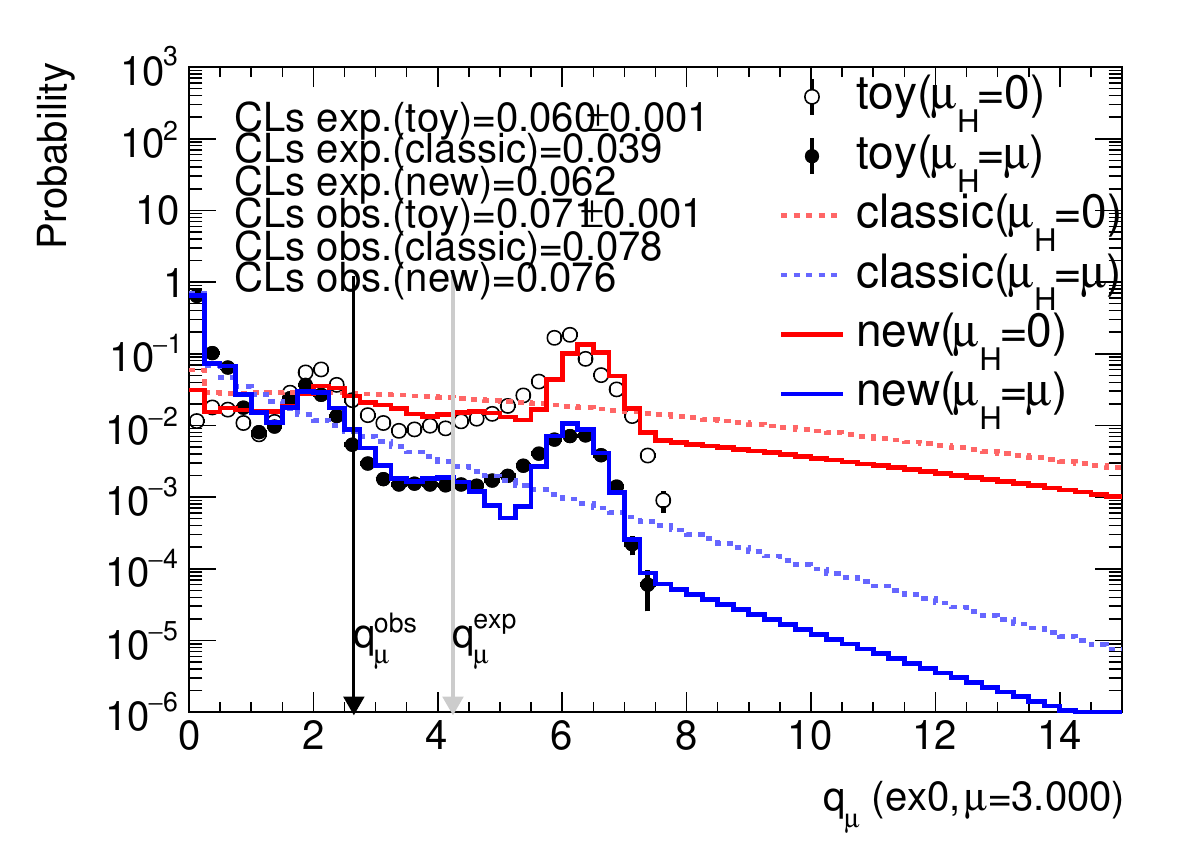}
     \includegraphics[width=0.45\textwidth]{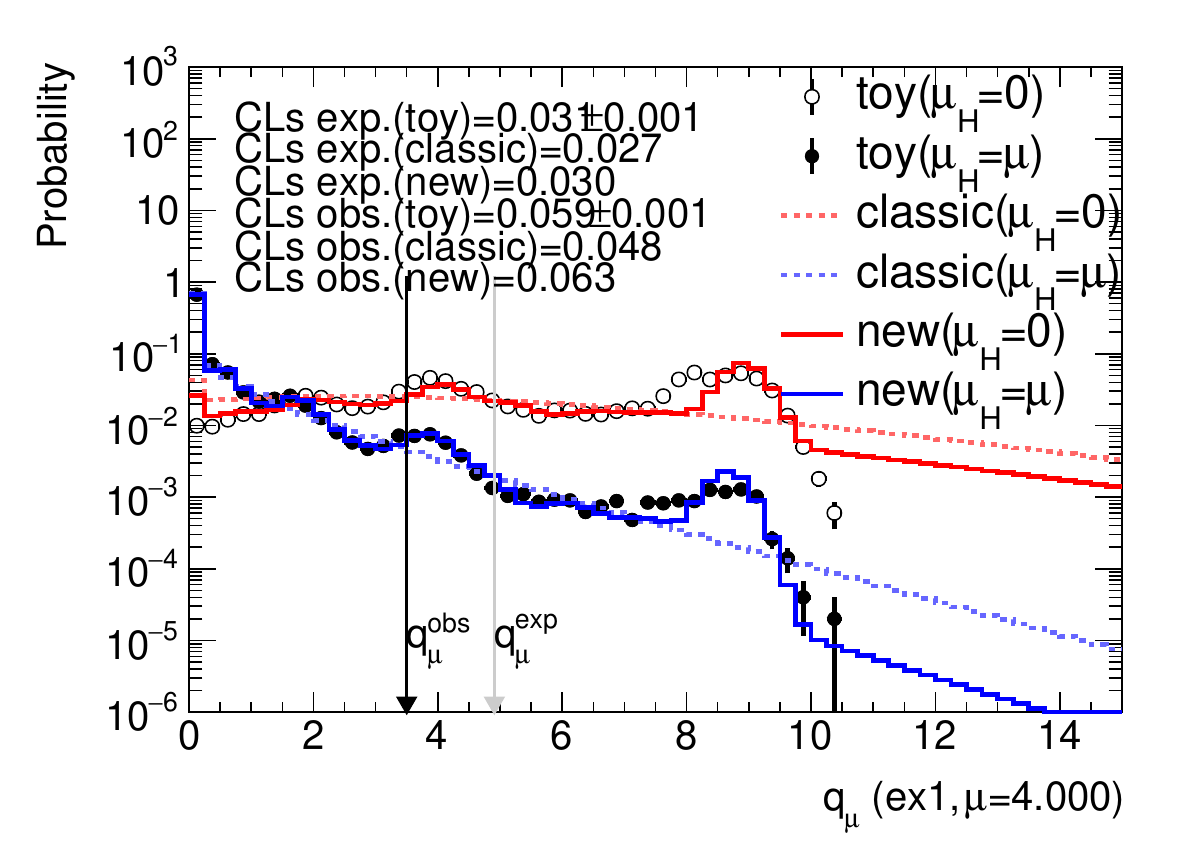}\\
     \includegraphics[width=0.45\textwidth]{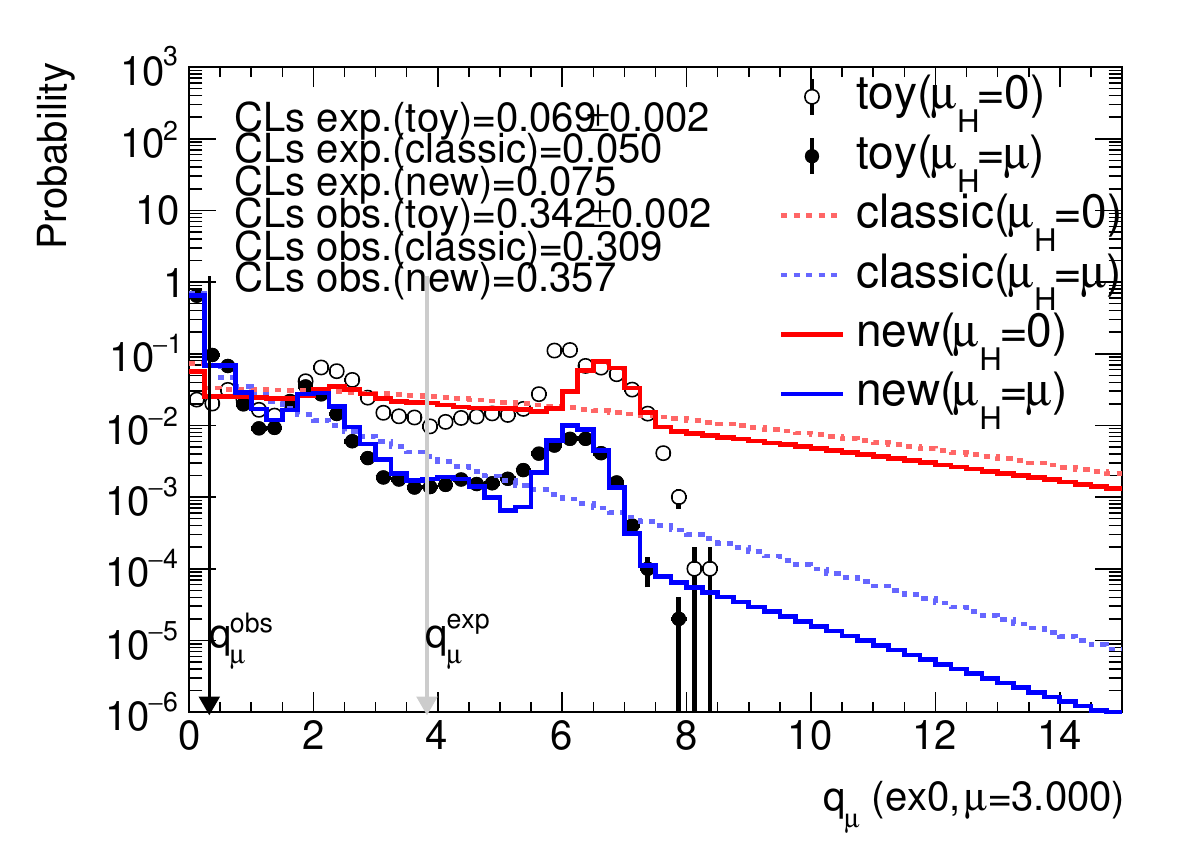}
     \includegraphics[width=0.45\textwidth]{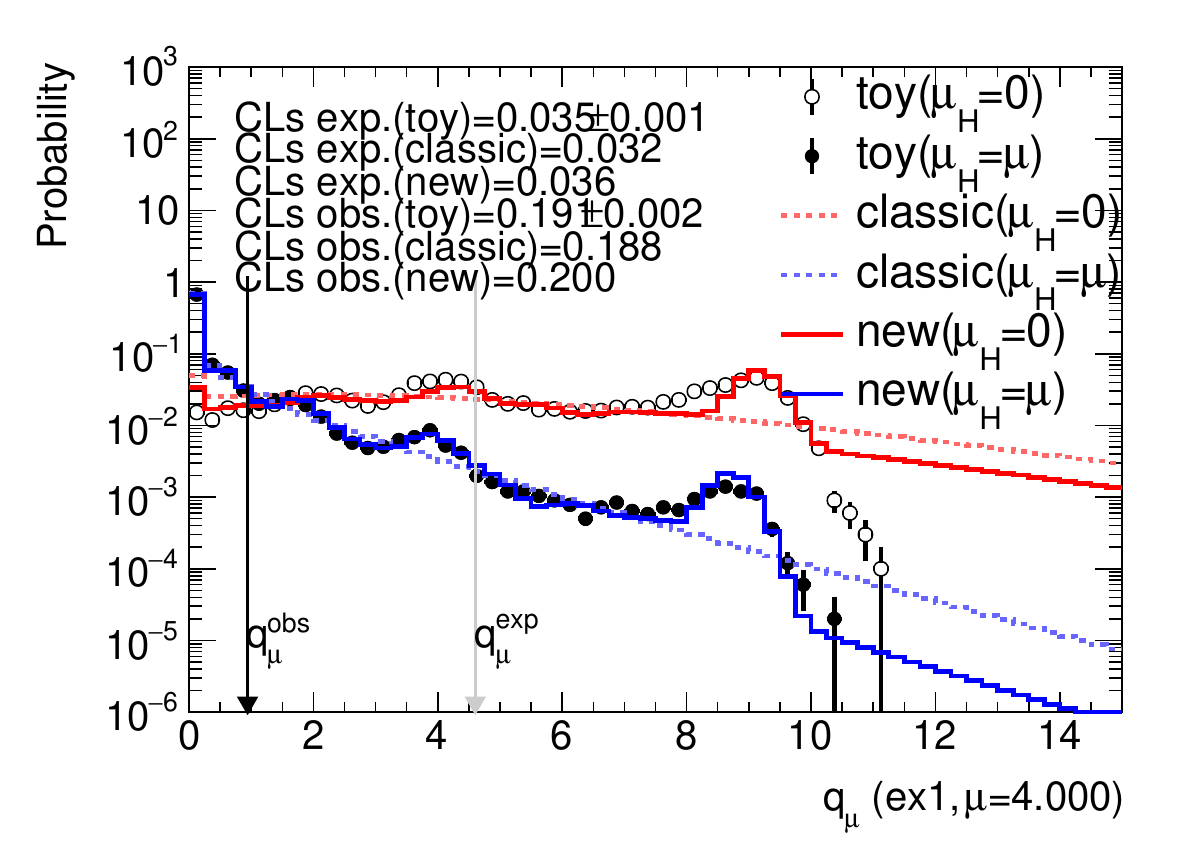}
     \caption{\label{fig:qmu}
     The probability distributions of $q_\mu$ in Ex.~0 (Left) and Ex.~1 (Right). From top to bottom, different data sets are used.
     The black dots and open circles represent the toy MC results. The blue/red solid histograms represent the new asymptotic formulae in this work while the blue/red dashed histograms represent the classic asymptotic  formulae from Wald's approximation. The black and gray arrows represent the observed and expected $q_\mu$, respectively.
     }
\end{figure}

\begin{figure}[htbp]
    \centering
     \includegraphics[width=0.45\textwidth]{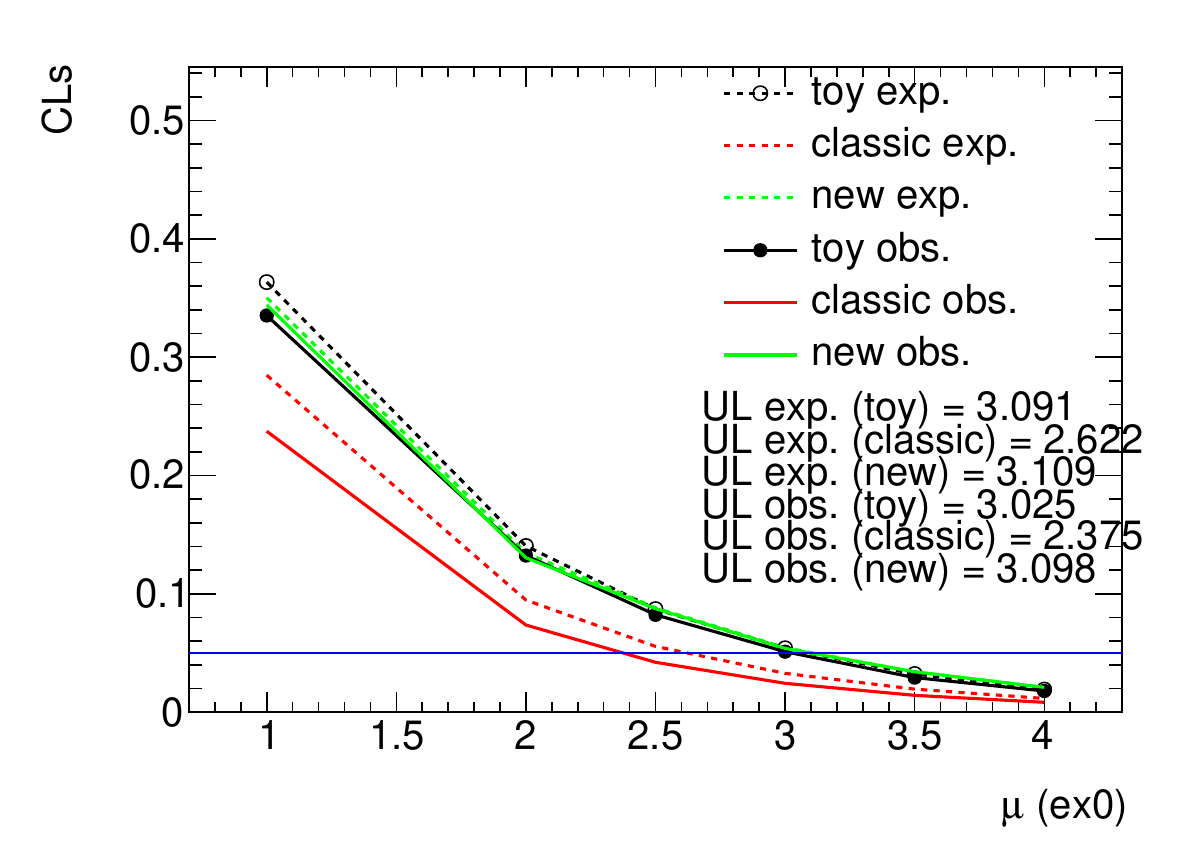}
     \includegraphics[width=0.45\textwidth]{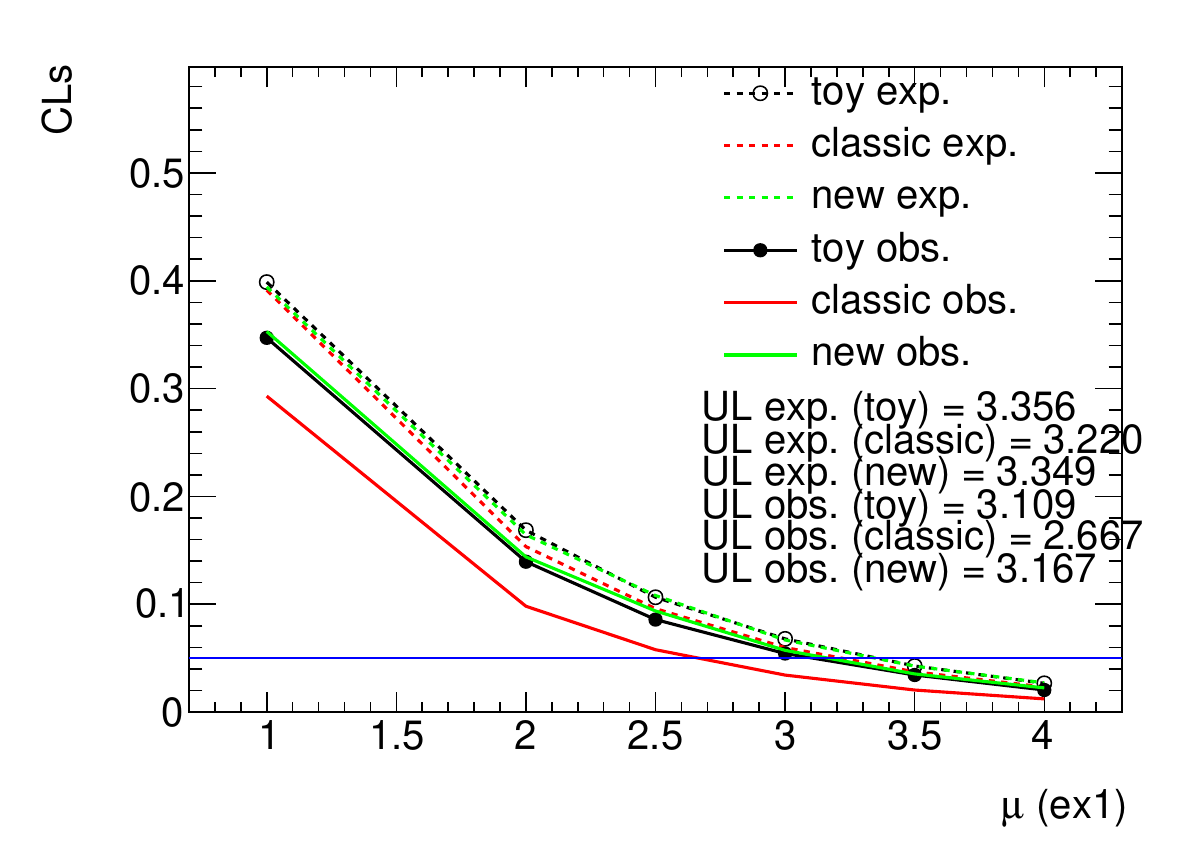}\\
     \includegraphics[width=0.45\textwidth]{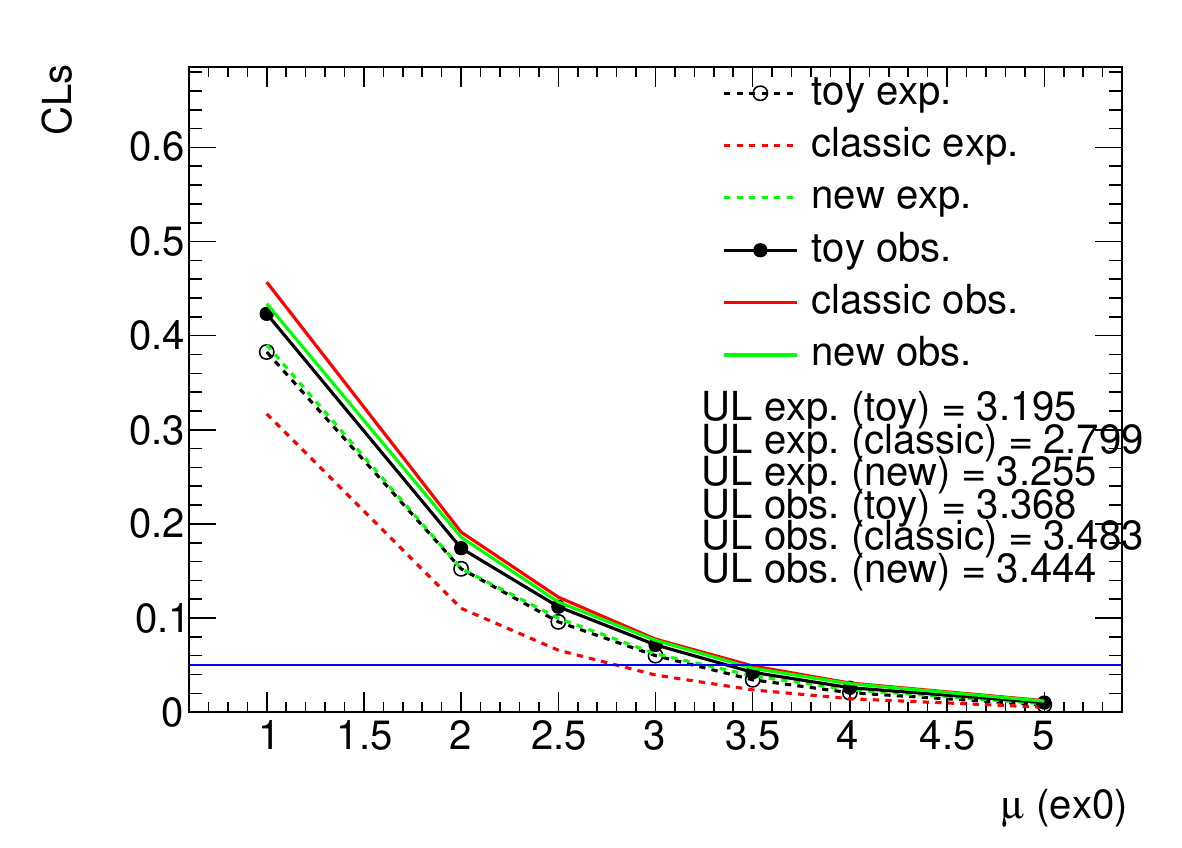}
     \includegraphics[width=0.45\textwidth]{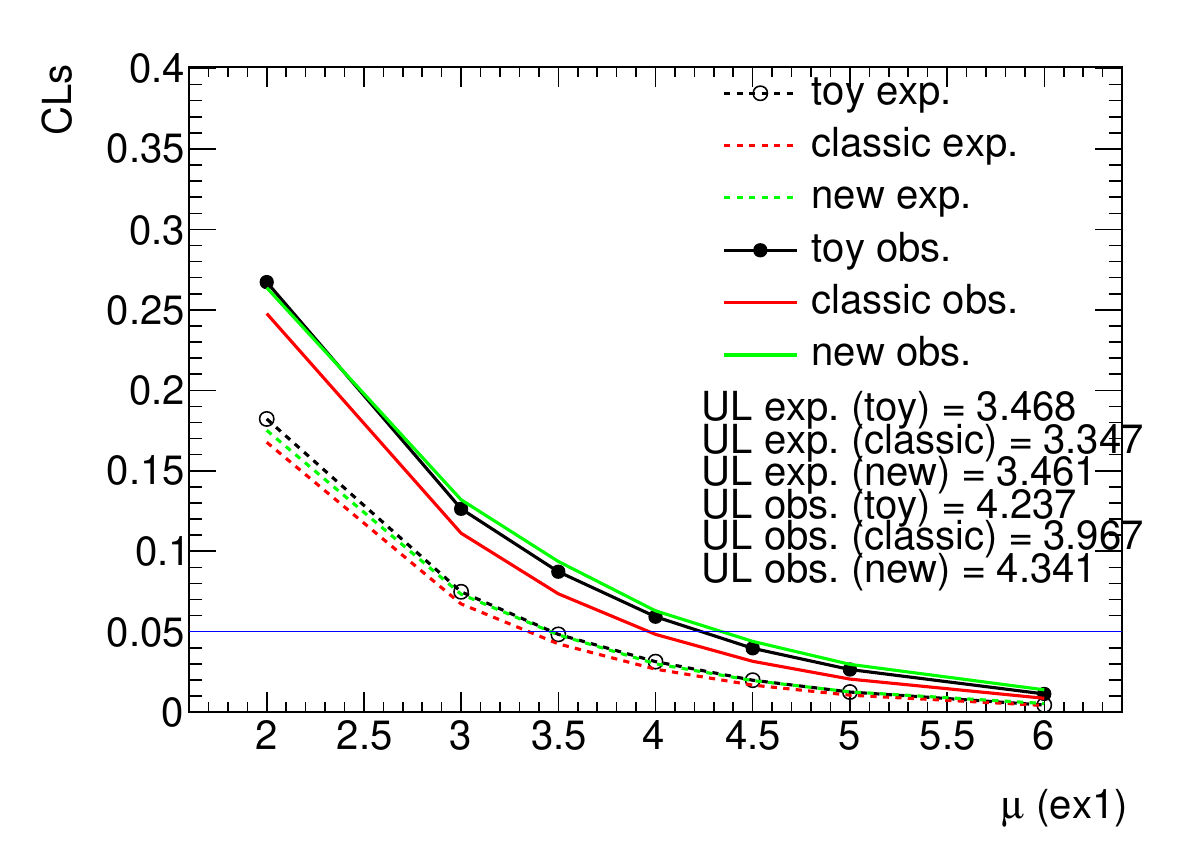}\\
     \includegraphics[width=0.45\textwidth]{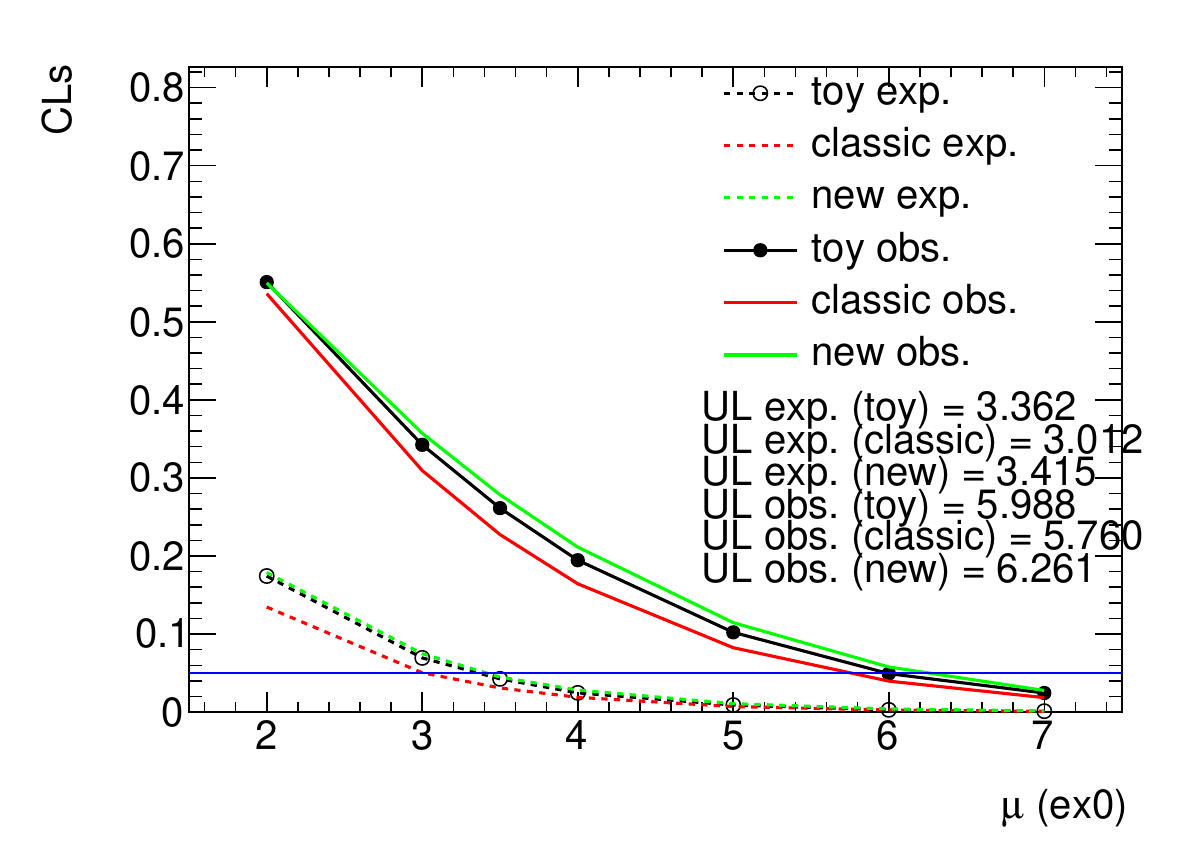}
     \includegraphics[width=0.45\textwidth]{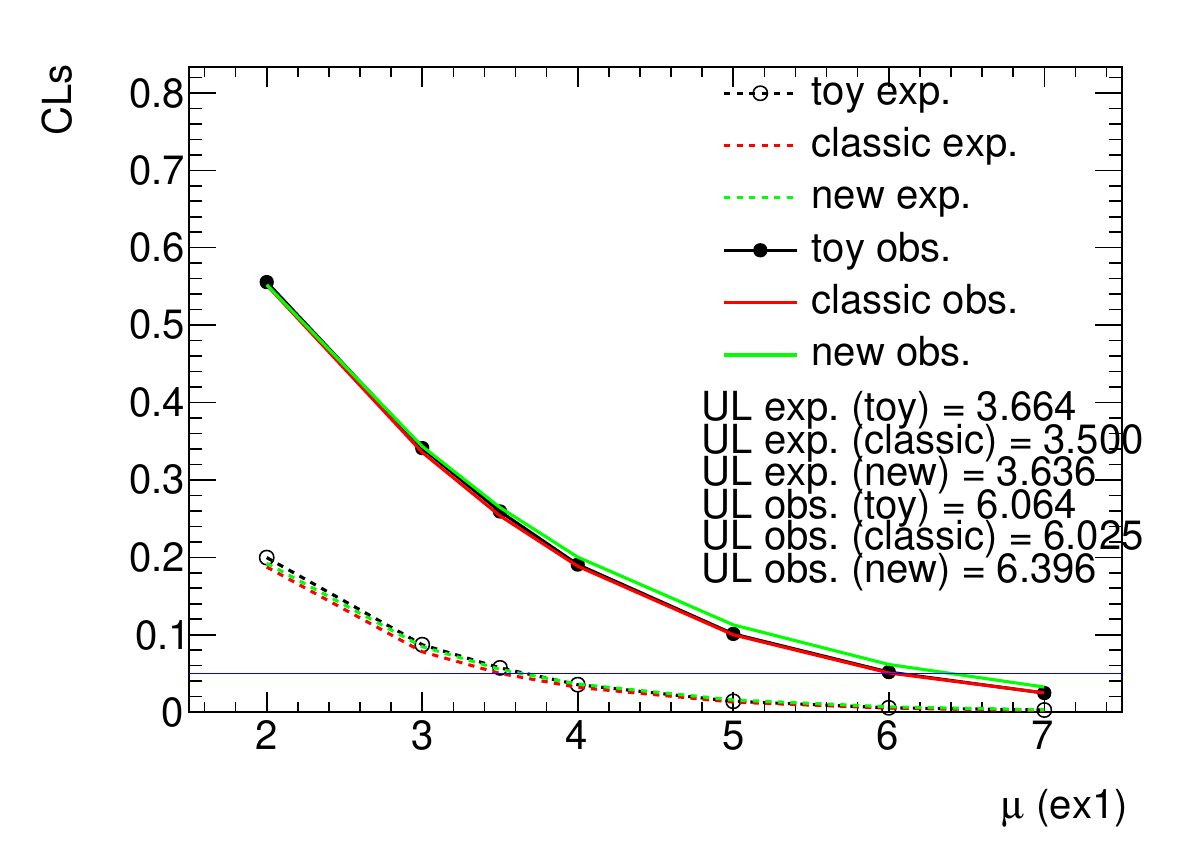}
     \caption{\label{fig:qmu_CLs}
     CLs as a function of $\mu$ in Ex.~0 (L) and Ex.~1 (R) using the test statistic $q_\mu$. From top to bottom, they represent different observed datasets. The black curves with markers show the toy MC results. The red and green curves are the predictions from the classic and new asymptotic formulae, respectively.  
     }
\end{figure}

In this section, we turn to using $q_\mu$ instead. 
It differs from $\qtil_\mu$ only when $\hatmu$ is negative.  
Figure~\ref{fig:qmu} shows some examples of $q_\mu$ distribution. 
Figure~\ref{fig:qmu_CLs} shows the upper limits. We can see the new formulae are generally better.

\subsection{The test statistic $q_0$}\label{sec:q0}
\begin{figure}[htbp]
    \centering
     \includegraphics[width=0.45\textwidth]{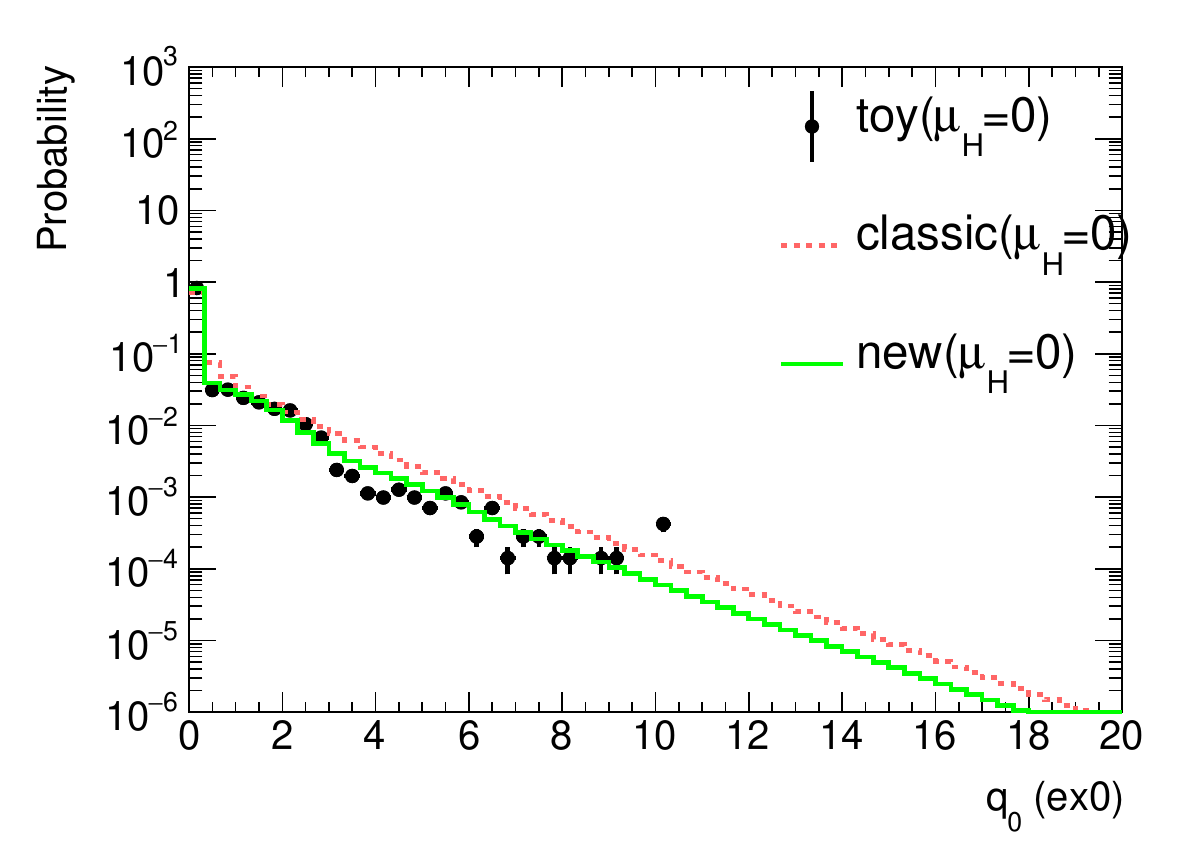}
     \includegraphics[width=0.45\textwidth]{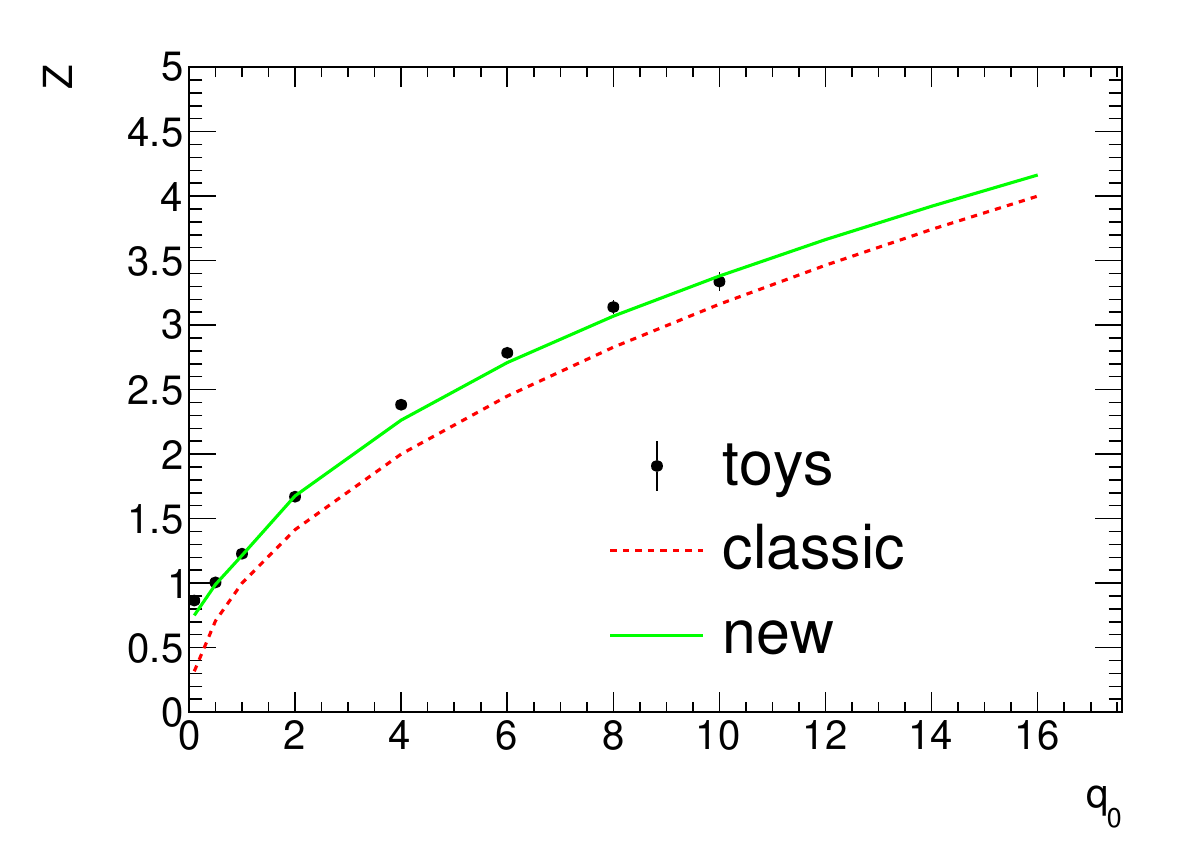}\\
     \includegraphics[width=0.45\textwidth]{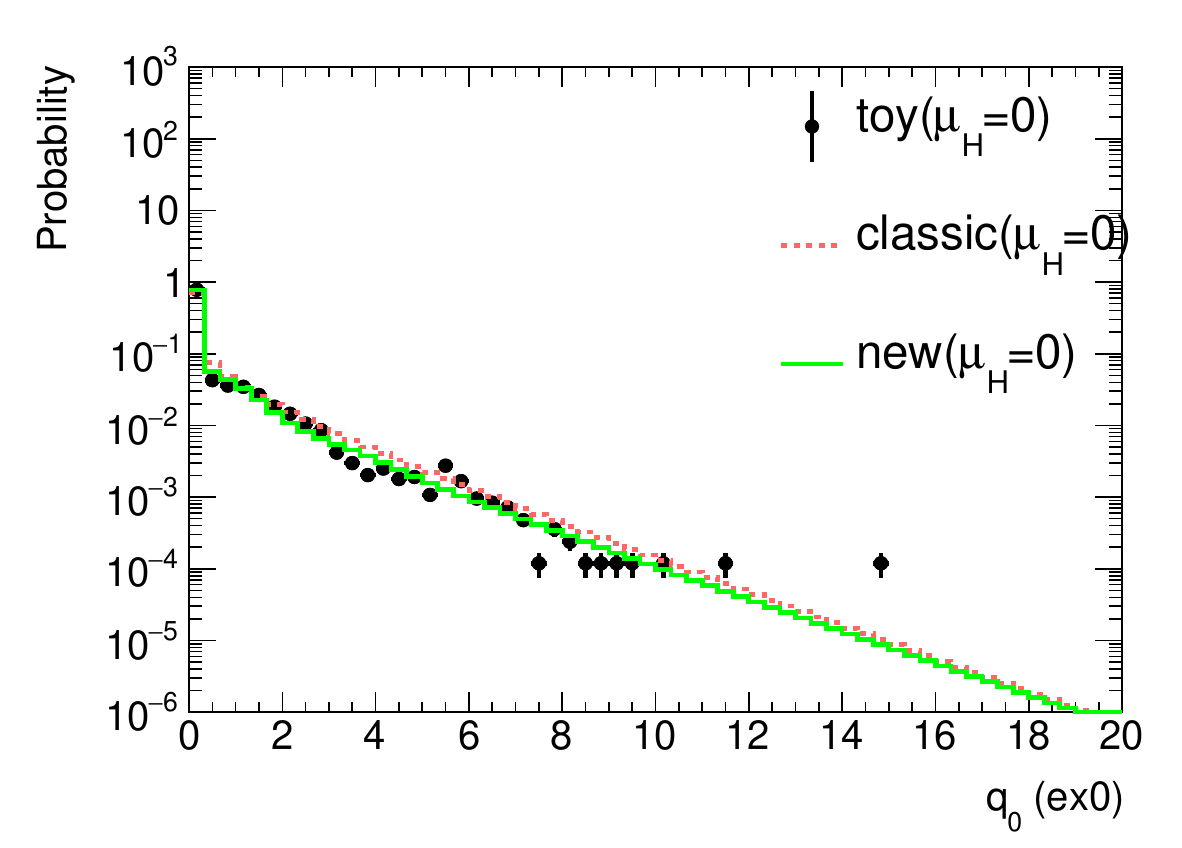}
     \includegraphics[width=0.45\textwidth]{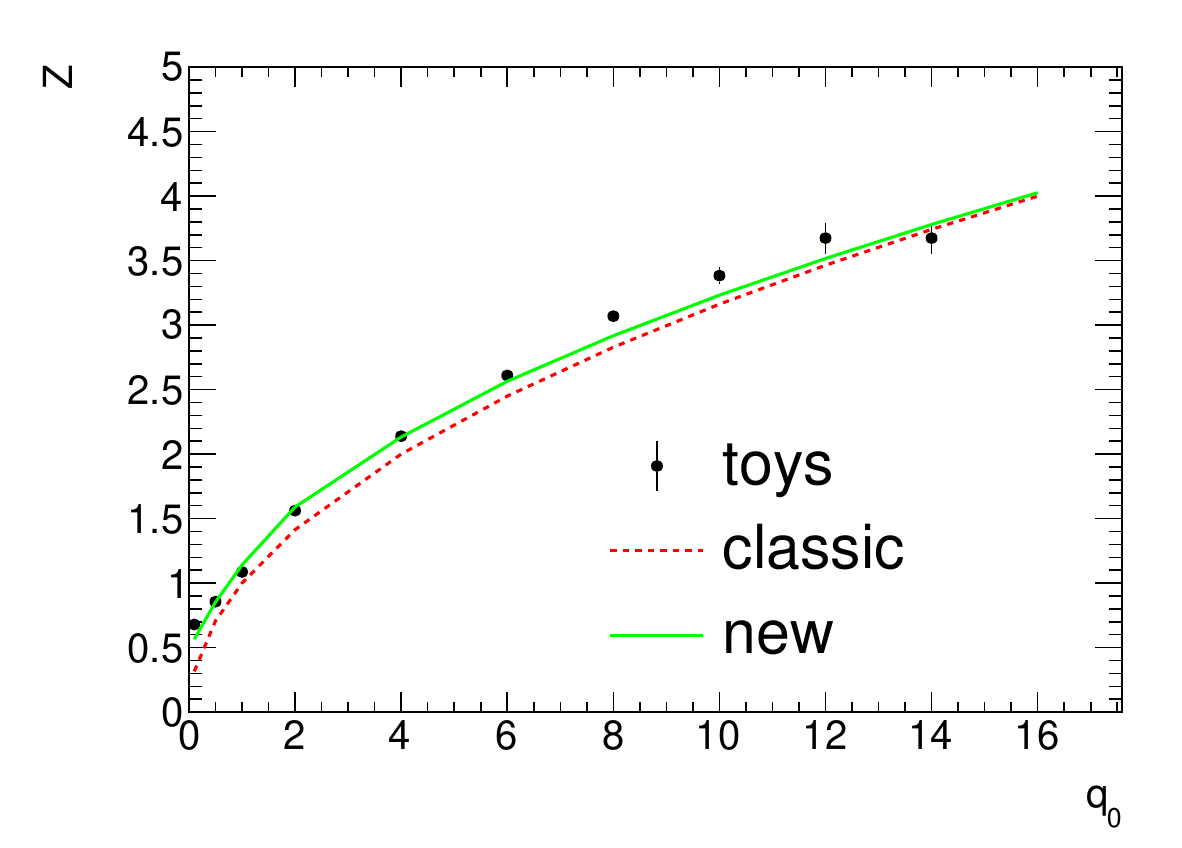}\\
     \includegraphics[width=0.45\textwidth]{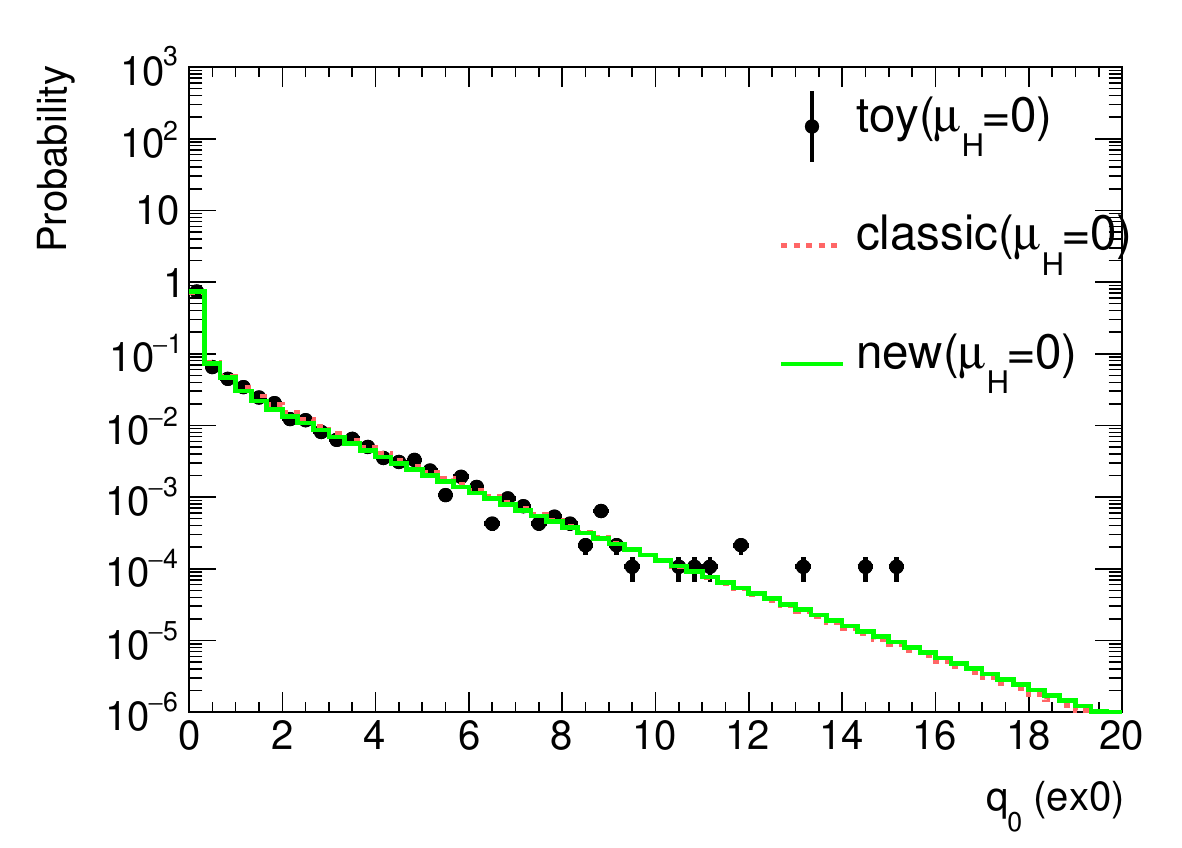}
     \includegraphics[width=0.45\textwidth]{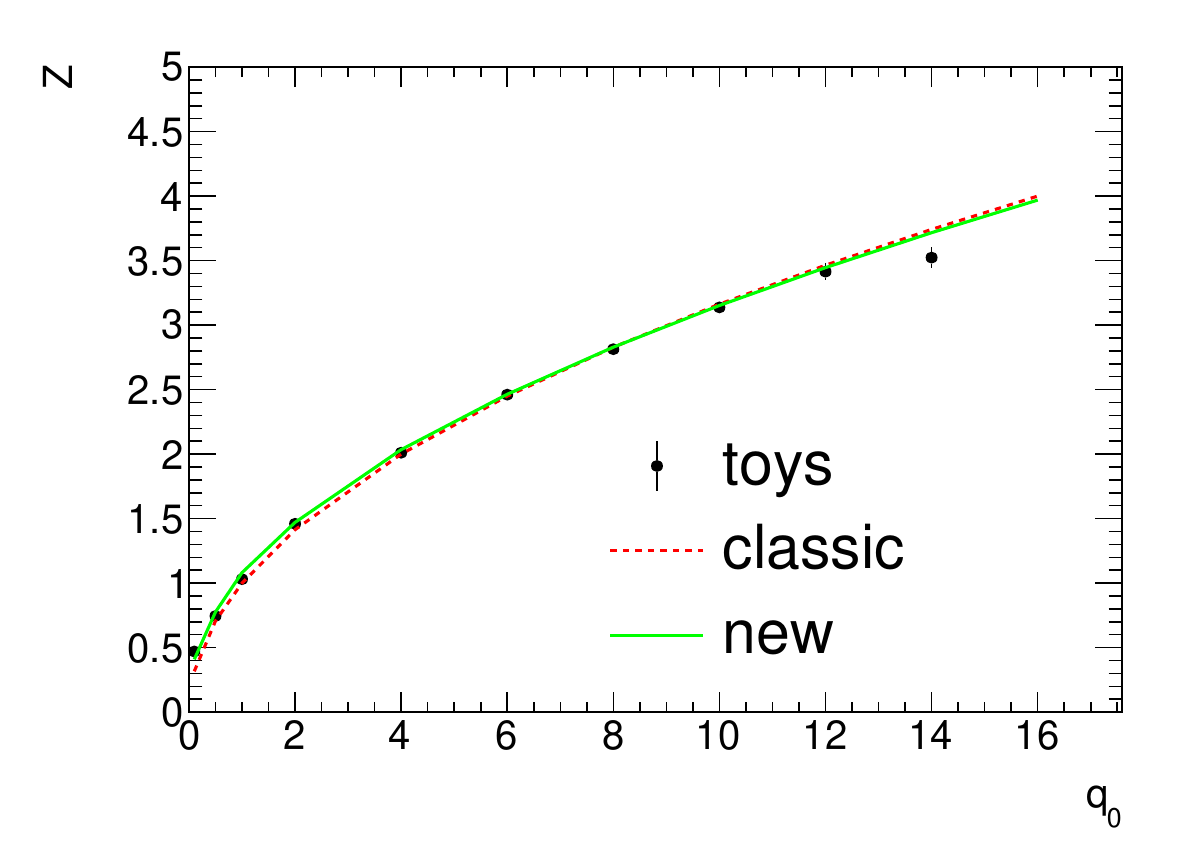}
     \caption{\label{fig:q0_ex0}
     Left column: the probability distributions of $q_0$ in Ex.~0.
     Right column: the significance $Z$ as a function of a possible observed value of $q_0$.
     From top to bottom, it represents different observed datasets with increasing input signal strength.
     The black dots represent the toy MC results. The green solid histograms represent the new asymptotic formulae in this work while the red dashed histograms represent the classic asymptotic formulae from Wald's approximation.
     }
\end{figure}

\begin{figure}[htbp]
    \centering
     \includegraphics[width=0.45\textwidth]{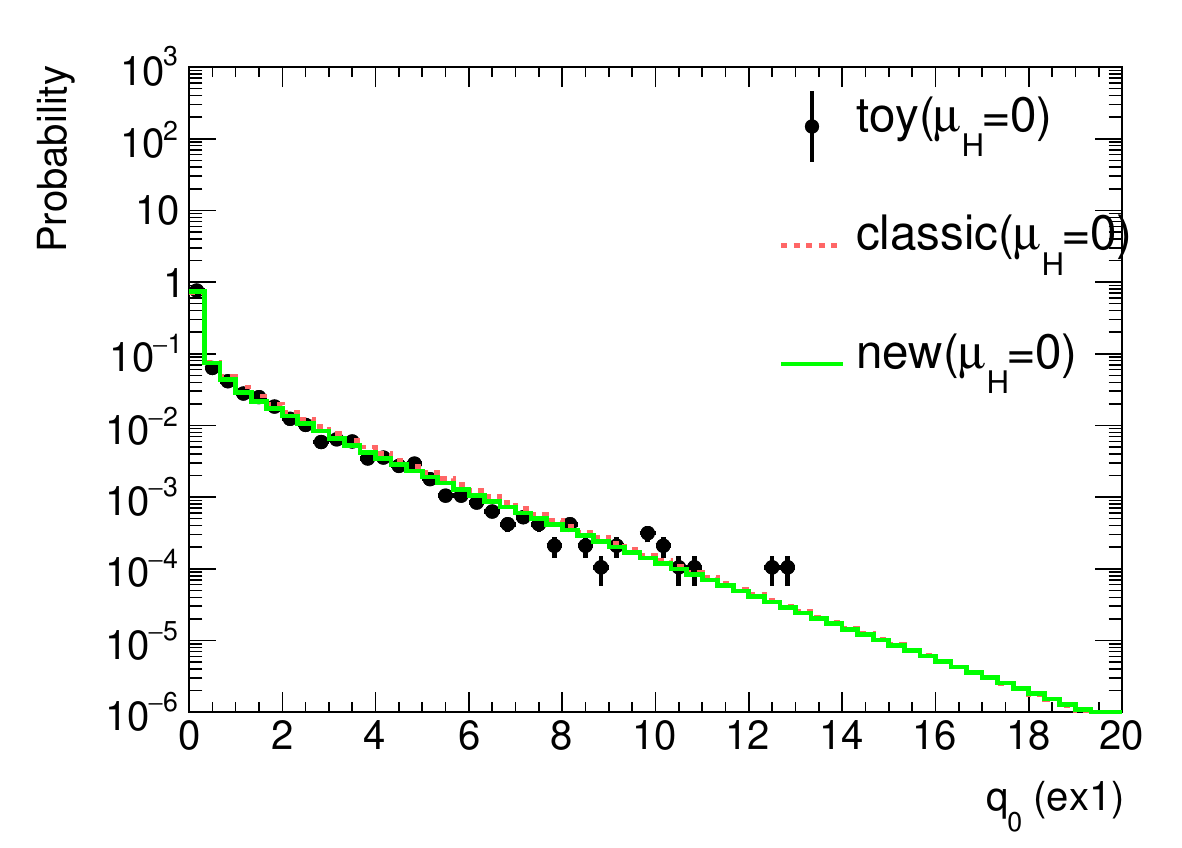}
     \includegraphics[width=0.45\textwidth]{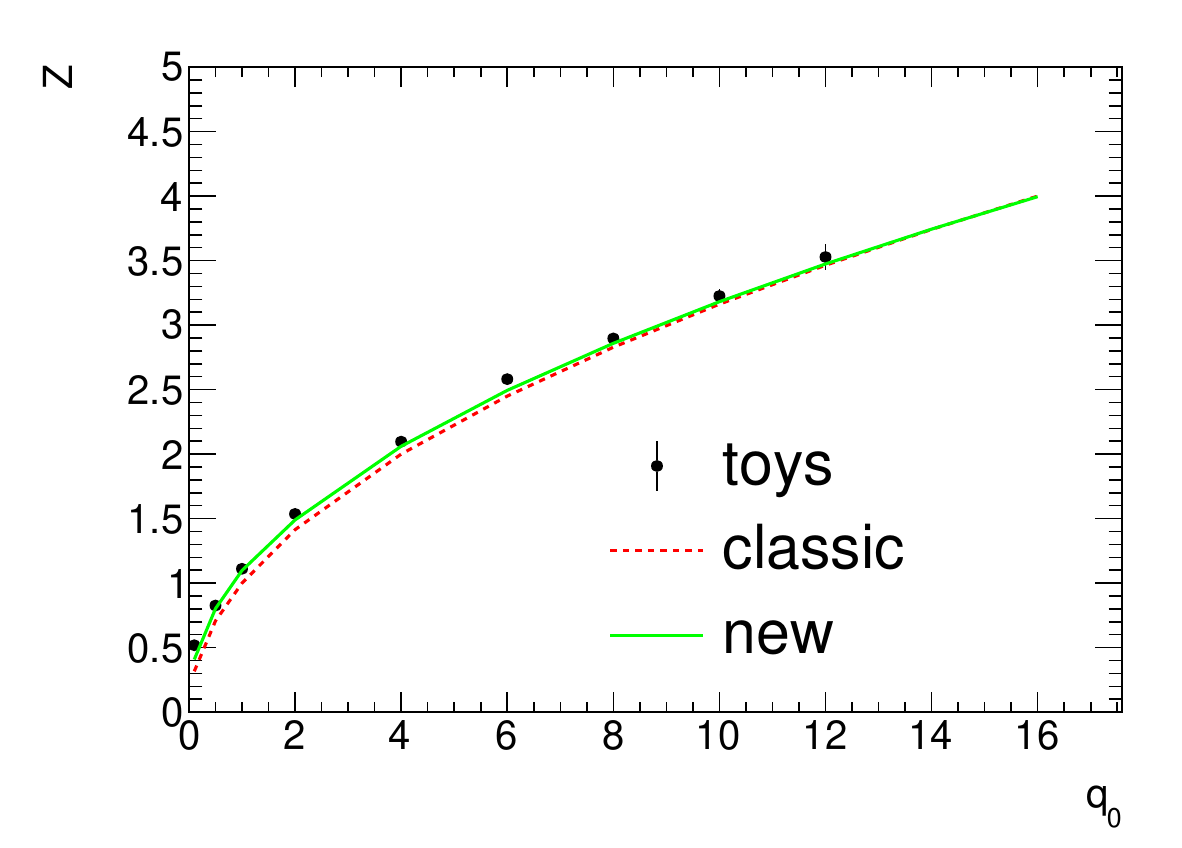}\\
     \includegraphics[width=0.45\textwidth]{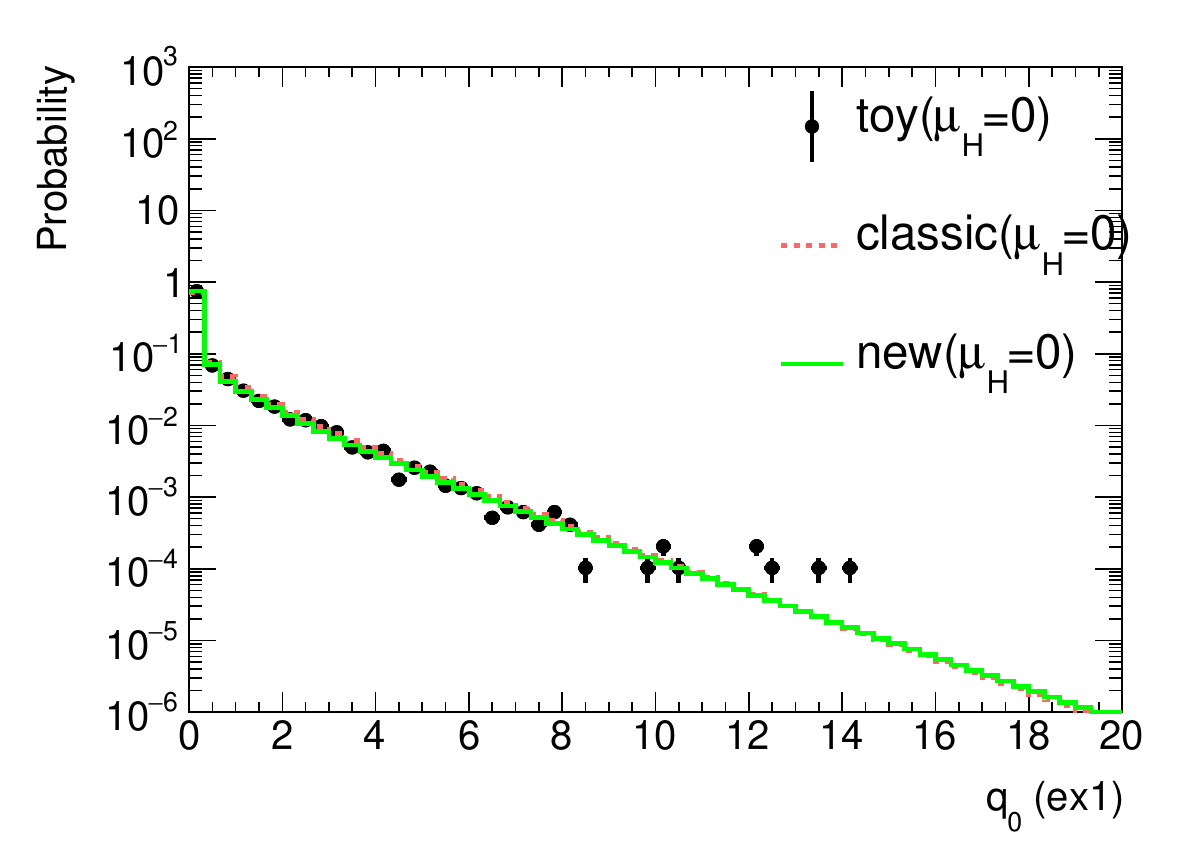}
     \includegraphics[width=0.45\textwidth]{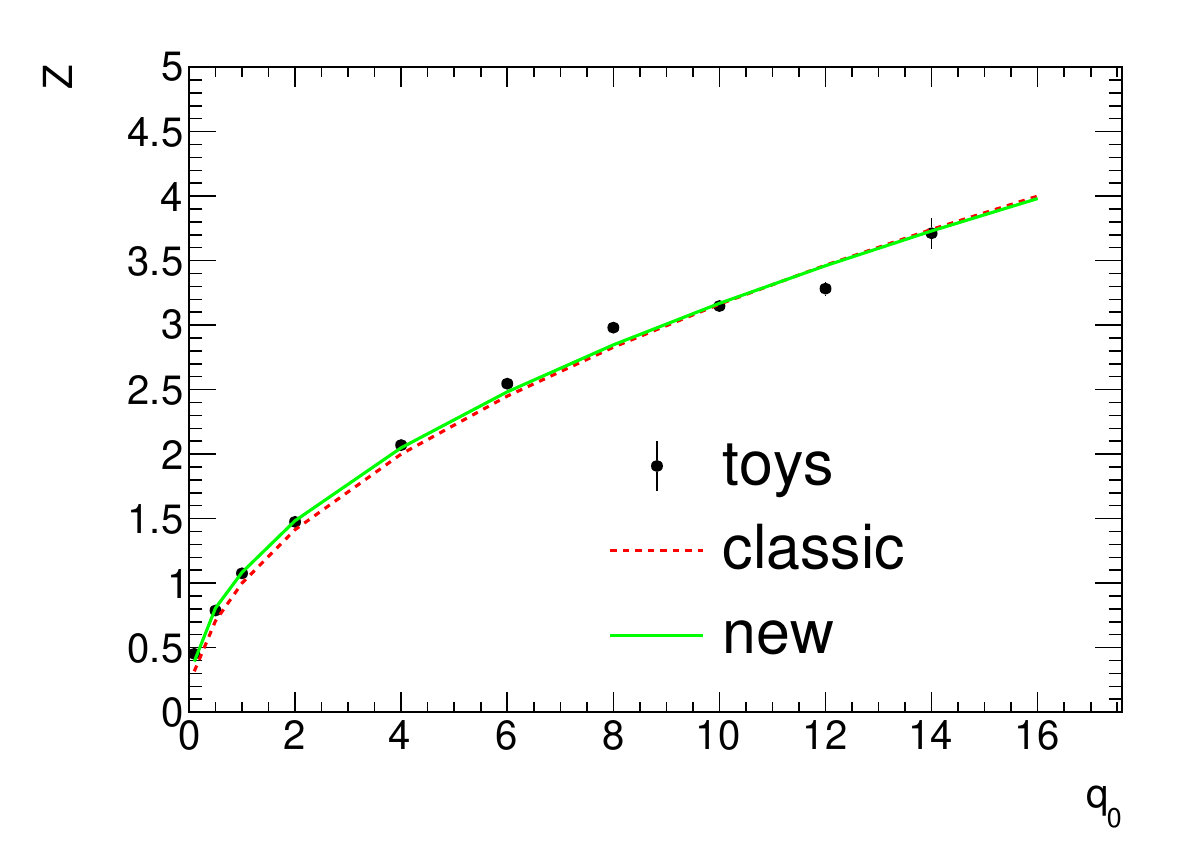}\\
     \includegraphics[width=0.45\textwidth]{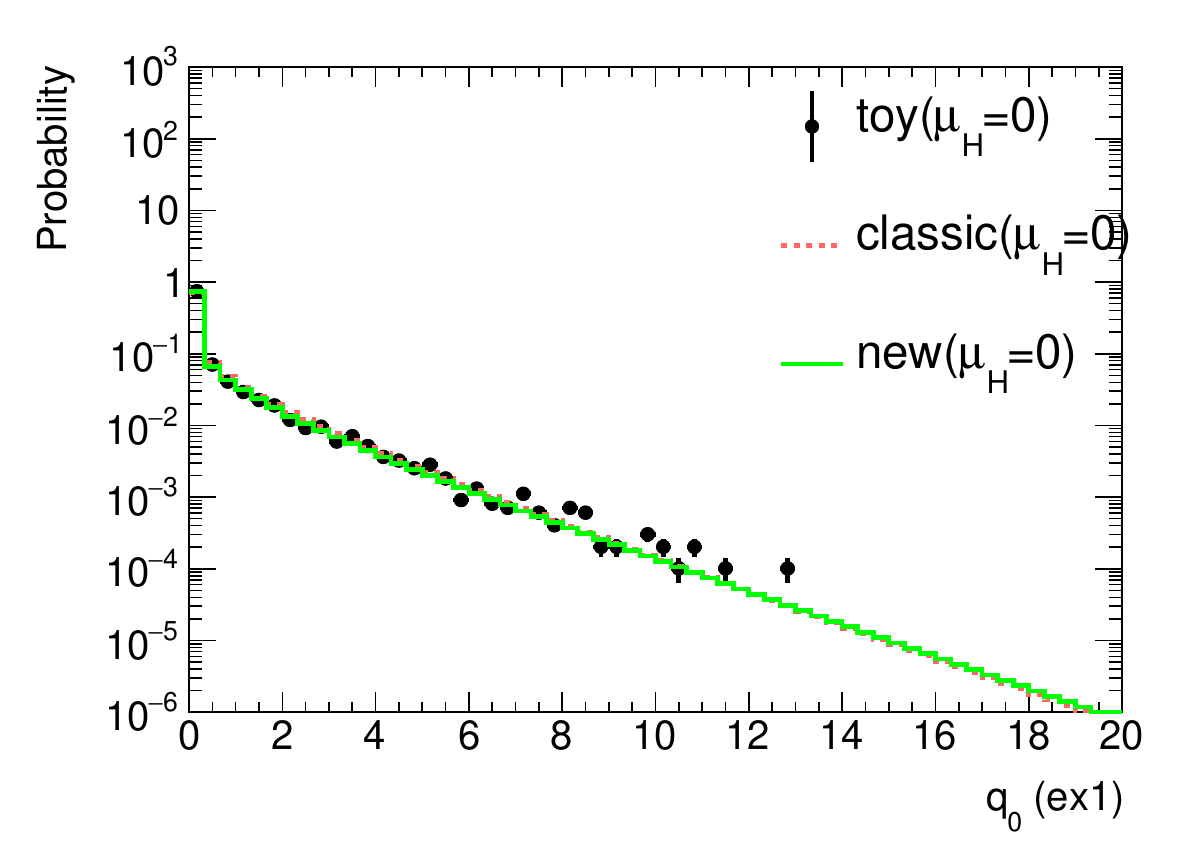}
     \includegraphics[width=0.45\textwidth]{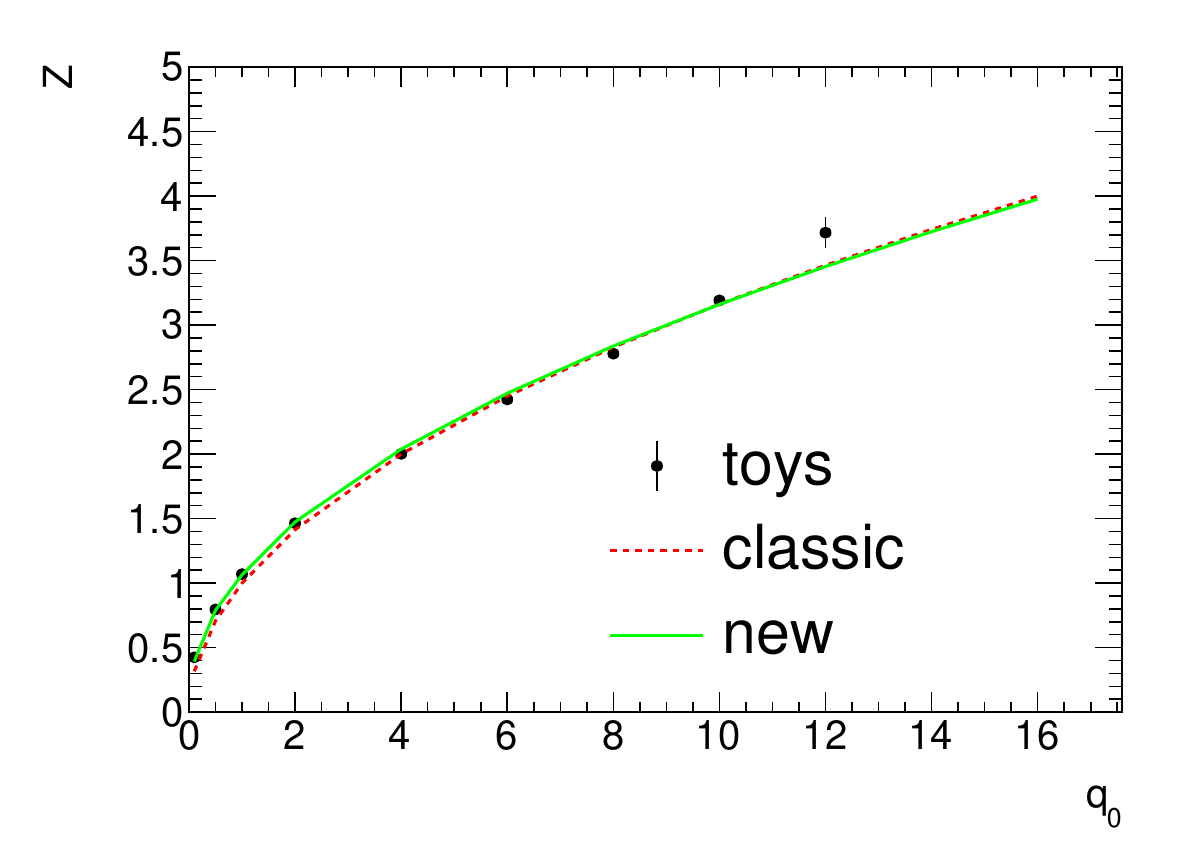}
     \caption{\label{fig:q0_ex1}
     Left column: the probability distributions of $q_0$ in Ex.~1.
     Right column: the significance $Z$ as a function of a possible observed value of $q_0$.
     From top to bottom, it represents different observed datasets with increasing input signal strength.
     The black dots represent the toy MC results. The green solid histograms represent the new asymptotic formulae in this work while the red dashed histograms represent the classic asymptotic formulae from Wald's approximation.
     }
\end{figure}

In this section, we consider the test statistic $q_0$, which is used to establish the discovery of a signal. Figure~\ref{fig:q0_ex0} and ~\ref{fig:q0_ex1} show the distribution of $q_0$ in Ex.~0 and Ex.~1 respectively for different observed datasets. For shape comparison, we also show the significance $Z$ as a function of possible observed value of $q_0$ although there is a unique observed value in each case. Using the classic asymptotic formulae, we have $Z=\sqrt{q_0}$.
From these plots, it is clear that the new formulae work better.

\section{Some remarks on different methods}\label{sec:advantages}
It is worthwhile to leave some remarks on different methods regarding the application. 
\begin{itemize}
    \item If the expected number of events is large, the classic asymptotic formulae are preferred due to their minimal assumptions and computation efficiency compared to either the new method or the toy MC simulations. 
    \item For fewer expected events, the new method proposed in this work is recommended, as it provides more accurate predictions than the classic approach while remaining significantly faster than the MC method. 
It should be noted that the computation time of a MC simulation (dominated by randomization and fitting) scales with the number of systematic uncertainties. 
        Thus in cases with limited statistics but many systematic uncertainties, the MC method remains computationally expensive.
Although the new method requires more time than the classic one, it typically completes an upper limit calculation within minutes, which is still practical.
A detailed comparison of computation times for these methods is provided in Appendix~\ref{app:real_phys_app}. 
\item Cross-validation with MC simulations is strongly advised in cases with small statistics. At least one signal model should be tested  to verify the asymptotic results from either the classic or new formulae.
\end{itemize}

\section{Summary}\label{sec:summary}
In this work, we attempt to refine the classic asymptotic formulae used to describe the probability distributions of the likelihood-ratio statistical tests which are widely used in the field of high energy physics. 
The idea is to split the probability distribution function into two parts. One part is described by the classic formulae with proper corrections. The other part is calculated by mimicking the process of toy MC simulation.
This approach successfully predicts the discrete features in small-statistics cases. Examples with different sample sizes and different ``observed'' datasets are presented. 
The new formulae are found to provide stable improvements to both the differential distribution of the test statistic, and the upper limit and significance calculations.

\acknowledgments{
L.G. Xia would like to thank Fang Dai for her encouragement and partial financial support. This work is supported by the Young Scientists Fund of the National Natural Science Foundation of China (Grant No. 12105140). }

\begin{appendix}
    \section{Motivation for the uncertainty breaking}\label{app:sigma}
    For a binned measurement with $\nbins$ bins, let $b_i$, $s_i$ and $n_i$ be the number of predicted background events, signal events and observed events in the $i$-th bin, respectively. Introducing one signal systematic uncertainty $\delta_i$ and one background systematic uncertainty $\Delta_i$ with the corresponding nuisance parameter $\alpha$ and $\beta$, the logarithmic likelihood function is
    \begin{equation}
        \ln\mL(\mu,\alpha,\beta) = \sum_{i=1}^{\nbins}[n_i\ln(b_i(1+\beta\Delta_i)+\mu s_i(1+\alpha\delta_i))-(b_i(1+\beta\Delta_i)+\mu s_i(1+\alpha\delta_i))] - \frac{\alpha^2}{2}-\frac{\beta^2}{2} \:, 
    \end{equation}
    where the last two terms are due to the Gaussian constraints. For an Asimov dataset $n_i = b_i+\mu s_i$, let the partial derivatives $\frac{\partial \ln\mL}{\partial\mu}$, $\frac{\partial\ln\mL}{\partial \alpha}$ and $\frac{\partial\ln\mL}{\partial\beta}$ vanish to reach the maximum likelihood. We obtain
    \begin{equation}
        \hatmu = \mu\:, \quad \hat{\alpha} = 0 \:, \quad \hat{\beta} = 0 \: .
    \end{equation}
    Now we evaluate the Hessian matrix elements at these optimal values. For simplicity, we introduce $\tilde{\Delta}_i\equiv b_i\Delta_i$, $\tilde{\delta}_i\equiv s_i\delta_i$, and the symbol $\otimes$, which is defined as
    \begin{equation}
        A\otimes B \equiv \sum_{i=1}^{\nbins} \frac{A_iB_i}{n_i} \:.
    \end{equation}
    The Hessian matrix elements are then
    \begin{eqnarray}
        && -\frac{\partial^2\ln\mL}{\partial\mu^2}= s\otimes s \:,\quad -\frac{\partial^2\ln\mL}{\partial\mu\partial\alpha}= (s\otimes \tilde{\delta})\mu\:,\quad  -\frac{\partial^2\ln\mL}{\partial\mu\partial\beta}= (s\otimes \tilde{\Delta})\nonumber\\
        && -\frac{\partial^2\ln\mL}{\partial\alpha^2}= (\tilde{\delta}\otimes \tilde{\delta})\mu^2 + 1\:, -\frac{\partial^2\ln\mL}{\partial\alpha\partial\beta}= (\tilde{\delta}\otimes \tilde{\Delta})\mu \:,  -\frac{\partial^2\ln\mL}{\partial\beta^2}= \tilde{\Delta}\otimes \tilde{\Delta} + 1\: .
    \end{eqnarray}
Let $\bH$ denote the Hessian matrix. It can be written as a sum of two matrices, $\bA$ and $\bB$. 
    \begin{equation}
        \bA = \begin{pmatrix}
            s\otimes s & 0 & 0 \\
            0 & 1 & 0\\
            0 & 0 & 1 \\
        \end{pmatrix}\: ,\quad
        \bB = \begin{pmatrix}
            0 & (s\otimes\tilde{\delta})\mu & s\otimes\tilde{\Delta} \\
            (s\otimes\tilde{\delta})\mu & (\tilde{\delta}\otimes \tilde{\delta})\mu^2 & (\tilde{\delta}\otimes \tilde{\Delta})\mu \\
              (s\otimes \tilde{\Delta}) & (\tilde{\delta}\otimes \tilde{\Delta})\mu & \tilde{\Delta}\otimes \tilde{\Delta}\\
        \end{pmatrix}\: .
    \end{equation}
    Let $\bV$ denote the covariance matrix for $\mu$, $\alpha$ and $\beta$. We have $\bV = \bH^{-1}=(\bA+\bB)^{-1}=(\bI+\bA^{-1}\bB)^{-1}\bA^{-1}$. Assuming all systematic uncertainties are small, we can approximate $\bV$ using the following trick~\cite{xia_constraint}
    \begin{equation}
        \bI = (\bI+\bx)(\bI-\bx+\bx^2-\cdots) \: ,
    \end{equation}
    where $\bI$ is the unit matrix.
Hence we have
    \begin{equation}
        \bV = (\bI + \bA^{-1}\bB)^{-1}\bA^{-1}\approx (\bI - \bA^{-1}\bB + (\bA^{-1}\bB)^2)\bA^{-1}\:, 
    \end{equation}
    and the uncertainty of the signal strength, $\sigma_\mu$, is approximately
    \begin{equation}
        \sigma_\mu^2 = \bV_{11} \approx \frac{1}{s\otimes s}+\left(\frac{s\otimes\tilde{\Delta}}{s\otimes s}\right)^2+\left(\frac{s\otimes \tilde{\delta}}{s\otimes s}\right)^2\mu^2 \:,
    \end{equation}
    where the first term is the statistical uncertainty, the second term is due to the background systematic uncertainty and the third term is due to the signal systematic uncertainty. This is the motivation for the form in Eq.~\ref{eq:sigma_form} and hence the following relation.
    \begin{equation}
         \sigma_0 = \frac{s\otimes\tilde{\Delta}}{s\otimes s} \:,\quad \kappa = \frac{s\otimes\tilde{\delta}}{s\otimes s} \: .
    \end{equation}
    It is of no difficulty to extend to the case of multiple signal and background systematic uncertainties, and the same conclusion holds in the sense that all systematic uncertainties are small. For better explanation in next appendix and supposing we have $\nsysts^S$ signal systematic uncertainties, $\kappa$ becomes
    \begin{equation}
        \kappa = \sqrt{\sum_{k=1}^{\nsysts^S}\left(\frac{s\otimes\tilde{\delta}^k}{s\otimes s}\right)^2 }\: ,
    \end{equation}
    where $s\otimes\tilde{\delta}^k = \sum_{i=1}^{\nbins} \frac{s_is_i\delta_i^k}{n_i}$ with $\delta_i^k$ being the effect of the $k$-th signal uncertainty on the $i$-th bin. If the uncertainties only affect the yield, we have $\delta_i^k$ is the same for $i$ (written as $\delta^k$) and 
    \begin{equation}\label{eq:kappa0}
        \kappa = \sqrt{\sum_{k=1}^{\nsysts^S}(\delta^k)^2} \: .
    \end{equation}
    
    \section{Likelihood-ratio tests in the case of 0 events}\label{app:zero_events}
    The probability of observing 0 events is significant in searching for new physics with low background. The likelihood-ratio tests may behave very differently in this extreme case. Therefore, we consider it dedicatedly in this appendix. For a binned measurement with $\nsysts^B$ background systematic uncertainties and $\nsysts^S$ signal systematic uncertainties, the logarithmic likelihood function is 
    \begin{eqnarray}
        \label{eq:logl_aux}
        \ln\mL(\mu,\theta_1,\alpha_1,\cdots) = && \sum_{i=1}^{\nbins}[n_i\ln \nu_i(\mu,\theta_1,\alpha_1,\cdots) - \nu_i(\mu,\theta_1,\alpha_1,\cdots)]\nonumber\\
        &&- \sum_{j=1}^{\nsysts^B}\frac{(\theta_j-\bar{\theta}_j)^2}{2} - \sum_{k=1}^{\nsysts^S}\frac{(\alpha_k-\bar{\alpha}_k)^2}{2}\:,
    \end{eqnarray}
    with
    \begin{equation}
        \nu_i(\mu,\theta_1,\alpha_1,\cdots) \equiv b_i(1+\sum_{j=1}^{\nsysts^B}\theta_j\Delta_i^j)+\mu s_i(1+\sum_{k=1}^{\nsysts^S}\alpha_k\delta_i^k) \: .
    \end{equation}
    Here $\nbins$ is the number of bins; $b_i$ and $s_i$ are the expected number of background and signal events, respectively; $n_i$ is the observed number of events; $\Delta_i^j$ ($\delta_i^k$) is the effect in the $i$-th bin due to the $j$-th background (the $k$-th signal) systematic uncertainty; $\theta_j$s and $\alpha_k$s are the nuisance parameters while $\bar{\theta}_j$s and $\bar{\alpha}_k$s are auxiliary data in the toy experiment generation. 
    The last two summations are due to Gaussian constraint.

    To reach the maximum likelihood, we investigate the partial derivatives $\frac{\partial \ln\mL}{\partial \theta_j}$, $\frac{\partial \ln\mL}{\partial \alpha_k}$ and $\frac{\partial \ln\mL}{\partial\mu}$. For the unconditional fit, we have
    \begin{eqnarray}
        &&\frac{\partial \ln\mL}{\partial \theta_j} = \sum_{i=1}^{\nbins}\frac{n_ib_i\Delta_i^j}{\nu_i(\mu,\theta_1,\alpha_1,\cdots)} - b_i\Delta_i^j - (\theta_j-\bar{\theta}_j) \:,\\ 
        &&\frac{\partial \ln\mL}{\partial \alpha_k} = \sum_{i=1}^{\nbins}\frac{n_i\mu s_i\delta_i^k}{\nu_i(\mu,\theta_1,\alpha_1,\cdots)} - \mu s_i\delta_i^k - (\alpha_k-\bar{\alpha}_k) \:,\\ 
        &&\frac{\partial \ln\mL}{\partial \mu} = \sum_{i=1}^{\nbins} \frac{n_is_i(1+\sum_{k=1}^{\nsysts^S}\alpha_k\delta_i^k)}{\nu_i(\mu,\theta_1,\alpha_1,\cdots)}-s_i(1+\sum_{k=1}^{\nsysts^S}\alpha_k\delta_i^k)  \:. 
    \end{eqnarray}
    We only consider the case of 0 events in all bins, namely, $n_1=n_2=\cdots=0$. $\frac{\partial \ln\mL}{\partial \mu}$ is assumed to be negative as it is true for small signal uncertainties. The optimal values satisfy
    \begin{eqnarray}
        && \hat{\theta}_j = \bar{\theta}_j -\sum_{i=1}^{\nbins} b_i\Delta_i^j\quad (j=1,2,\cdots,\nsysts^B) \: , \\
        && \hat{\alpha}_j = \bar{\alpha}_k -\hat{\mu}\sum_{i=1}^{\nbins} s_i\delta_i^k \quad (j=1,2,\cdots,\nsysts^S) \: , \\
        && \hatmu = \max\{-\frac{b_1(1+\sum_{j=1}^{\nsysts^B}\hat{\theta}_j\Delta_1^j)}{s_1(1+\sum_{k=1}^{\nsysts^S}\hat{\alpha}_k\delta_i^k)}, -\frac{b_2(1+\sum_{j=1}^{\nsysts^B}\hat{\theta}_j\Delta_2^j)}{s_2(1+\sum_{k=1}^{\nsysts^S}\hat{\alpha}_k\delta_i^k)},\cdots\} \label{eq:muhat_0event}\: . 
    \end{eqnarray}
Since $\frac{\partial \ln\mL}{\partial \mu}<0$, we choose $\hatmu$ to be smallest value to make the expected number of events non-negative in all bins.
    For the conditional fit with $\mu$ fixed, the optimal values are
    \begin{eqnarray}
        && \hat{\hat{\theta}}_j(\mu) = \hat{\theta}_j \:, \\
        && \hat{\hat{\alpha}}_k(\mu) = \bar{\alpha}_k + \frac{\mu}{\hatmu}(\hat{\alpha}_k-\bar{\alpha}_k) \quad (j=1,2,\cdots,\nsysts^S) \: .
    \end{eqnarray}
We find that $\hat{\hat{\theta}}_j(\mu)$ is the same as $\hat{\theta}_j$ and independent on the value of $\mu$. This is essentially different from the case with non-vanishing observed events.
    
    The tests, $q_\mu$ and $\qtil_\mu$, are then
    \begin{eqnarray}
        q_\mu&=& -2\ln\frac{\mL(\mu,\hat{\hat{\theta}}_1(\mu),\hat{\hat{\alpha}}_k(\mu),\cdots)}{\mL(\hatmu,\hat{\theta}_1, \hat{\alpha}_1,\cdots)} \nonumber\\
        &=& 2(\mu-\hatmu) \sum_{i=1}^{\nbins}s_i(1+\sum_{k=1}^{\nsysts^S}\delta_i^k\bar{\alpha}_k) - (\mu^2-\hatmu^2)\sum_{k=1}^{\nsysts^S}(\sum_{i=1}^{\nbins}s_i\delta_i^k)^2\:, \label{eq:qmu_0event}\\
        \qtil_\mu&=& -2\ln\frac{\mL(\mu,\hat{\hat{\theta}}_1(\mu),\hat{\hat{\alpha}}_k(\mu),\cdots)}{\mL(0,\hat{\hat{\theta}}_1(0), \hat{\hat{\alpha}}_1(0),\cdots)} \nonumber\\
        &=& 2 \mu \sum_{i=1}^{\nbins}s_i(1+\sum_{k=1}^{\nsysts^S}\delta_i^k\bar{\alpha}_k) - \mu^2\sum_{k=1}^{\nsysts^S}(\sum_{i=1}^{\nbins}s_i\delta_i^k)^2\:. \label{eq:qmutil_0event}
    \end{eqnarray}
We can see that the effect of background systematic uncertainties is vanishing for $\qtil_\mu$ and its effect on $q_\mu$ is via $\hatmu$ and also greatly reduced because of a single bin with the least background-to-signal ratio in Eq.~\ref{eq:muhat_0event}. 
It means that the distribution of the tests is mainly due to signal systematic uncertainties. 

Assuming all signal systematic uncertainties are small, we neglect the last term in Eq.~\ref{eq:qmu_0event} and treat the auxiliary data $\bar{\alpha}_k$s as independent random variables abiding by a normal distribution. The standard derivation of $q_\mu$, denoted as $\Delta q_\mu$, would be approximately
\begin{equation}
    \frac{\Delta q_\mu}{q_\mu} = \sqrt{\sum_{k=1}^{\nsysts^S}\left(\frac{\sum_{i=1}^{\nbins}s_i\delta_i^k}{\sum_{i=1}^{\nbins}s_i}\right)^2} \:, 
\end{equation}
We can see that only the yield-related uncertainties matter here (otherwise $\sum_{i=1}^{\nbins}s_i\delta_i^k=0$). If all uncertainties affect the yield only, we have (writing $\delta_i^k$ as $\delta^k$)
\begin{equation}
    \frac{\Delta q_\mu}{q_\mu} = \sqrt{\sum_{k=1}^{\nsysts^S}(\delta^k)^2} \:, 
\end{equation}
We can see that it is equal to $\kappa$ in Eq.~\ref{eq:kappa0}.
Therefore, in the case of observing 0 events, we assume that the standard derivation of the tests is due to signal related systematic uncertainties only and
\begin{equation}\label{eq:dqmu}
    \frac{\Delta q_\mu}{q_\mu}=\frac{\Delta\qtil_\mu}{\qtil_\mu}=\kappa \:,
\end{equation}
where $\kappa$ is defined in Eq.~\ref{eq:k}. 

\begin{figure}[htbp]
     \centering
     \includegraphics[width=0.45\textwidth]{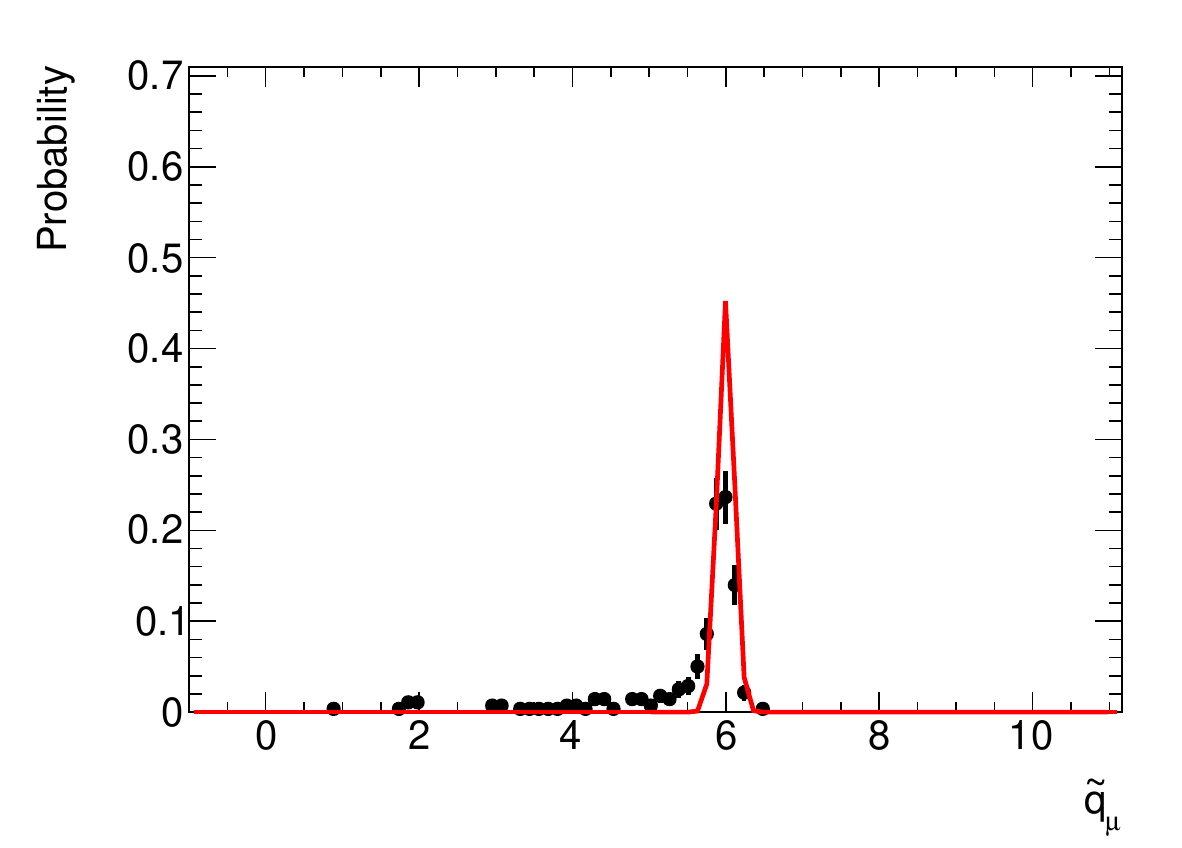}
     \includegraphics[width=0.45\textwidth]{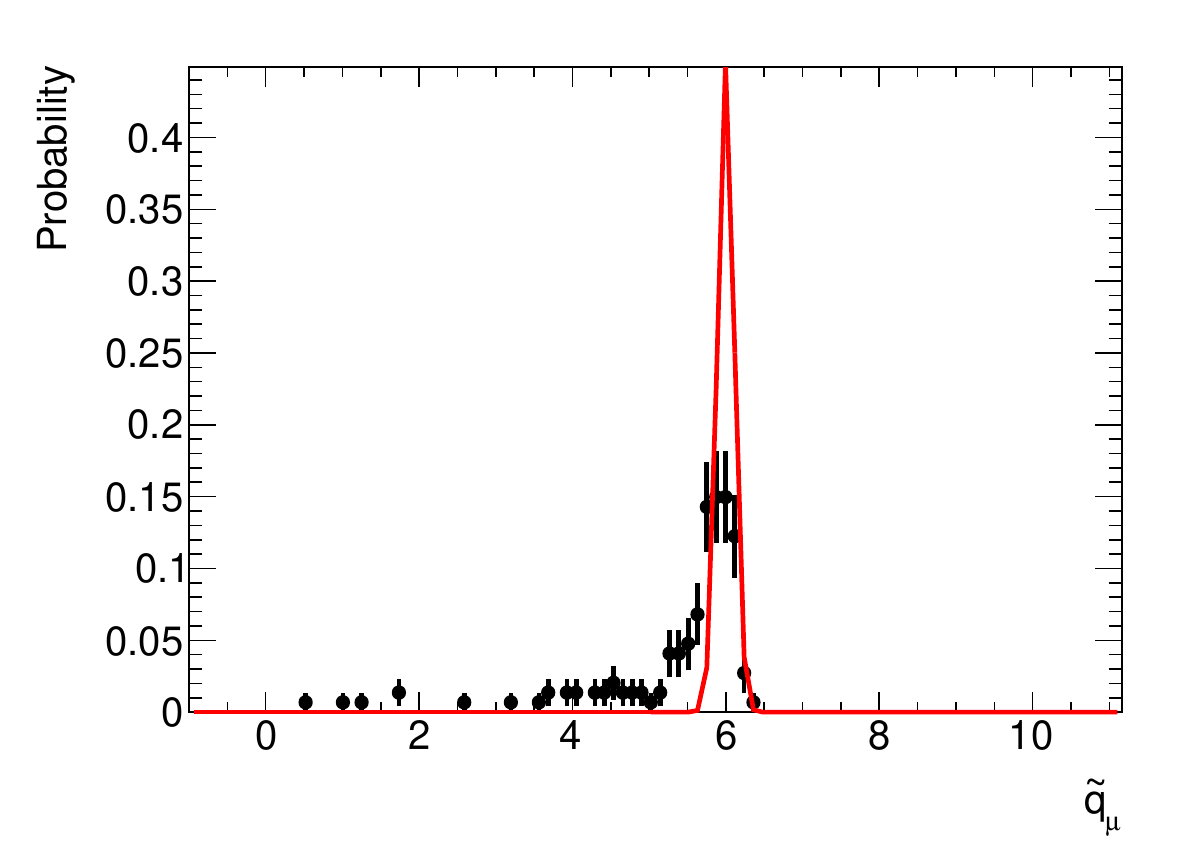}
     \caption{\label{fig:qmu_0event}
     The distribution of $\qtil_\mu$ in Ex.~0 from the toy experiments under the hypothesis $\muH=0$ (Left) and $\muH=\mu=3$ (Right). The red curve represents the prediction in this work. 
     }
\end{figure}

Taking the Ex.~0 in Sec.~\ref{sec:example} as example, Fig.~\ref{fig:qmu_0event} shows the distribution of $\qtil_\mu$ in the case of 0 events in the toy experiments as well as the prediction in this work.
We have also checked several real measurements and found that the assumption above is reasonable.
In the 6-bin model, the last bin has the least signal-to-background ratio and a negligible signal expectation. Therefore, in Eq.~\ref{eq:f_SS} we adopt the following $\sigma(\hatmu)$ if $k_0=k_1=k_2=k_3=k_4=0$, to make the assumption in Eq.~\ref{eq:dqmu} hold. 
\begin{equation}\label{eq:sigma_key}
    \sigma(\hatmu) = \left\{
        \begin{matrix}
            \kappa\frac{\mu-2\hatmu}{2}\:, &\text{for}\: \qtil_\mu \\
            \kappa\frac{\mu-\hatmu}{2}\:, &\text{for}\: q_\mu \\
        \end{matrix}
        \right.
\end{equation}
As shown above, the optimal value, $\hatmu$, is at its lowest bound in the case of observing 0 events. But it also happens that $\hatmu$ is at the lowest bound with non-vanishing events. The key is that the first-order derivative of the logarithmic likelihood function is non-vanishing and will contribute to the likelihood-ratio tests. 
The binned model in current work does not include any nuisance parameter or freely-floating parameter and cannot predict the spread of the test statistic's distribution well. Therefore, in such cases, we propose to use the following $\sigma(\hatmu)$.
\begin{equation}\label{eq:sigma_negative}
    \sigma(\hatmu) = \sqrt{\sigma_0^2+(\kappa\hatmu)^2+(\kappa\frac{\mu-c\hatmu}{2})^2}
\end{equation}
Here $c=1$ for $q_\mu$ and $c=2$ for $\qtil_\mu$. This is just the combination of the uncertainty in the usual case in Eq.~\ref{eq:sigma_form} and that in the 0-event case in Eq.~\ref{eq:sigma_key}. 

\section{Two applications in the searches for new physics}\label{app:real_phys_app}
In this appendix, we would like to apply the new asymptotic formulae to real physics analyses. Some codes about the application are shared in the GitHub repository~\cite{githubxia}, which may be helpful for people who want to try the new formulae. Two physics searches reported in Ref.~\cite{sbottom_ATLAS2019} and Ref.~\cite{alp_ATLAS2024} are considered. 

\subsection{Search for the lightest bottom quark supersymmetric partner}\label{app:sbottom}
In Ref.~\cite{sbottom_ATLAS2019}, a search for the lightest bottom quark supersymmetric partner ($\tilde{b}_1$) using the decay chain $\tilde{b}_1 \to b+\tilde{\chi}_2^0,\tilde{\chi}_2^0\to h+\tilde{\chi}_1^0$ has been performed in the ALTAS experiment. The
experimenters have designed 3 signal regions. If using all of them in the physics interpretation, we believe the classic asymptotic formulae will work well. 
Hence we will use only one signal region (``SRA'' in Ref.~\cite{sbottom_ATLAS2019}) and consider the specific signal model with the mass of $\tilde{b}_1,\tilde{\chi}_2^0,\tilde{\chi}_1^0$ being 600~GeV, 280~GeV and 150~GeV, respectively. Compared to the toy examples in the main text, a real measurement usually has much more nuisance parameters corresponding to various systematic uncertainties. The one considered here has 59 nuisance parameters. 

\begin{figure}[htbp]
    \centering
    \includegraphics[width=0.45\textwidth]{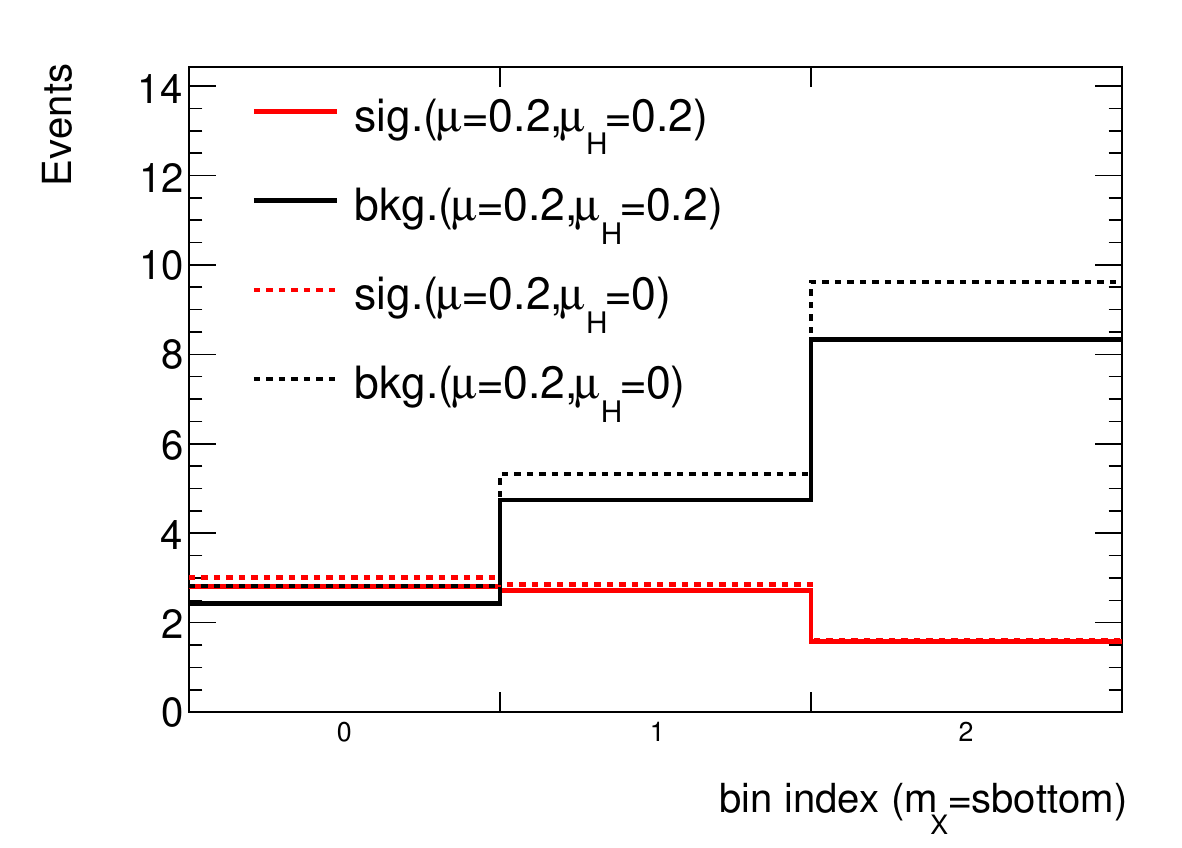}
    \caption{\label{fig:sbottom_sb}
    The 3-bin model for the signal and background in different hypotheses. Here the signal strength is fixed at the test value of 0.2. The observable distribution difference between different hypotheses is due to the other nuisance parameters or free parameters. 
    }
\end{figure}

The toy simulations based on the public workspace~\cite{sbottom_hepdata} are produced using the PyHF~\cite{pyhf} framework. The observable distribution in the signal region has only 3 bins. Thus the SS part in the new asymptotic formulae is estimated based on a 3-bin model. Figure~\ref{fig:sbottom_sb} shows the 3-bin model for the hypothesis $\mu_H=0$ and $\mu_H=0.2$. 
Since the number of total background events is around 15, the predictions using the classic asymptotic formulae are expected to be not bad. We use the statistic $\qtil_\mu$ and consider difference $\nsmall$ choices.

\begin{figure}[htbp]
    \centering
    \includegraphics[width=0.45\textwidth]{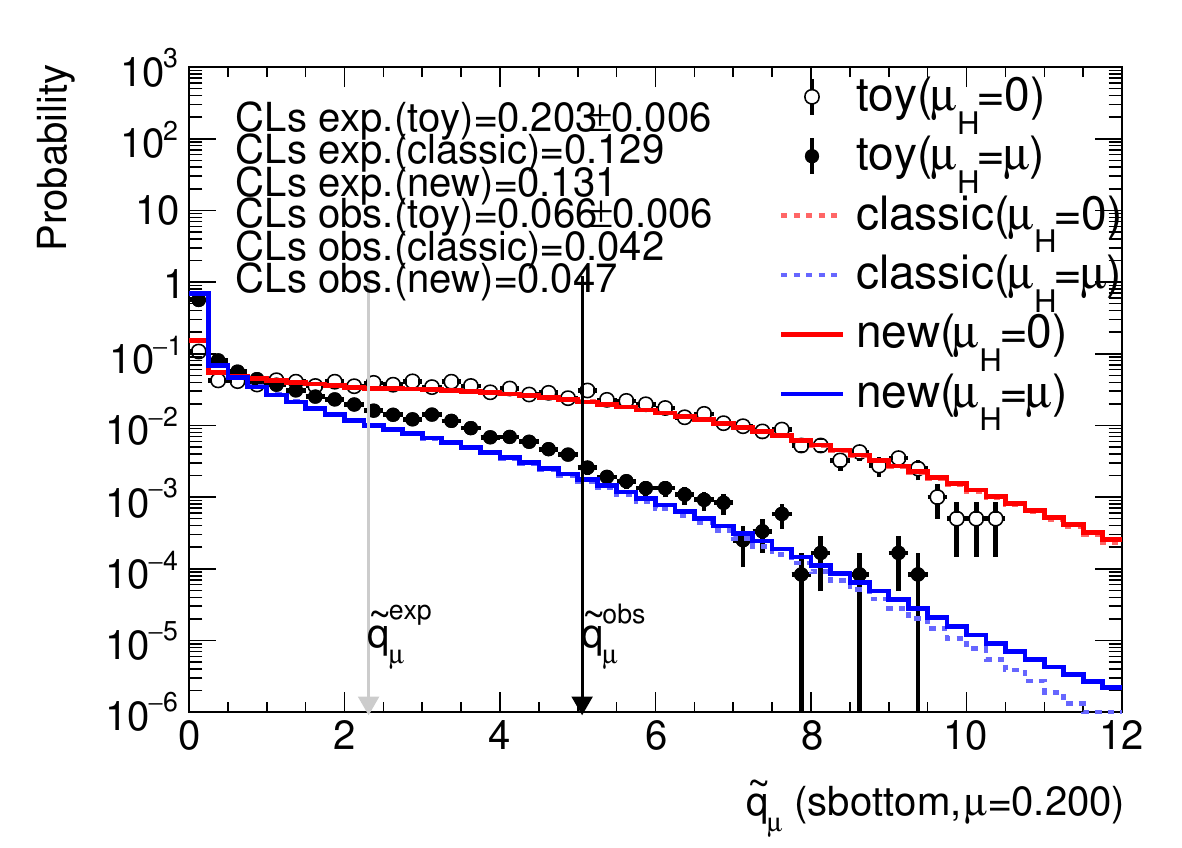}
    \includegraphics[width=0.45\textwidth]{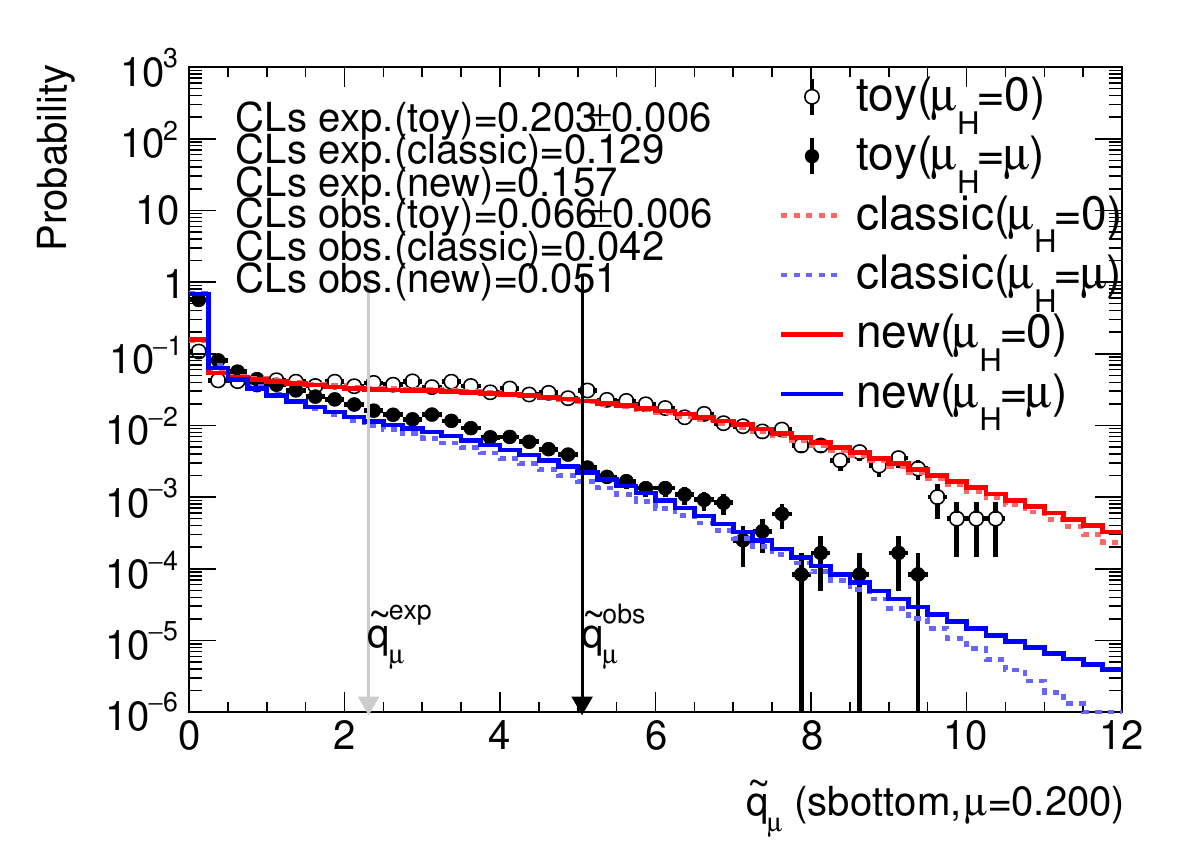}
    \caption{\label{fig:sbottom_qmu}
     The probability distributions of $\qtil_\mu$ for $\nsmall=\nsmall^{0}$ (L) and $\nsmall=\nsmall^{0}+5$ (R).
     The black dots and open circles represent the toy MC results. The blue/red solid histograms represent the new asymptotic formulae in this work while the blue/red dashed histograms represent the classic asymptotic  formulae from Wald's approximation. The black and gray arrows represent the observed and expected $\tilde{q}_\mu$, respectively.
    }
\end{figure}

\begin{figure}[htbp]
     \centering
     \includegraphics[width=0.45\textwidth]{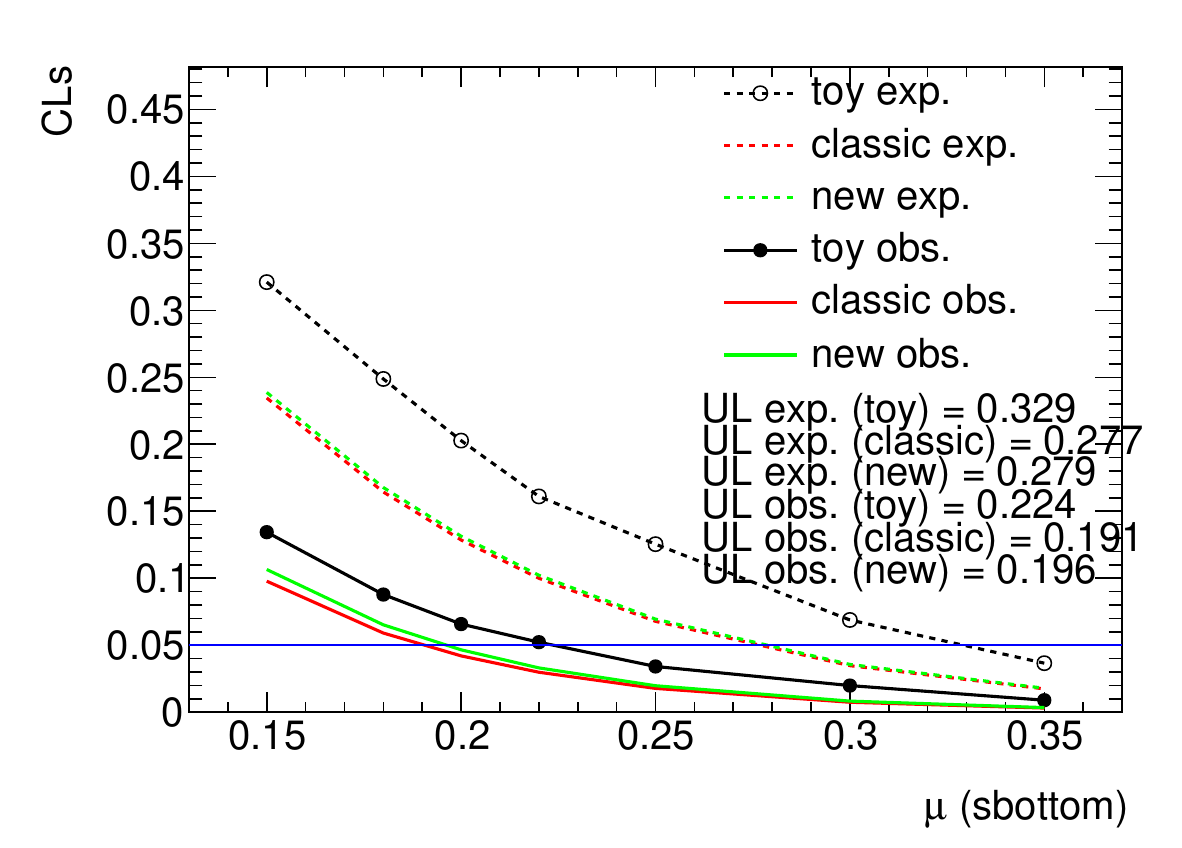}
     \includegraphics[width=0.45\textwidth]{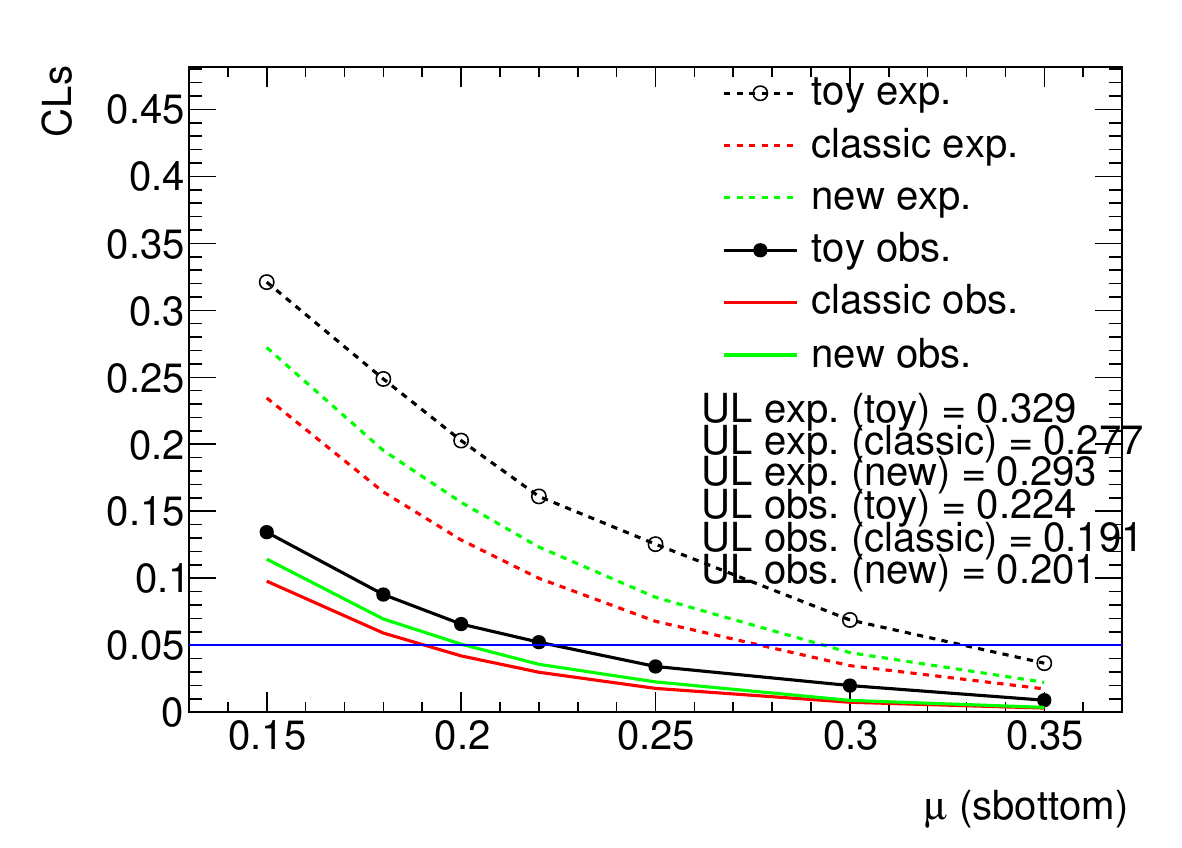}
     \caption{\label{fig:sbottom_CLs}
     CLs as a function of $\mu$ for $\nsmall=\nsmall^{0}$ (L) and $\nsmall=\nsmall^{0}+5$ (R) using the test statistic $\qtil_\mu$. From top to bottom, they represent different observed datasets. The black curves with markers show the toy MC results. The red and green curves are the predictions from the classic and new asymptotic formulae, respectively.  
     }
 \end{figure}

\begin{figure}[htbp]
    \centering
     \includegraphics[width=0.45\textwidth]{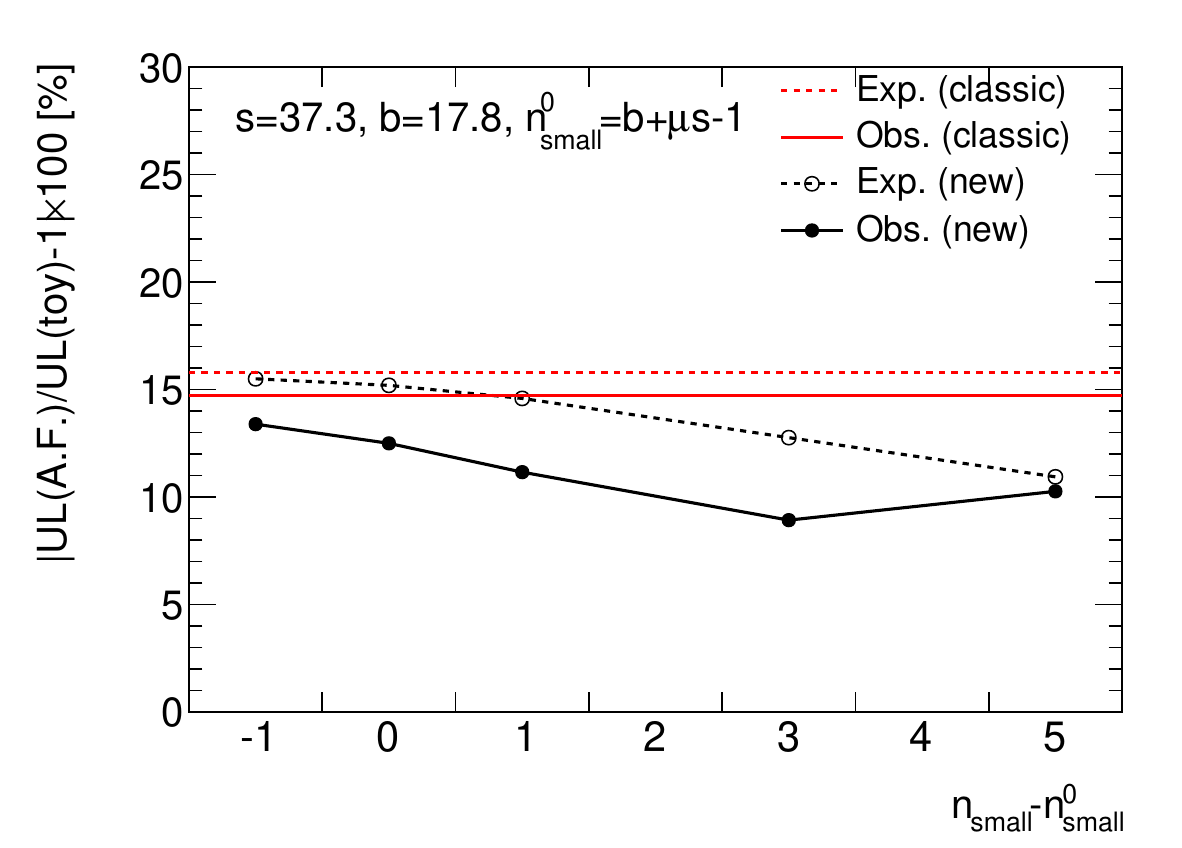}
     \caption{\label{fig:sbottom_nsmall}
     The relative difference of upper limit compared to the toy results as a function of the choice of $\nsmall$.
     }
\end{figure}

Figure~\ref{fig:sbottom_qmu} shows an example of the $\qtil_\mu$ distribution for $\nsmall=\nsmall^{0}$ and $\nsmall=\nsmall^{0}+5$. The CLs curves are shown in Fig.~\ref{fig:sbottom_CLs}. The upper limits from the toy simulations and the asymptotic formulae are compared in Fig.~\ref{fig:sbottom_nsmall}. The new formulae work better according to these plots. 

In the end, let us take this application as example to compare the time consumption for different methods. For each test value of the signal strength, we produce 12000 toys for the signal-plus-background hypothesis and 4000 toys the each background-only hypothesis. They could determine the observed CLs at around $5\%$ with a precision of about $10\%$ and their production takes more than 5 hours. We also have to produce toys for a few test values to determine the upper limit.
For comparison, the new method takes about 5 minutes while the classic method takes less than 1 minute.

\subsection{Search for axion-like particles}
\begin{figure}[htbp]
    \centering
     \includegraphics[width=0.45\textwidth]{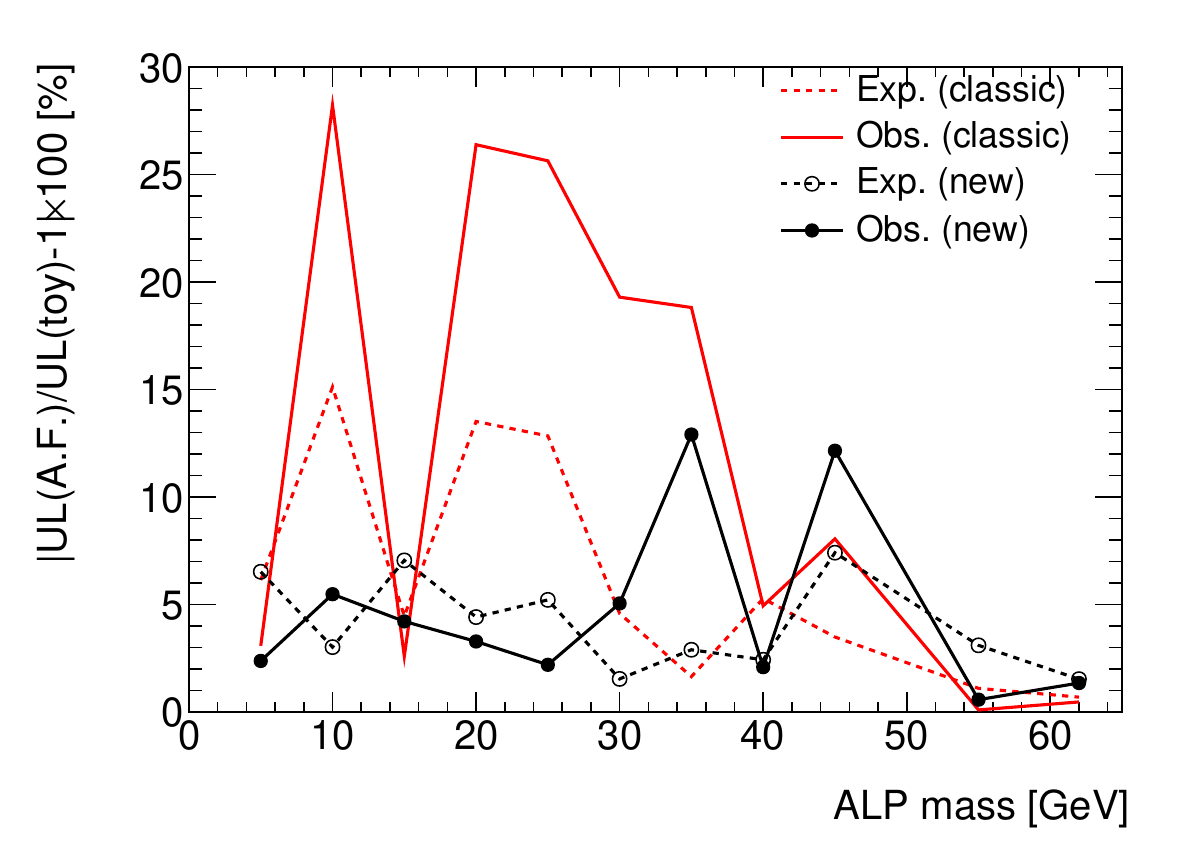}
     \caption{\label{fig:alp_perf}
     The relative difference of upper limit compared to the toy results in the search for prompt ALPs with mass ranging from 5~GeV to 62~GeV. 
     }
\end{figure}

\begin{figure}[htbp]
    \centering
     \includegraphics[width=0.45\textwidth]{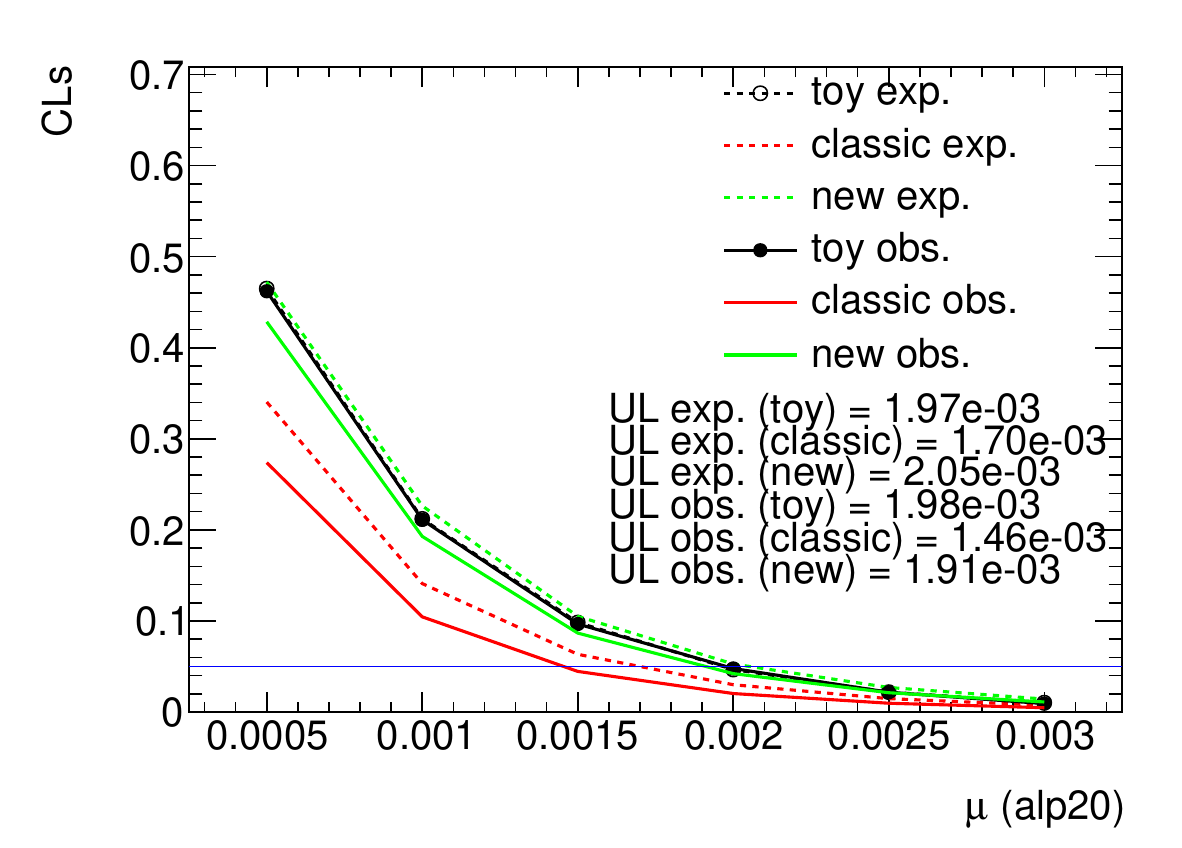}
     \includegraphics[width=0.45\textwidth]{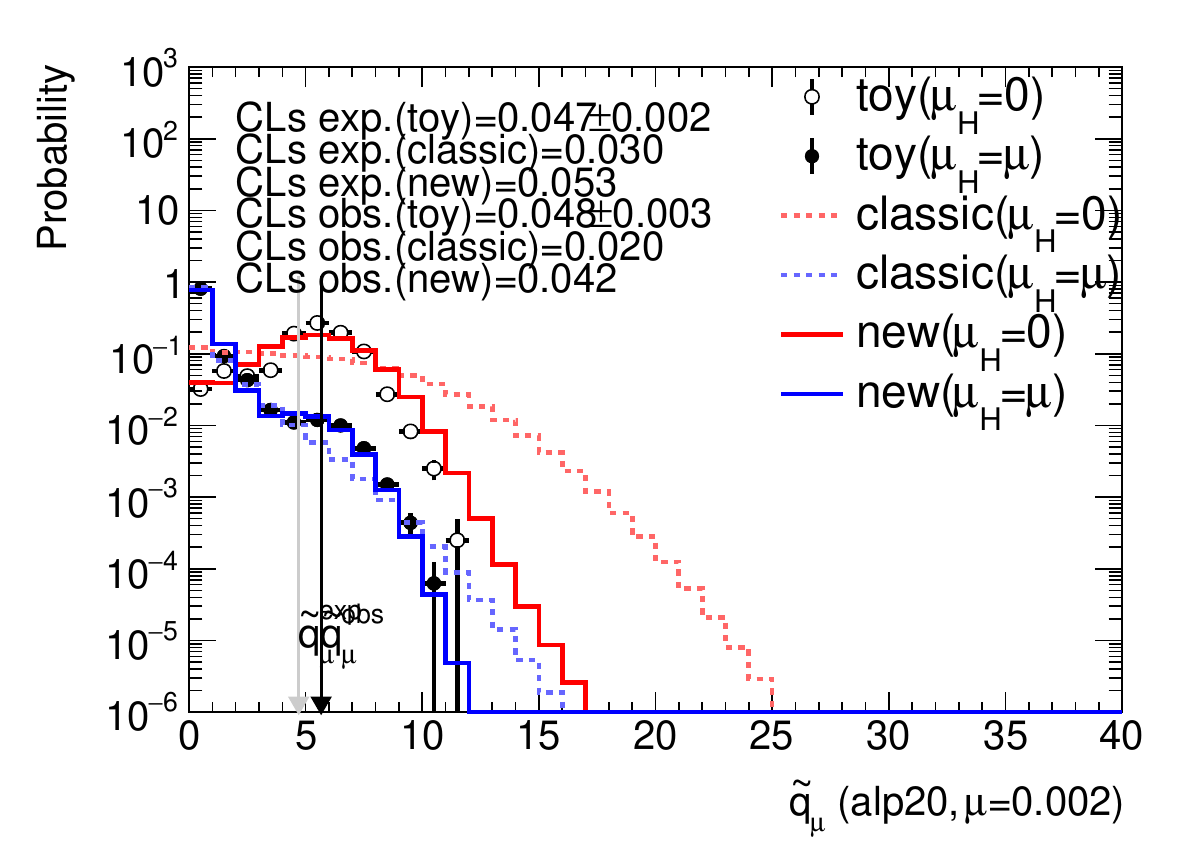}
     \caption{\label{fig:alp_cls_qmu}
     The CLs curve as a function of the test signal strength (L) and the $\qtil_\mu$ distributions (R) for the test signal strength close to the 95~\% upper limit for the ALP signal with mass of 20~GeV. 
     }
\end{figure}

In Ref.~\cite{alp_ATLAS2024}, a search for axion-like particles (ALPs) in $H\to aa \to 4\gamma$ decays with the ATLAS experiment has been conducted. Here we only consider the search for prompt ALPs with mass ranging from 5~GeV to 62~GeV. The full likelihoods for the statistical interpretation can be found in~\cite{alp_hepdata}. Table~\ref{tab:alp} is the overview of the number of observed events in comparison to the expected number of background events for different ALP signals quoted from Table~2 in Ref.~\cite{alp_ATLAS2024}.
Few events are observed for most of the signals. It seems appropriate to compare the two asymptotic approaches.
The upper limits on the branching fraction $B(H\to aa\to 4\gamma)$ are derived using toy simulations and asymptotic formulae. The results are compared in Fig.~\ref{fig:alp_perf}. 
Taking the ALP signal with mass of 20~GeV as example, the CLs curves and the $\qtil_\mu$ distributions are compared in Fig.~\ref{fig:alp_cls_qmu}. 
As shown in these figures, the performance of the new method is generally better. 

\begin{table}
     \centering
     \caption{\label{tab:alp}
     Overview of the number of data events in comparison to the expected number of background events from the search for prompt ALPs.
     }
     \begin{tabular}{l l l}
         \hline\hline
         ALP mass [GeV]  & Data & Background \\
         \hline
         5 & 2 & $0.35\pm0.27$\\
         10 &0 & $0.24\pm0.18$\\
         15 & 1 & $0.24\pm0.28$\\
         20 & 0 & $0.24\pm0.28$\\
         25 & 0 & $0.24\pm0.28$\\
         30 & 0 & $0.60\pm0.27$\\
         35 & 0 & $1.31\pm0.76$\\
         40 & 2 & $1.55\pm0.94$\\
         45 & 2 & $3.23\pm0.95$\\
         55 & 7 & $2.84\pm1.51$\\
         62 & 4 & $4.52\pm1.50$\\
         \hline
         \hline
     \end{tabular}
 \end{table}

\end{appendix}


\begin{thebibliography}{99}

\bibitem{higgs1}
    ATLAS Collaboration, Phys. Lett. B 716 (2012) 1, arXiv:1207.7214.
\bibitem{higgs2}
    CMS Collaboration, Phys. Lett. B 716 (2012) 30, arXiv:1207.7235.
\bibitem{asimov}
    G. Cowan, K. Cranmer, E. Gross, and O. Vitells, Eur. Phys. J. C 71 (2011) 1554, Eur. Phys. J. C 73 (2013) 2501 (Erratum), arXiv:1007.1727
\bibitem{Wald}
    A. Wald, \emph{Tests of Statistical Hypothesis Concerning Several Parameters When the Number of Observations is Large}, Transactions of the American Mathematical Society, Vol. {\bf 54}, No. 3, pp. 426-482.
\bibitem{higher_order}
    E. Canonero, A. R. Brazzale, and G. Cowan, Eur. Phys. J. C 83 (2023) 1100.
\bibitem{githubxia}
    \url{https://github.com/xialigang/new_asymptotic_formulae}
\bibitem{LQxia}
    L.-G. Xia, JHEP {\bf08} (2021) 071, arXiv:2012.15618, version~1.
\bibitem{book_cowan}
    G. Cowan, Statistical Data Analysis, Clarendon Press, Oxford, 1998.
\bibitem{xia_constraint}
    L.-G. Xia, J. Phys. {\bf G 46} (2019) 085004, arXiv:1805.03961.
\bibitem{beyond}
    S. Algeri, J. Aalbers, K. D. Mor\r{a}, and J. Conrad, Nature Rev.Phys. 2 (2020) 5, 245-252, arXiv: 1911.10237.
\bibitem{Hyy2022}
    ATLAS Collaboration, JHEP 07 (2023) 088, arXiv: 2207.00348.
\bibitem{CLs_Zech}
    G. Zech, Nucl. Instrum. Meth. {\bf A 277} (1989) 608.
\bibitem{CLs}
    A. L. Read, J. Phys. {\bf G 28} (2002) 2693.
\bibitem{sbottom_ATLAS2019}
    ATLAS Collaboration, JHEP {\bf12} (2019) 060, arXiv:1908.03122.
\bibitem{sbottom_hepdata}
    \url{https://www.hepdata.net/record/ins1748602}
\bibitem{alp_ATLAS2024}
    ATLAS Collaboration, Eur. Phys. J. {\bf C 84} (2024) 742, arXiv:2312.03306.
\bibitem{alp_hepdata}
    \url{https://www.hepdata.net/record/ins2731621}
\bibitem{pyhf}
    L. Heinrich, M. Feickert, G. Stark, and K. Cranmer, J. Open Source Softw. {\bf6} (2021) 58, 2823.


\end{thebibliography}
\end{document}